\documentclass[sigconf]{acmart}
\usepackage{amsmath,amssymb,amsfonts,dsfont,multirow,multicol,epsfig,url,array,makecell,balance,color,epstopdf}
\usepackage{algorithmic}
\usepackage[ruled, vlined, linesnumbered]{algorithm2e}
\usepackage{hhline}
\usepackage{array}
\usepackage{enumerate}
\usepackage{enumitem}
\usepackage{subfigure}
\usepackage{booktabs}
\usepackage{xcolor,colortbl}
\setlist[itemize]{leftmargin=*}

\def\header{\vspace{1.5mm} \noindent}

\def\tblcapup{\vspace{-1mm}}
\def\tblcapdown{\vspace{-1mm}}

\newcommand{\pushright}[1]{\ifmeasuring@#1\else\omit\hfill$\displaystyle#1$\fi\ignorespaces}
\newcommand{\pushleft}[1]{\ifmeasuring@#1\else\omit$\displaystyle#1$\hfill\fi\ignorespaces}

\def\e{\varepsilon}

\def\E{\mathrm{E}}
\def\Var{\mathrm{Var}}

\def\dout{d_{out}}

\def\a{\alpha}

\def\epi{\hat{\pi}}

\def\pprfull{Personalized PageRank\xspace}

\AtBeginDocument{%
  \providecommand\BibTeX{{%
    \normalfont B\kern-0.5em{\scshape i\kern-0.25em b}\kern-0.8em\TeX}}}

\allowdisplaybreaks
\copyrightyear{2020}
\acmYear{2020}
\setcopyright{acmlicensed}\acmConference[KDD '20]{Proceedings of the 26th ACM SIGKDD Conference on Knowledge Discovery and Data Mining}{August 23--27, 2020}{Virtual Event, CA, USA}
\acmBooktitle{Proceedings of the 26th ACM SIGKDD Conference on Knowledge Discovery and Data Mining (KDD '20), August 23--27, 2020, Virtual Event, CA, USA}
\acmPrice{15.00}
\acmDOI{10.1145/3394486.3403108}
\acmISBN{978-1-4503-7998-4/20/08}

\begin{document}
\fancyhead{}
\title{Personalized PageRank to a Target Node, Revisited}
\subtitle{[Technical Report]}

\author{Hanzhi Wang}
	\email{hanzhi_wang@ruc.edu.cn}
	\affiliation{
		\institution{School of Information\\ Renmin University of China}
	}
	
	\author{Zhewei Wei}
	\authornote{Zhewei Wei is the corresponding author. Work partially done at Beijing Key Laboratory of Big Data Management and Analysis Methods, and at Key Laboratory of Data Engineering and Knowledge Engineering, MOE, Renmin University of China.}
	\email{zhewei@ruc.edu.cn}
	\affiliation{
		\institution{Gaoling School of Artificial Intelligence\\ Renmin University of China}
	}
	
	\author{Junhao Gan}
	\email{junhao.gan@unimelb.edu.au}
	\affiliation{
		\institution{School of Computing and Information Systems\\University of Melbourne}
	}
	
	\author{Sibo Wang}
	\email{swang@se.cuhk.edu.hk}
	\affiliation{
		\institution{Department of Systems Engineering and Engineering Management\\ The Chinese University of Hong Kong}
	}
	
	\author{Zengfeng Huang}
	\email{huangzf@fudan.edu.cn}
	\affiliation{
		\institution{School of Data Science\\ Fudan University}
	}

%\renewcommand{\shortauthors}{Wang, et al.}

\begin{abstract}
  Personalized PageRank (PPR) is a widely used node proximity measure
  in graph mining and network analysis. Given a source node $s$ and a target node $t$, the PPR value
  $\pi(s,t)$ represents the probability that a random walk from $s$ terminates
  at $t$, and thus indicates the bidirectional importance between $s$ and $t$. 
The majority of the existing work focuses on the  single-source
queries, which asks for the PPR value of  a given source node $s$ and
every node $t \in V$. However, the single-source query only reflects the
importance of each node $t$ with respect to $s$. In this paper, we
consider the {\em single-target PPR query}, which measures the opposite
direction of importance for PPR.  Given a
target node $t$, the single-target PPR query asks for the PPR 
value of every node $s\in V$ to a given target node
$t$. We propose RBS, a novel algorithm that answers approximate
single-target queries with optimal computational complexity. We
show that RBS improves three
concrete applications:  heavy hitters PPR
query, single-source SimRank computation, and scalable graph neural
networks. We conduct experiments to demonstrate that RBS
outperforms the state-of-the-art algorithms in terms of both
efficiency and precision on real-world benchmark datasets.

% Single-target query has various concrete applications, including
% heavy hitters PPR query, single-source SimRank computation and
% scalable graph neural networks.

%   of a target node $t$ with respect to a source node $s$ is defined as the probability that {\em an $\alpha$-discounted} random walk from node $s$ terminates at $t$. 

%   PPR has widespread applications in the area of data mining,
%   including  web search~\cite{JehW03}, social
%   networks~\cite{gupta2013wtf}, graph neural
%   networks~\cite{Xu2019PPRGo,klicpera2019GDC}, and graph
%   representation
%   learning~\cite{ou2016asymmetric,tsitsulin2018verse,wei2019strap},
%   and thus has drawn increasing attention during the past years.

%   In this
%   paper, we consider

\end{abstract}

%%% Local Variables:
%%% mode: latex
%%% TeX-master: "paper"
%%% End:




\maketitle

\vspace{+6mm}
\section{Introduction} \label{sec:intro}

\begin{table*} [t]
	\centering
	\renewcommand{\arraystretch}{1.5}
	\begin{small}
		%\vspace{-5mm}
		%\tblcapup
		\caption{Complexity of single-source and single-target PPR queries.}\label{tbl:intro-compare}
		\vspace{-3mm}
		%\tblcapdown
		\begin{tabular} {|c|c|c|c|c|c|} \hline
			%	&  \multirow{3}{*}{ Single-Source } &\multicolumn{2}{c|}{Query Time (Power-Law Graphs)} & Space Overhead& Preprocessing Time \\ \hline
			~ &  \multirow{2}{*}{ Single-Source PPR} & \multicolumn{4}{c|}{Single-Target PPR} \\ \cline{3-6}
			~ & 														& \multicolumn{2}{c|}{Random target node} & \multicolumn{2}{c|}{Worst case} \\ \cline{2-6}
			~ & 	Monte-Carlo~\cite{fogaras2005towards}													& Backward Search~\cite{lofgren2013personalized} & Ours & Backward Search~\cite{lofgren2013personalized} & Ours  \\ \hline
			Relative error & $\tilde{O} \left( \frac{1}{\delta} \right)$ & $O\left( \frac{\bar{d}}{\delta} \right)$  & $\tilde{O} \left( \frac{1}{\delta} \right)$ &  $O \left( \sum_{u \in V} \frac{d_{out}(u) \cdot \pi(u,t)}{\delta} \right)$ & $\tilde{O} \left( \frac{n \pi(t)}{\delta} \right)$  \\ \hline
			Additive error & $\tilde{O}\left(\frac{1}{\e^2} \right)$ & $O\left( \frac{\bar{d}}{\e} \right)$ & $\tilde{O}\left(\frac{\sqrt{\bar{d}}}{\e} \right)$ &  $O\left(\sum_{u \in V} \frac{d_{out}(u) \cdot \pi(u,t)}{\e} \right)$ & $\tilde{O}\left( \sum_{u \in V} \frac{\sqrt{d_{out}(u)} \cdot \pi (u,t)}{\e} \right)$ \\ \hline
		\end{tabular}
	\end{small}
	%\normalsize
	%\tbldown
	\vspace{-3mm}
 \end{table*}

\textit{\pprfull(PPR)}, as a variant of PageRank~\cite{page1999pagerank}, focuses on the relative significance of a target node with respect to a source node in a graph.
Given a directed graph $G=(V,E)$ with $n$ nodes and $m$ edges, the PPR value $\pi(s,t)$ of a target node $t$ with respect to a source node $s$ is defined as the probability that {\em an $\alpha$-discounted} random walk from node $s$ terminates at $t$. Here an $\alpha$-discounted random walk represents a random traversal that, at each step, either terminates at the current node with probability $\alpha$, or  moves to a random out-neighbor with probability $1-\alpha$.
% probability or move to one of current node's out-neighbors randomly. %with the left $(1-\alpha)$ probability.
For a given source node $s$, the PPR value of each node $t$ sum up to $\sum_{t \in V} \pi(s,t) =1$, and thus $\pi(s,t)$ reflects the significance of node $t$ with respect to the source node $s$. On the other hand, PPR to a target node can  be related to PageRank: the summation of PPR from each node $s\in V$ to a given target node $t$ is $\sum_{s \in V}\pi(s,t)=n\cdot \pi(t)$, where $\pi(t)$ is the PageRank of $t$~\cite{page1999pagerank}.
Large $\pi(s,t)$ also shows the great contribution $s$ made for $t$'s PageRank, the overall importance of $t$. Therefore, $\pi(s,t)$ indicates bidirectional importance between $s$ and $t$.

PPR has widespread applications in the area of data mining, including  web search~\cite{JehW03}, spam detection~\cite{andersen2008robust}, social networks~\cite{gupta2013wtf}, graph neural networks~\cite{Xu2019PPRGo,klicpera2019GDC}, and graph representation learning~\cite{ou2016asymmetric,tsitsulin2018verse,wei2019strap}, and thus has drawn increasing attention during the past years.
%However, the definition formula of PPR
%Meanwhile, it has been widely used as a fundamental tool in massive algorithms, such as SimRank computation~\cite{wei2019prsim}, heavy hitters detection~\cite{wang2018heavyhitters}, graph embeddings~\cite{tsitsulin2018verse,wei2019strap}, GNN learning~\cite{Xu2019PPRGo, Klicpera2018APPNP} and so forth.
%Hence, the improvement of PPR calculation efficiency affects comprehensive work and application fields.
Studies on PPR computations can be broadly divided into four categories: 
1) single-pair query, which asks for the PPR value of a given source node $s$ and a given target node $t$; 
2) single-source query, which asks for the PPR value of  a given source node $s$ to every node $t \in V$ as the target; 
3) single-target query, which asks for the PPR value of every node $s\in V$ to a given target node $t$. 
4) all-pairs query, which asks for the PPR value of each pair of nodes. 
While single-pair and single-source queries have been extensively studied~\cite{Lofgren2014FASTPPR,lofgren2015personalized,wei2018topppr,WYXWY17}, single-target PPR query is less understood due to its hardness. 
In this paper, we study the problem of efficiently computing the single-target PPR query with error guarantee. We demonstrate that this problem is a primitive of both practical and theoretical interest.

% Reviewing previous work, we can classify them into three categories: single-pair PPR with two given nodes, single-source/target PPR with fixed source/target node and all-pairs PPR for any pair of nodes.
% Among them, all-pairs PPR is generally derived by repeating single source/target PPR for each node.
% Single-pair PPR's calculation is always divided into single source and single target PPR's combination.

% However, most previous work focus on single-

% Therefore, single-source/target PPR plays a vital role in PPR's relative researches.
% In this paper, we concentrate on single-target PPR's computation and propose an algorithm RBS to improve its calculation efficiency.
% Our motivations and contributions are summaries in section~\ref{subsec:motivations} and ~\ref{subsec:contributions}.

\vspace{-5mm}
\subsection{Motivations and Concrete Applications}
\label{subsec:motivations}
We first give some concrete applications of the single-target PPR query. We will elaborate on how to use our single-target PPR algorithm to improve the complexity for these applications in Section~\ref{sec:applications}.
% Although not as extensively studied as the single-source query, We now provide three concrete applications.

% Single-target PPR has numerous application fields such as SimRank calculation, heavy-hitters finding, all-pairs PPR in graph embeddings/GNN and so on.

\vspace{+0.25mm}
\header{\bf Approximate heavy hitters in PPR.}
The  heavy hitters PPR problem~\cite{wang2018heavyhitters} asks for all nodes $s \in V$ such that  $\pi(s,t)>\phi \cdot n \pi(t)$ with a given node $t$ and a parameter $\phi$.
As opposite to the single-source PPR query, which asks for the important nodes for a given source node $s$, heavy hitters PPR query asks for the nodes $s \in V$ for which $t$ is important.
The motivation of the heavy hitters PPR query is to consider the opposite direction of importance as a promising approach to enhance the effectiveness of recommendation. 
Intuitively, the single-target query is a generalization of heavy hitters PPR query.

% that while single-source PPR queries asks for the important nodes for an given source node $s$, heavy hitters query asks for the nodes

% The goal is to  return  $\forall u\in V$ satisfying $\pi(s,t)>\phi \cdot n \pi(t)$ with given target node $t$ and parameter $\phi$.
% The main idea given in ~\cite{wang2018heavyhitters} is to derive the single-target PPR and the PageRank of given target node firstly.
% Then pick out the qualified node as $\phi$-heavy hitters with regard to target node $t$.
% Hence, in the heavy hitters finding problem, we need to obtain the results of single-target PPR inevitably.

\vspace{+0.25mm}
\header{\bf Approximate single-source SimRank.}
SimRank is a widely used node similarity measure proposed by Jeh and Widom\cite{JW02}. Compared with PPR, SimRank is symmetric and thus is of independent interest in various graph mining tasks~\cite{Li2013MapReduce,LLink,Jin11,Liben-NowellK03,SpirinH11}. 
A large number of works~\cite{fogaras2005towards,jiang2017reads,KMK14,LeeLY12,LiFL15,liu2017probesim,MKK14,SLX15,TX16,YuPartial-pairs} focus on the single-source SimRank query, which asks for the SimRank similarity between a given node $u$ and every other node $v \in V$.  
Following~\cite{SLX15}, we can formulate SimRank  in the framework of $\alpha$-discounted random walks. 
In particular, if we revert the direction of every edge in the graph, the SimRank similarity $s(u,v)$ of node $u$ and $v$ equals to the probability that  two $\alpha$-discounted random walks from $u$ and $v$ visit at the same node $w$ with the same steps. 
As a result, it is shown in~\cite{wei2019prsim} that the bottleneck of the computational complexity of single-source SimRank depends on  how fast we can compute the single-target PPR value for each node $v$ and the target node $w \in V$. 
Hence,  by improving the complexity of single-target PPR query, we can also improve the performance of the state-of-the-art single-source SimRank algorithms. % We will provide a detailed discussion on single-source SimRank in  Section~\ref{sec:applications}.

% PRSim~\cite{wei2019prsim}, the state-of-the-art SimRank calculation method to our knowledge, rewrites the SimRank formula as corresponding PPR's multiplication.
% One of product terms is in the format of single-target PPR.

\vspace{+0.25mm}
\header{\bf Approximate PPR matrix and graph neural networks.}
In recent years, graph neural networks have drawn increasing attention due to their applications in various machine learning tasks. Graph neural networks focus on  learning a low-dimensional latent representation for each node in the graph from the structural information and the node features. 
Many graph neural network algorithms are closely related to the approximate PPR matrix problem, which computes the approximate PPR value for every pair of nodes $s,t\in V$.  
For example,  a few unsupervised  graph embedding methods, such as HOPE~\cite{ou2016asymmetric}  and STRAP~\cite{wei2019strap}, suggest that directly computing and decomposing the approximate PPR matrix into low-dimensional vectors achieves satisfying performance in various downstream tasks.  
On the other hand, several recent works on semi-supervised  graph neural networks, such as APPNP~\cite{Klicpera2018APPNP}, PPRGo~\cite{Xu2019PPRGo},  and GDC~\cite{klicpera2019GDC},  propose to use the (approximate) PPR matrix to smooth the node feature matrix. 
It is shown~\cite{klicpera2019GDC} that the approximate PPR matrix  outperforms spectral methods, such as GCN~\cite{kipf2016GCN} and GAT~\cite{velikovi2017GAT}, in various applications.

The computation bottleneck for these graph learning algorithms is the computation of the approximate PPR matrix, as the power method takes at least $O(n^2)$ time and space and is not scalable on large graphs. 
On the other hand, there are two alternative approaches to compute the approximate PPR matrix:  issue a single-source query to every source node $s \in V$ to compute $\pi(s,t), t \in V$, or issue a single-target query to every target node $t \in V$ to compute $\pi(s,t), s \in V$. 
As we shall see in Section~\ref{sec:applications}, the later approach is superior as it can provide the absolute error guarantee. 
Therefore, by proposing a faster single-target PPR algorithm, we also improve the computation time of the approximate PPR matrix.  In particular, we show that our new  single-target PPR algorithm computes the approximate PPR matrix in time sub-linear to the number of edges in the graphs, which significantly improves the scalability of various graph neural networks.
%

% PPRGo and GDC~\cite{PPRGo, APPNP, GDC} suggests applying the PPR matrix rather than the spectral convolutional matrix to the feature matrix $X$ improves the performance of the graph neural networks. Thus, the computational bottleneck for these algorithms is the compassion of the (approximate) PPR matrix.

% We claim that the computation time of approximate PPR matrix is closely connected to the problem of designing scalable algorithms in graph neural networks.

% Hence, our  single-target PPR algorithm can also be employed in these graph learning algorithms.

% such that various downstream machine learning algorithms can take the representation rather than the entire graph as the input.

% I

% Thirdly, in graph embedding problems, the basic framework of most of embedding methods can be summarized into two steps, choosing a proximity measure $P(u,v)$ for $u,v \in V$ and training the embedding vector $s_u$ of each node $u \in V$ according to $s_u \cdot s_v \sim P(u,v)$.
% When PPR is selected as the proximity measure such as \cite{ tsitsulin2018verse,wei2019strap,zhou2017scalable}, all-pairs PPR is required to train the embedded vectors.
% As mentioned before, all-pairs PPR is generally derived by repeating the single-source/target PPR methods $n$ times for each node, where $n$ denote
% es the number of nodes.

\vspace{+0.25mm}
\header{\bf Theoretical motivations.}
Unlike the single-source PPR query, the complexity of the single-target PPR query remains an open problem. In particular, given a source node $s$, it is known that a simple Monte-Carlo algorithm can approximately find all nodes $t \in V$  such that $\pi(s,t) \ge \delta$  with constant probability  in $\tilde{O}(1/\delta)$ time (see Section~\ref{sec:pre} for a detailed discussion), where $\tilde{O}$ denotes the Big-Oh notation ignoring the log factors. Note that there are at most $O(1/\delta)$ nodes $t$ with $\pi(s,t) \ge \delta$, which implies that there is a lower bound of $\Omega(1/\delta)$ and thus the simple Monte Carlo algorithm is optimal. On the other hand, given a random target node $t$, the state-of-the-art single-target PPR algorithm finds all nodes $s\in V$ with $\pi(s,t) \ge \delta$ in $\tilde{O}(\bar{d}/\delta)$ time, where $\bar{d}$ is the average degree of the graph. Thus, there is an $O(\bar{d})$ gap between the upper bound and lower bound for the single-target PPR problem. For dense graphs such as complete graphs, the $O(\bar{d})$ gap is significant.  Therefore, an interesting open problem is: is it possible to achieve the same optimal complexity  as the single-source PPR query for the single-target PPR query?

\subsection{Problem defintion and Contributions }
\label{subsec:contributions}

\header{\bf Problem definition. }
In this paper,  we consider the problem of efficiently computing approximate single-target PPR queries.
Following~\cite{bressan2018sublinear}, the approximation quality is determined by relative or additive error. More specifically,
we define approximate single-target PPR with additive error as follows.

% given a target node $t$ and a threshold $\delta$, a single-target PPR query with relative error returns $\epi(s,t)$ for each node $s \in V$ such that $\left|\epi(s,t)-\pi(s,t)\right| \le \e_r \pi(s,t)$, where $\e_r$ is a constant. Given a target node $t$ and an error parameter  $\e$, a single-target PPR query with additive error returns $\epi(s,t)$ for each node $s \in V$ such that $\left|\epi(s,t)-\pi(s,t)\right| \le \e$. For both queries, we also allow a small constant failing probability $p_f$.

\begin{definition}[Approximate Single-Target PPR with additive error]
	%\caption{Node Incode, Edge Saving and Edge Expense in forwardPush}
  Given a directed graph $G=(V,E)$, a target node $t$, an additive error bound $\e$,
  an approximate single-target PPR query with additive error returns an estimated PPR value $\epi(s,t)$ for each $s \in V$, such that
	\begin{align}
	\left|\epi(s,t) - \pi(s,t)\right| \leq \e
	\end{align}
	holds with a constant probability.
\end{definition}
For  single-target PPR query with relative error, we follow the definition of~\cite{bressan2018sublinear}.

\begin{definition}[Approximate Single-Target PPR with relative error]
	%\caption{Node Incode, Edge Saving and Edge Expense in forwardPush}
  Given a directed graph $G=(V,E)$, a target node $t$, and a threshold $\delta$,
  % an constant relative error parameter $\e_r$, and a constant failure probability $p_f$,
  an approximate single-target PPR with relative error returns an estimated PPR value $\epi(s,t)$ for each $s \in V$, such that for any $\pi(s,t)>\delta$,
  \begin{equation}
      \begin{aligned}
	\left|\epi(s,t)-\pi(s,t)\right| \leq {1\over 10} \cdot \pi(s,t)
	%\epi(u,t) \leq (1 \pm O(1)) \cdot \pi(u,t)
	\end{aligned}
  \end{equation}
	holds with a constant probability.
\end{definition}

Note that to simplify the presentation, we assume that the relative error parameter and success probability are constants following~\cite{bressan2018sublinear}.  
We can boost the success probability to  arbitrarily close to $1$ with the Median-of-Mean trick~\cite{charikar2002finding}, which only adds a log factor to the running time. 
For these two types of error, we propose Randomized Backward Search (RBS), a unified algorithm that achieves optimal complexity for the single-target PPR query. 
We summarize the properties of the RBS algorithm as follows.

% that achieves the following properties: $\tilde{O}({1 / \delta})$ complexity
%Motivated by this, we revise the Backward Push algorithm and achieve the same time cost as single-source PPR computation does when the target node is chosen randomly.
%Even in the worst case that the target node is chosen according to their degrees,
%we can still spare at least $\sqrt{d}$ time cost compared with traditional Backward Search method.
%\header{\bf Time complexity reduction for single-target PPR calculation.}

 %Time complexity reduction for single-target PPR calculation
\begin{itemize}
  \item
Given a target node $t$, RBS answers a single-target PPR query with constant relative error for all $\pi(s,t) \ge \delta$ with constant probability using  $\tilde{O} \left( \frac{n \pi(t)}{\delta} \right)$ time. 
This result suggests that RBS achieves optimal time complexity for the single-target PPR query with relative error, as there may be $O\left( \frac{n\pi(t)}{\delta}\right)$ nodes with $\pi(s,t) \ge \delta$.

\item
Given a  random target node $t$, RBS answers a single-target PPR query with an additive error  $\e$ with constant probability using  $\tilde{O} \left( \frac{\sqrt{\bar{d}}}{\e} \right)$ time. This query time complexity improves previous bound for single-targe PPR query with additive error by a factor of $\sqrt{\bar{d}}$. Table~\ref{tbl:intro-compare} presents a detailed comparison between RBS and the state-of-the-art single-target PPR algorithm.
\end{itemize}

We demonstrate that the RBS algorithm improves the complexity of single-source SimRank computation, heavy hitters PPR query, and PPR-related graph neural networks in Section~\ref{sec:applications}.
We also conduct an empirical study to evaluate the performance of RBS.
The experimental results show that  RBS outperforms the state-of-the-art single-target PPR algorithm on real-world datasets.

\section{Preliminary} \label{sec:pre}

\begin{table} [t]
	\centering
	\renewcommand{\arraystretch}{1.3}
	\begin{small}
		\tblcapup
		\caption{Table of notations.}\label{tbl:def-notation}
		\vspace{-1mm}
		%\tblcapdown
		%p{2.3in}
		\begin{tabular} {|l|p{2.2in}|} \hline
			{\bf Notation} &  {\bf Description}  \\ \hline
			$n, m$      &   the numbers of nodes and edges in $G$                            \\ \hline
			$N_{in}(u), N_{out}(u)$	& 	the in/out neighbor set of node $u$	\\ \hline
			$d_{in}(u), d_{out}(u)$ & the in/out degree of node $u$ \\ \hline
			%$\pi(s,t), \epi(s,t)$	& 	the true and estimated PPR values of node $t$ with respect to $s$. \\ \hline
			$\pi(s,t), \epi(s,t)$	& 	the true and estimated PPR values of node $s$ to $t$. \\ \hline
            %$\pi_\ell(s,t), \epi_\ell(s,t)$ & the true and estimated $\ell$-hop PPR values of node $t$ with respect to $s$. \\ \hline
            $\pi_\ell(s,t), \epi_\ell(s,t)$ & the true and estimated $\ell$-hop PPR values of node $s$ to $t$. \\ \hline
			%$\pi_{\ell}(s,t), r_{\ell}(s,t)$	& 	the reserve and residue of $t$ during $\ell$-hop PPR push from $s$.\\ \hline
			%$X_{\ell}(u,v)$	&	the backward push's increments from node $u$ (in level $\ell$) to node v (in level $\ell+1$)\\ \hline
			%$C_{\ell}(u,v)$	&	cost in the backward push from node $u$ (in level $\ell$) to node $v$	\\ \hline
			%$r_s^f(u), r_t^b(u)$	&	the Node Income	of $u$ in forward / backward Push start from node $s / t$\\ \hline
			%$r_s^f(u,v), r_t^b(u,v)$	&	the Edge Saving	of edge ($u,v$) in forward / backward Push start from node $s / t$\\ \hline
			%$p_s^f(u,v), p_t^b(u,v)$	&	the Edge Expense of edge ($u,v$) in forward / backward Push start from node $s / t$\\ \hline			
			$\alpha$	& the teleport probability that a random walk terminates at each step \\ \hline
			%$c$          &   the decay factor in the definition of SimRank                   \\ \hline
			$\e$         &   the additive error parameter     \\ \hline
			%$\e_r$         &  relative error parameter \\ \hline
			$\delta$         &  the relative error threshold \\ \hline
			$\bar{d}$		& the average degree,
                                           $\bar{d}=\frac{m}{n}$ \\
                  \hline
                  $\tilde{O}$&  the Big-Oh notation ignoring the log factors \\ \hline
			%$P$, $D$   & the transition matrix and the diagonal correction matrix\\ \hline
			%$\vec{\pi}_i, \vec{\pi}_i^\ell,$   & the Personalized PageRank and $\ell$-hop Personalized PageRank vectors of node $v_i$\\ \hline
			%$ \vec{h}_i^\ell$   &  the $\ell$-hop Hitting Probability vector of $v_i$\\	\hline
			
			%     $\rf(s,t)$, $\pif(s,t)$ & The reserve and residue of $t$ from $s$ in the forward search \\
			%     \hline
			%     $\frsum$ & The sum of all nodes' residues during in the forward
			%                search from $s$\\
			% \hline
			%   $h^{\l}(v_i, v_j)$ & the hitting probability (HP) from node $v_i$ to node $v_j$ at step $\l$ (see Section~\ref{sec:our-overview}) \\ \hline
		\end{tabular}
	%\vspace{-2mm}
	\end{small}
\end{table}

\subsection{Existing Methods} \label{sec:main_competitors}

% \begin{comment}
% \header{\bf Subset Sampling.}
% %~\cite{fogaras2005towards}
% 	Given a set $S=\{x_1,x_2,x_3,...,x_s\}$ of $s$ elements, each $x_i \in S$ has a selected probability $p(x_i)$ and $\mu=\sum_{i=1}^{s} p(x_i)$.
% 	In the sampling process, each $x_i$ need to be chosen with probability $p(x_i)$.
% 	The basic method to fulfill this sampling process is to generate random numbers for each elements in set $S$.
% 	This costs $O(n)$ time and is treated as an upper bound.
% 	%Subset Sampling wants to

% \end{comment}

%\header{\bf Monte-Carlo.}
%~\cite{fogaras2005towards}

%\header{\bf Forward Search.}
%~\cite{AndersenCL06}

\header{\bf Power Method} is an iterative method for computing
single-source and single-target PPR queries~\cite{page1999pagerank}. Recall that, at each step, an
$\alpha$-discounted random walk terminates at the current node with
$\alpha$ probability or moves to a random out-neighbor with $(1-\alpha)$ probability.
This process can be expressed as the iteration formula with single-source PPR vector. % an iterative multiplication formula shown as below.
%According to the definition of PPR,
\begin{align}
\label{eqn:power_method_ss}
\vec{\pi_s}=(1-\alpha)\vec{\pi_s}\cdot \mathbf{P}+\alpha \cdot \vec{e_s},
\end{align}
where $\vec{\pi_s}$ denotes the PPR vector with respect to a given
source node $s$,  $\vec{e_s}$ denotes the one-hot vector with
$\vec{e_s}(s)=1$, and
$\mathbf{P}$ denotes the transition matrix where
%\vspace{-1mm}
\begin{equation}%\nonumber
\label{eqn:transition_matrix_ss}
\begin{aligned}
P(i,j)
&=\left\{
\begin{array}{ll}
\frac{1}{d_{out}(v_i)}, &if \quad v_j \in N_{out}(v_i),  \\
0, &otherwise.
\end{array}
\right.\\
\end{aligned}
\end{equation}
Reversing this process, we can also compute single-target PPR values with the given target node $t$.
The iteration formula should be adjusted correspondingly:%as below:
\vspace{-1mm}
\begin{align}
\label{eqn:power_method_st}
\vec{\pi_t}=(1-\alpha)\vec{\pi_t}\cdot \mathbf{P}^\top+\alpha \cdot \vec{e_t}.
%\vspace{-4mm}
\end{align}

% \begin{align}
% \label{eqn:power_method_st}
% \vec{\pi_t}=(1-\alpha)\vec{\pi_t}\cdot \mathbf{P_t}+\alpha \cdot \vec{e_t}
% \end{align}
% $\mathbf{P_t}$ is the transition matrix and
% \begin{equation}%\nonumber
% \label{eqn:transition_matrix_st}
% \begin{aligned}
% P(i,j)
% &=\left\{
% \begin{array}{ll}
% \frac{1}{d_{out}(v_j)}, &if \quad v_j \in N_{in}(v_i)  \\
% 0, &otherwise
% \end{array}
% \right.\\
% \end{aligned}
% \end{equation}
%Repeat the iteration more than times to obtain approximate single-target PPR with $\e$ additive error.
Power Method can be used to compute the ground truths for
the single-source and single-target query. After $\ell= \log_{1-\alpha}(\e)$ iterations, the absolute error can be
bounded by $(1-\alpha)^\ell = \e$. Since each iteration takes $O(m)$
time, it follows that the Power Method computes the approximate
single-target PPR query with additive error in $O\left( m\cdot \log{1
    \over \e} \right)$ time. Note that the dependence on the error
parameter $\e$ is logarithmic, which implies that the Power Method can
answer single-target PPR queries with high precision. However, the
query time also linearly depends on the number of edges, which limits
its scalability on large graphs.

% The above two equations cost the same time and the analysis of time complexity is the same.
% Taking single-target PPR as an example, the cost of one multiplication is $O(m)$.
% Meanwhile, equation~\eqref{eqn:power_method_st} is required to repeat $\log_{1-\alpha}(\e)$ times to derive the approximate single-target PPR with $\e$ additive error.
% Therefore, the time complexity of the two iteration method is $O\left( m\cdot \log{1 \over \e} \right)$, which is infeasible on large graph.
% with $\e$ additive error.

\header{\bf Backward Search}~\cite{lofgren2013personalized} is a local search method that efficiently
computes the single-target PPR query on large
graphs.
Algorithm~\ref{alg:bp} illustrates the pseudo-code of Backward
Search. We use residue
$r^b(s,t)$ to denote  the probability mass to be distributed at
node $s$, and reserve $\pi^b(s,t)$ denotes the probability mass that will stay at
$s$ permanently.   %PPR between $s$ and $u$.
%For initialization, Backward Search sets the residue $r^b(t,t) =1$, and  $r^b(u,t)=\pi^b(u,t)=0$ for $\forall u \in V$.
For initialization, Backward Search sets $r^b(u,t)=\pi^b(u,t)=0$ for $\forall u \in V$, except for the residue $r^b(t,t) =1$.
In each push operation, it picks the node $v$ with the largest residue
$r^b(v,t)$, and transfer a fraction of $\alpha$ to $\pi^b(v,t)$, the
reserve of $v$. Then the algorithm transfers the other $(1-\alpha)$
faction to the in-neighbors of $v$. 
For each in-neighbor $u$ of $v$,
the residue  $r^b(u,t), u \in N_{in}(v)$ is incremented  by
$\frac{(1-\alpha)r^b(v,t)}{d_{out}(u)}$. After all in-neighbors are
processed, the algorithm sets the residue of $v$ to be $0$. The
process ends when the maximum residue descends below the error
parameter $\e$. Finally, Backward Search uses the reserve $\pi^b(s,t)$
as the estimator for $\pi(s,t)$, $s\in V$.
Backward Search utilizes the following property of the single-target
PPR vector.

% transmit $(1-\alpha)r^b(u,t)$ to its in-neighbors.
% Each $r^b(v,t), v \in N_{in}(u)$ is incremented  by $\frac{(1-\alpha)r^b(u,t)}{d_{out}(v)}$.
% Meanwhile, we transfer $\pi^b(u,t)$ is added by $\alpha \cdot r^b(u,t)$.
% Repeat the above process until $r^b(u,t) <\e$ for each $u \in V$.
% It regards $\pi^b(u,t), \forall u \in V$ as the estimator of $\pi(u,t)$.

\begin{algorithm}[t]
%\renewcommand{\arraystretch}{1.3}
%\begin{small}
\caption{Backward Search~\cite{lofgren2013personalized} } \label{alg:bp}
\BlankLine
\KwIn{Graph $G = (V, E)$, target node $t$, teleport probability $\a$, additive error parameter $\e$ }
\KwOut{Reserve $\pi^b(s,t)$ for all $s\in V$}
\For {each $u \in V$}
{
  $r^b(u,t), \pi^b(u,t) \gets 0$\;
}
$r^b(t, t) \gets 1$\;
\While {The largest $r^b(v,t)> \e$}
{
  $\pi^b(v,t) \gets \pi^b(v,t) + \alpha  \cdot r^b(v, t)$\;
    \For {each $u \in \mathcal{N}^{in}(v)$}
    {
       $r^b(u, t) \gets r^b(u, t) + (1-\alpha) \cdot \frac{ r^b(v, t)}{d_{out}(u)}$
    }
    $r^b(v,t) \gets 0$\;
  }
  \Return   $\pi^b(s,t)$ as the estimator for   $\pi(s,t), s\in V$\;
%\end{small}
\end{algorithm}

\vspace{-1 mm}
\begin{proposition}
	\label{pro:onestepPPR}
	Denote $I\{u=v\}$ as the indicator variable such that
        $I\{u=v\}=1$ if $u=v$.
	For $\forall s,t \in V$, $\pi(s,t)$ satisfies that
	\begin{equation}
	\begin{aligned}
	\pi(s,t)=\sum_{u \in N_{in}(t)}\frac{1-\alpha}{d_{out}(u)} \cdot \pi(s,u) +\alpha \cdot I\{s=t\}.
	\end{aligned}	
	\end{equation}
	\vspace{-3 mm}
\end{proposition}
%\vspace{-2 mm}
%In the algorithm, it maintains variables $r^{b}(u,t)$ and $\pi(u,t)$ for each $u \in V$.
Utilizing this property, it is shown in
~\cite{lofgren2013personalized} that the residues and reserves of 
Backward Search satisfies the following invariant:
\begin{align}\label{eqn:BS_error}
\vspace{-1 mm}
\pi(s,t)=\pi^b(s,t)+\sum_{u \in V} r^b(u,t) \cdot \pi(s,u).
\vspace{-1 mm}
\end{align}
Note that when the Backward Search algorithm terminates, all residues
$r^b(u,t)  \le \e$. 
It follows that $\pi^b(s,t) \le \pi(s,t) \le \pi^b(s,t)+ \e \sum_{u \in V}\pi(s,u) = \pi^b(s,t)+ \e$, 
where $\sum_{u \in V}\pi(s,u)=1$. 
Therefore, Backward Search ensures an additive error of $\e$. 
It is shown in~\cite{lofgren2013personalized} that the running time of Backward Search is bounded by  
%Despite the running time bound of Backward Search $$ given in~\cite{lofgren2013personalized}, 
%$O \left( \sum_{u \in V} \frac{d_{out}(u)\cdot \pi(u,t)}{\e} \right)$. 
$O \left( \sum_{u \in V} \frac{d_{in}(u)\cdot \pi(u,t)}{\e} \right)$.
%We claim that the running time of Backward Search can also be bounded by $O \left( \sum_{u \in V} \frac{d_{out}(u)\cdot \pi(u,t)}{\e} \right)$. 
We claim that the running time of Backward Search can also be bounded in term of out degree given in the following lemma.
We defer the proof of Lemma~\ref{lem:BS_bound} to the appendix.
\begin{lemma}\label{lem:BS_bound}
\vspace{-1 mm}
	The running time of the algorithm Backward Search given in~\cite{lofgren2013personalized} can also be bounded by $O \left( \sum_{u \in V} \frac{d_{out}(u)\cdot \pi(u,t)}{\e} \right)$. 	
\end{lemma}
%Note that in each push operation from one of node $u$'s out-neighbors to $u$, the reserve $\pi^b(u,t)$ will be increased by $\frac{\e \cdot \alpha \left( 1-\alpha\right)}{d_{out}(u)}$ at least. 
%So the number of push operation conducted on node u will be $\frac{d_{out}(u)\cdot \pi(u,t)}{\e\cdot \alpha \left( 1-\alpha\right)}$ at most. 
%Hence, the total running time of Backward Search can be rewritten as $O \left( \sum_{u \in V} \frac{d_{out}(u)\cdot \pi(u,t)}{\e} \right)$.
According to Lemma~\ref{lem:BS_bound}, if the target node $t$ is
randomly selected, the complexity becomes $O(\frac{\bar{d}}{\e})$,
where $\bar{d}$ is the average degree of the graph. For relative
error, we can set $\delta = \Theta(\e)$ and obtain a worst-case complexity of $O \left( \sum_{u \in V} \frac{d_{out}(u)
    \cdot \pi(u,t)}{\delta} \right)$  and an average complexity of
$O(\frac{\bar{d}}{\delta})$, respectively.

\header{\bf Single-source algorithms.}
%\header{\bf Monte-Carlo.} % for single-source PPR
%The algorithm utilizing Monte-Carlo method in \cite{fogaras2005towards} can only be used to compute single-source PPR.
%Based on Monte-Carlo method,
%tends to estimate single-source PPR by sampling abundant random walks from source node $s$.
The {\em Monte-Carlo algorithm}~\cite{fogaras2005towards} computes the
approximate single-source PPR query by sampling abundant random
walks from source node $s$ and using the proportion of the
random walks that terminate at $t$ as the estimator of $\pi(s,t)$.
According to Chernoff bound, the number of random walks required for an
additive error $\e$ is $\tilde{O}(\frac{1}{\e^2})$, while  the number
of random walks required to ensure constant relative error for all PPR
larger than $\delta$ is $\tilde{O}(\frac{1}{\delta})$.
This simple method is optimal for single-source PPR
queries with relative error, as there are at most
$O(\frac{1}{\delta})$ nodes $t$ with PPR $\pi(s,t) \ge
\delta$. However, the  Monte-Carlo algorithm does not work for
single-target queries, as there lacks of a mechanism for sampling source
nodes from a given target node. Moreover, it remains an open problem
whether it is possible to achieve the same optimal $O(\frac{1}{\delta})$
complexity for the single-target query.

% obtain approximate single-source PPR is $O(\frac{1}{\e^2})$ with $\e$ additive error or $O(\frac{1}{\delta})$ with constant relative error.
% The expected length of each walk is $\frac{1}{\alpha}$ and $\alpha$ is a constant in our problem.
% Hence, the time complexity of Monte-Carlo based methods is $O\left( \frac{1}{\e^2} \right)$ or $O\left( \frac{1}{\delta} \right)$ for additive or relative error parameters.

% However, this Monte-Carlo method cannot be applied to single-target PPR unless repeating the sampling process from each node and derive the PPR with target node $t$.
% This complexity will extend to $O\left( \frac{n}{\e^2} \right) $ for additive error, which is infeasible when $n$ is large.

%\header{\bf Forward Search.}
{\em Forward Search}~\cite{AndersenCL06} is the analog of Backward
Search, but for single-source PPR queries. Similar to Backward Search, Forward Search uses residue $r^f(s,u)$ to
denote the probability mass to be distributed at node $u$, and
$\pi^f(s,u)$ to denote the probability mass  that will stay at node
$u$ permanently. 
%For initialization, Forward Search sets the residue $r^f(s,s) =1$, and  $r^f(s,u)=\pi^f(s,u)=0$ for $\forall u \in V$.
For initialization, Forward Search sets $r^f(s,u)=\pi^f(s,u)=0$ for $\forall u \in V$, except for the residue $r^f(s,s) =1$.
In each push operation, it picks the node $u$ with the largest
residue/degree ratio
$r^f(s,u)/d_{out}(u)$, and transfer a fraction of $\alpha$ to $\pi^f(s,u)$, the
reserve of $u$. Then  the algorithm transfers the other $(1-\alpha)$
faction to the out-neighbors of $u$. For each out-neighbor $v$ of $u$,
the residue  $r^f(s,v)$ is incremented  by
$\frac{(1-\alpha)r^f(s,u)}{d_{out}(u)}$. After all in-neighbors are
processed, the algorithm sets the residue of $u$ to be $0$. The
process ends when the maximum residue/degree ratio descends below the error
parameter $\e$. 
Finally, Forward Search uses the reserve $\pi^f(s,u)$ as the estimator for $\pi(s,u)$, $u\in V$.

As shown in~\cite{AndersenCL06},  Forward Search runs in $O(1/\e)$
time. However, the major problem with Forward Search is that it can
only ensure an additive error of $\e d_{out}(u)$ for each PPR value
$\pi(s, u)$ on undirected graphs. Compared to the $\e$ error bound by Backward Search, this weak error guarantee makes Forward Search
unfavorable when we need to compute the approximate PPR matrix in
various graph neural network applications. We also note that there are
a few works~\cite{wei2018topppr,WYXWY17,Wang2016HubPPR} that combines Forward Search, Backward Search and
Monte-Carlo to answer single-single PPR queries. However,
these methods are not applicable to the single-target PPR queries.

% proposes to derive the estimators of single-source PPR $\pi(s,u), \forall u \in V$ with given source node $s$ based on the below identical equation.
% \begin{align}
% \pi(s,t)=\pi^f(s,t)+\sum_{u \in V} r^f(s,u) \cdot \pi(u,t)
    %   \end{align}

\vspace{-2 mm}
\subsection{Other related work} \label{sec:other_relatedwork}
PPR has been extensively studied for the past
decades~\cite{JehW03,jung2017bepi,WuJZ14,WYXWY17,coskun2016efficient,AndersenBCHMT07,AndersenCL06,FujiwaraNSMO13,
  FujiwaraNSMO13sigmod,FujiwaraNYSO12,YuM16,FujiwaraNOK12,YuL13,fogaras2005towards,BackstromL11,SarmaMPU15,maehara2014computing,ZhuFCY13,ShinJSK15,BahmaniCX11,Chakrabarti07,
  GuoCCLL17,Ren2014clude,BahmaniCG10,OhsakaMK15,ZhangLG16,GuptaPC08,lofgren2014fast,lofgren2015personalized,LofgrenBG15}. Existing
work has studied other variants of PPR queries. Research work on exact
single-source
queries~
\cite{page1999pagerank,maehara2014computing,ZhuFCY13,ShinJSK15,jung2017bepi,Chakrabarti07}
aims at improving the efficiency and scalability of the Power Method.
Research work on single-source top-$k$
queries~\cite{ZhuFCY13,maehara2014computing,lofgren2014fast,lofgren2015personalized,LofgrenBG15,WYXWY17,WangTXYL16}
focuses on (approximately) returning $k$ nodes with the highest PPR values to a given
source node.
Single-pair PPR queries are studied by
\cite{fogaras2005towards,lofgren2014fast,lofgren2015personalized,LofgrenBG15,WangTXYL16}, which
estimates the PPR value of a given pair of nodes. PPR computation has
also been studied on dynamic graphs
\cite{Ren2014clude,BahmaniCG10,OhsakaMK15,ZhangLG16,YuM16,Chakrabarti07,reittu2019regular}
and in the distributed environment~\cite{BahmaniCX11,GuoCCLL17}. These studies, however, are orthogonal to our work.
Table~\ref{tbl:def-notation} summaries the notations used in this paper.

\section{The RBS Algorithm}
% As mentioned before, previous work which aims at single-target PPR is slower than the calculation of single-source PPR.
% %In order to speed up single-target PPR's computation method a
% So as to improve the calculation efficiency of single-target PPR, 
% we revise the traditional Backward Search and ultimately achieve the
% same time complexity as single-source PPR does when the target node is
% chosen randomly.

\header{\bf High-level ideas.} In this section, we present {\em Randomized Backward Search (RBS)},
an algorithm that achieves optimal query cost for the single-target
query. Compared to the vanilla Backward Search
(Algorithm~\ref{alg:bp}), we employ two novel techniques. First of all,
we decompose  the PPR value $\pi(s,t)$ into  the {\em $\ell$-hop Personalized PageRank} $\pi_\ell(s,t)$,
which is defined as the probability that an $\alpha$-discounted random
walk from $s$ terminates at $t$ at exactly $\ell$ steps. For different
$\ell$, such events are mutually exclusive, and thus  we can compute
the original PPR value by
$\pi(s,t)=\sum_{\ell=0}^{\infty}\pi_\ell(s,t)$.  Furthermore,  we can
truncate the summation to $\epi(s,t)=\sum_{\ell=0}^{L}\pi_\ell(s,t)$
where $L = \log_{1-\alpha}\theta$, which only adds a small additive
error $\theta$ to the final estimator and $\theta=\tilde{O}\left(\e\right)$.
Secondly, we
introduce randomization into the push operation of the Backward Search
algorithm to reduce the query cost. Recall
that in the vanilla Backward Search algorithm, each push operation transfers an
$(1-\alpha)$ faction of the probability mass from the current
node $u$ to each of its in-neighbors. This operation is expensive
as  it touches all the in-neighbors of $v$ and thus leads to the
$\bar{d}$ overhead. We avoid this complexity by pushing the
probability mass to  a small random subset of $v$'s in-neighbors. The
probability for including  an in-neighbor $u$ depends on the
out-degree of $u$. We show that this randomized push operation is
unbiased and has bounded variance, which enables us to derive
 probabilistic bounds for both additive error and relative error.

% \begin{comment}
% And each push operation will increase each $u \in N_{in}(v)$ by at least $\frac{\e}{d_{out}(u)}$.
% As shown in figure , node $u \in N_{in}(v)$ will receive the increment report from its out-neighbors and
% We can summarize this process into a count-tracking problem as expressed in~\cite{huang2012randomized}.
% There is $d_{out}(u)$ counters which is initialized by $0$ and a collecting counter to calculate the sum of its out-neighbors' .
% If we want to guarantee the additive error of the sum result obtained by the collecting counter will not exceed $\theta$, 
% each counter has to report their change to the collecting center so long as its exceeds $\frac{\e}{d_{out}(u)}$.
% However, according to the conclusion in ~\cite{huang2012randomized}, 
% we can import randomness to reduce the report times. 
% \end{comment}

%Algorithm~\ref{alg:vbs2} illustrates the pseudocode of the $\e
%\pi_\ell(s,t)$-variance $\ell$-hop backward search.
\header {\bf Sorted adjacency lists.}
Before presenting our algorithm, we assume the in-adjacency list of
each node  is sorted in the ascending order of the
out-degrees. More precisely, let $\{u_1, \ldots,u_d\}$ denote the
in-adjacency list of $v$. We assume that $d_{out}(u_1) \le \ldots, \le
d_{out}(u_d) $. We claim that it is possible to sort all the in-adjacency
lists in $O(m+n)$ time, which is asymptotically the same as
reading the graph into the memory. This  means we can pre-sort the graph as we
read the graph into the main memory without increasing the asymptotic
cost. More specifically, we construct a tuple $(u,v,d_{out}(u))$ for
each edge $(u,v) \in E$, and
use counting sort to sort $(u,v,d_{out}(u))$ tuples in the ascending
order of $d_{out}(u)$. Since each $d_{out}(u)$ is bounded by $n$, and
there are $m$ tuples, the cost of counting sort is bounded by
$O(m+n)$. 
Finally, for each $u,v, d_{out}(u)$, we append $u$ to the end of $v$'s
 in-adjacency list. This preprocessing algorithm runs in $O(m+n)$
 time, which is asymptotically the same as reading the graph
 structure. 

\begin{figure}[t]
	\begin{small}
		\centering
		%\vspace{-3mm}
		%    \begin{footnotesize}
		\begin{tabular}{c}
			%\multicolumn{4}{c}{\hspace{-4mm} \includegraphics[height=5mm]{./Figs/legend_large.eps}} \vspace{-1mm} \\
			\hspace{-3mm} \includegraphics[width=85mm]{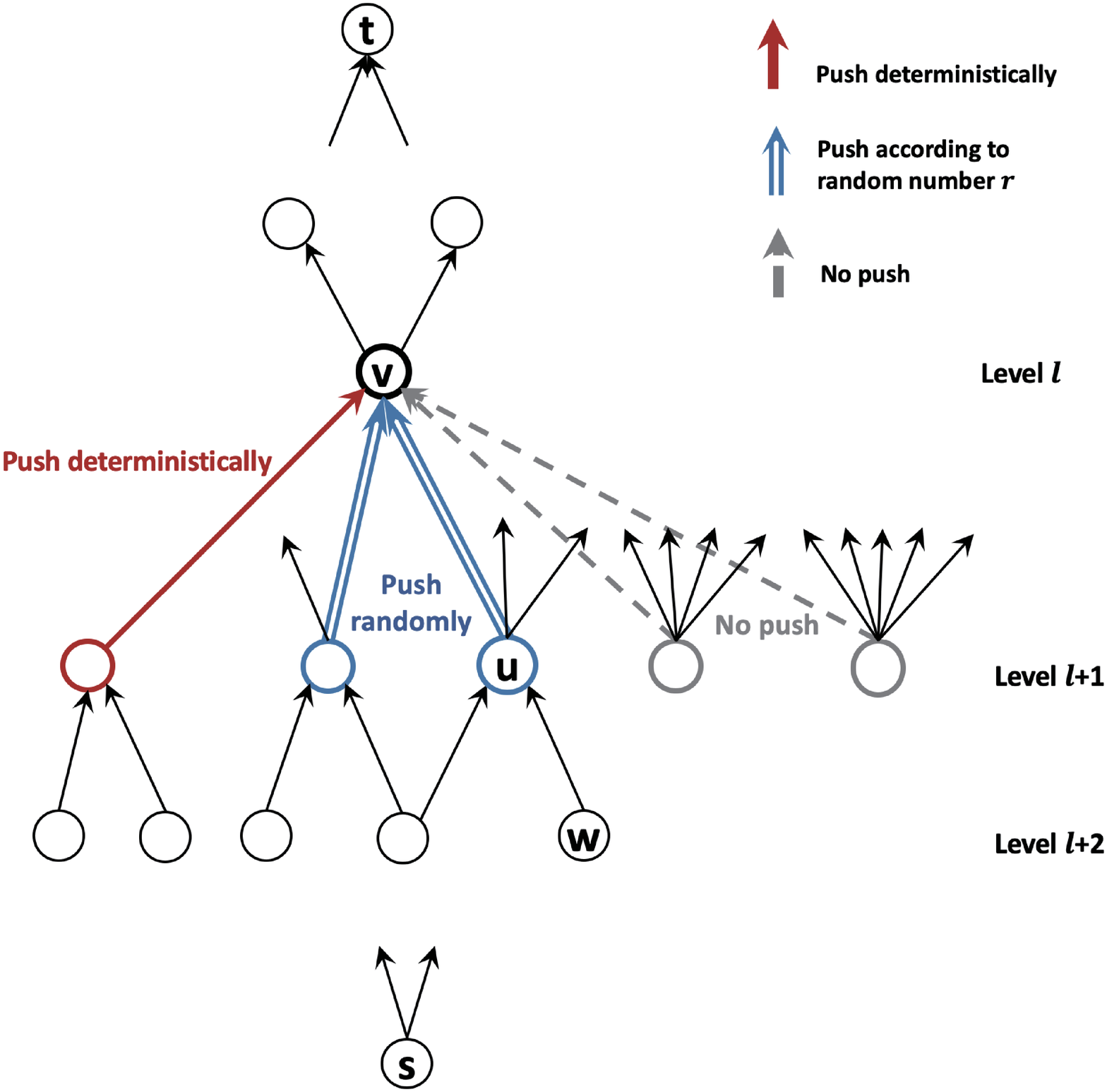} 
			%\hspace{-3mm} \includegraphics[width=80mm]{./Figs/Figure_1.png} 
			%\hspace{-3mm} \includegraphics[width=80mm]{./Figs/lena_new_sz.npy} 
		\end{tabular}
		\vspace{-3mm}
		\caption{Sketch of Randomized Backward Search} 
		\label{fig:backwardPush}
		\vspace{-3mm}
	\end{small}
\end{figure}

\header  {\bf Algorithm description.}    
Algorithm~\ref{alg:vbs2} illustrates the pseudocode of the RBS algorithm.
Consider a directed graph $G=(V,E)$, target node $t$, and a teleport probability $\alpha$. The algorithm takes in two additional parameters:  error parameter $\theta$ and sampling function $\lambda(u)$.  We can
manipulate these two parameters to obtain the additive or relative
error guarantees. Recall that  $\tilde{O}$ denotes the Big-Oh notation
ignoring the log factors. For
single-target PPR query with
additive error, we set  $\theta$ to be  $\tilde{O}(\e)$, 
the maximum additive error allowed, and $\lambda(u) = \sqrt{d_{out}(u)}$. For relative error, we set $\theta
=\tilde{O}(\delta)$, the threshold for constant relative error guarantee, and
$\lambda(u) = 1$.

% By setting different
% sampling function $\lambda(u)$, the 
% we can finish the $\ell$-hop Backward Search.
% Concretely, construct the in-adjacency list for each node in the
% ascending order of $d_{out}(u)$ where $u \in N_{in}(v)$ (Line 1-3).
For initialization, we set the maximum number of hops $L$ to be $\log_{1-\alpha}\theta$
(line 1).
We then initialize the estimators $\epi_\ell(s,t)=0$ for $\forall s \in V$ and $\forall \ell \in [0,L]$ except for $\pi_0(t,t)=\alpha$ (Line 2-3).
We iteratively push the probability mass from level $0$ to $L-1$ (line 4). 
At level $\ell$, we pick a node $v \in V$ with non-zero estimator
$\epi_\ell(v,t)$, and push $\epi_\ell(v,t)$ to a subset of its
in-neighbors (line 5). More precisely, for an in-neighbor $u$ with
out-degree  $\dout(u) \le {\lambda(u) \cdot (1-\alpha)\epi_\ell(v, t)
  \over \alpha \theta}$, we deterministically push a probability mass of
${(1-\alpha)\epi_{\ell}(v, t) \over \dout(u)}$ to the $(\ell+1)$-hop
estimator $\epi_{\ell+1}(u,t)$ (lines 6-7). 
Recall that the
in-neighbors of $v$ are sorted according to their out-degrees. 
Therefore, we can sequentially scan the in-adjacency list
of $v$ until we encounter the first in-neighbor $u$ with $\dout(u) > {\lambda(u) \cdot (1-\alpha)\epi_\ell(v, t)
  \over \alpha \theta}$. 
For  in-neighbors
with higher  out-degrees, we generate a random number $r$ from
$(0,1)$, and push a probability mass of $\frac{ \alpha \theta
}{\lambda(u)}$ to $\epi_{\ell+1}(u,t)$ for each in-neighbor $u$ with  $
\dout(u)\le {\lambda(u) \cdot (1-\alpha)\epi_{\ell}(v, t) \over r
  \alpha \theta}$ (lines 8 -10).
Similarly, we can sequentially scan the in-adjacency list
of $v$ until we encounter the first in-neighbor $u$ with $
\dout(u)>{\lambda(u) \cdot (1-\alpha)\epi_{\ell}(v, t) \over r
  \alpha \theta}$.  Finally, after all $L$ hops are
processed, we return $\epi(s, t) = \sum_{\ell=0}^L
\epi_\ell(s,t)$ as the estimator for each  $\pi(s,t), s\in V$. The sketch map of the above process is shown in Figure~\ref{fig:backwardPush}.

\begin{algorithm}[t]
	\caption{Randomized Backward Search\label{alg:vbs2}}
	\KwIn{Directed graph $G=(V,E)$ with sorted adjacency lists, target node $t \in V$, teleport probability $\a$, error parameter $\theta$,
          sampling function  $\lambda(u)$\\}
	\KwOut{Estimator $\epi(s, t)$ for each $s\in V$\\}
	% Construct a tuple $(u,v,d_{out}(u))$ for each edge $(u,v) \in E$\; 
	% Use counting sort to sort $(u,v,d_{out}(u))$ tuples in the
        % ascending order of $d_{out}(u)$\;
        % \For{each ($u,v, d_{out}(u)$)}
	% {
	% 	Append $u$ to the end of $v$'s in-adjacency list\;
	% }
        $L\gets \log_{1-\alpha}\theta$\;
	$\epi_\ell(s, t) \gets 0$ for $\ell=0,\ldots, L$, $s\in V$\;
	$\epi_0(t,t) \gets \alpha$\;
	\For{$\ell=0$ to $L-1$}
	{
		\For{each $v \in V$ with non-zero $\epi_{\ell}(v,t)$ }
		{
			\For{each $u \in N_{in}(v)$ and $\dout(u) \le
				{\lambda(u) \cdot (1-\alpha)\epi_\ell(v, t) \over \alpha \theta}$}
			{
				$\epi_{\ell+1}(u, t) \gets \epi_{\ell+1}(u,t) + {(1-\alpha)\epi_{\ell}(v, t) \over \dout(u)}$;
			}
			$r \gets rand(0,1)$\;
			\For{each $u \in N_{in}(v)$ and $ {\lambda(u) \cdot (1-\alpha)\epi_{\ell}(v, t) \over \alpha \theta}< \dout(u)\le
				{\lambda(u) \cdot (1-\alpha)\epi_{\ell}(v, t) \over r \alpha \theta}$}
			{
				$\epi_{\ell+1}(u, t) \gets \epi_{\ell+1}(u,t) + \frac{ \alpha \theta }{\lambda(u)}$;
			}
		}
	}
	
	\Return all non-zero $\epi(s, t) = \sum_{\ell=0}^L \epi_\ell(s,t)$ for each $s\in V$\;
\end{algorithm}

%%% Local Variables:
%%% mode: latex
%%% TeX-master: "paper"
%%% End:

%\vspace{-1mm}
%\vspace{-1 mm}
\section{Analysis}\label{sec:analysis}
% \vspace{+1mm}

% %\hspace*{\fill} \\
% %\hspace*{\fill} \\
% \vspace{+1mm}
In this section, we analyze the theoretical property of the RBS
algorithm.
Recall that  for
single-target PPR query with
additive error, we set  $\lambda(u)$ to be $\sqrt{d_{out}(u)}$ and
$\theta=\tilde{O}(\e)$ to be the error bound. For relative error, we
set $\lambda(u) = 1$ and $\theta=\tilde{O}(\delta)$, the threshold for constant relative error
guarantee. Recall that $\tilde{O}$ denotes the Big-Oh notation
ignoring the log factors.
Theorem~\ref{thm:relative} and
~\ref{thm:additive}  provide the theoretical results of running time and error guarantee for the RBS
algorithm with additive
and relative error, respectively. 
% are given in {\newblue Proposition}~\ref{pro:lambda1} and
% ~\ref{pro:lambda2}, respectively. 

\begin{theorem}
%\vspace{-2 mm}
\label{thm:relative}
	By setting $\lambda(u)=1$ and $\theta = \tilde{O}(\delta)$,
        Algorithm~\ref{alg:vbs2} answers the single-target PPR queries
        with a relative error threshold $\delta$ with high
        probability. The expected worst-case
        time cost is bounded by  $\tilde{O}\left(\frac{n \pi(t)}{\delta}
        \right)$.
	If the target node $t$ is chosen uniformly at random from $V$,
        the time cost becomes $\tilde{O}\left( \frac{1}{ \delta} \right)$. 
\end{theorem}
%\vspace{-3 mm}
\begin{theorem}
\label{thm:additive}
	By setting $\lambda(u)=\sqrt{d_{out}(u)}$ and $\theta = \tilde{O}(\e)$,
        Algorithm~\ref{alg:vbs2} answers the single-target PPR queries
        with an additive error parameter $\e$ with high
        probability. The expected worst-case
        time cost is bounded by
        \vspace{-1 mm} $$\vspace{-1 mm} \E[Cost] = \tilde{O} \left(\frac{1}{\e}
          \sum_{u \in V} \sqrt{d_{out}(u)}\cdot \pi(u,t) \right).$$ 
	If the target node $t$ is chosen uniformly at random from $V$,
        the time cost becomes $\tilde{O}\left(
          \frac{\sqrt{\bar{d}}}{ \e} \right)$,
        where $\bar{d}$  is the average degree of the graph. 
      \end{theorem}
 %\vspace{-2 mm}

To prove Theorem~\ref{thm:relative} and
~\ref{thm:additive}, we need several technical lemmas. In particular,
% Utilizing the result of each level's $\ell$-hop PPR, sum them up and derive the estimator of $\pi(s,t)$ for $\forall s \in V$ based on the below formula: 
% \begin{align}
% \label{eqn:lhop_ppr_sum}
% \epi(s,t)=\sum_{\ell=0}^{L}\epi_\ell(s,t).
% \end{align}
% %However, the number of summation entries in equation~\eqref{eqn:lhop_ppr_sum} is infinite which is unpractical. 
% If we allow $\e$ additive error, let $L$ denote the maximum level which should be considered. Then 
% \begin{align}
% \pi(s,t)-\epi(s,t)\leq \sum_{i=L+1}^{\infty}\alpha(1-\alpha)^i \leq (1-\alpha)^{(L+1)} \leq \e.
% \end{align}
% Hence, if $L \ge \log_{1-\alpha}(\e)$, the additive error of
% $\epi(s,t)$ will not exceed $\e$.
% Besides, we can choose appropriate $\lambda(u)$ according to the
% specific application scenarios.
we first prove that Algorithm~\ref{alg:vbs2} provides an unbiased
estimator for the $\ell$-hop PPR values $\pi_{\ell}(s,t)$.
\begin{lemma}
%\vspace{-1 mm}
	\label{lem:vbs2_unbiasedness}
	Algorithm~\ref{alg:vbs2} returns an estimator $\epi_\ell(s,t)$
        for each $\pi_{\ell}(s,t)$ such that $\E\left[ \epi_{\ell}(s,t) \right]=\pi_{\ell}(s,t)$
	%\begin{align*}
	%\E\left[ \epi_{\ell}(s,t) \right]=\pi_{\ell}(s,t)
	%\end{align*}
	holds for $\forall s \in V$ and $\ell \in \{0,1, 2, ... , L\}$.
\end{lemma}

Next, we bound the variance of the $\ell$-hop estimators.
\begin{lemma}
	\label{lem:vbs2_variance}
	For any $s \in V$ and $\ell \in \{0,1, 2, ... , L\}$, the
        variance of each estimator $\epi_{\ell}(s,t)$ obtained by 
        Algorithm~\ref{alg:vbs2} satisfies that:

        1) If  we set $\lambda(u) =1$, then
        $$\Var \left[ \epi_{\ell}(s,t) \right] \leq \theta
        \pi_{\ell}(s,t).$$
        
        2) If  we set $\lambda(u) = \sqrt{d_{out}(u)}$, then
        $$\Var \left[ \epi_{\ell}(s,t) \right] \leq \alpha \theta^2.$$
	% \begin{align}
	% \Var \left[ \epi_{\ell}(u,t) \right] \leq \frac{\theta}{\lambda(u)} \cdot \pi_{\ell}(u,t)
	% \end{align}
	% where 
\end{lemma}

        % we can rewrite the variance of estimator $\epi_{\ell}(s,t)$ for each $s \in V$ as below.
	% \begin{align}
	% \Var \left[ \epi_{\ell}(u,t) \right] \leq \frac{\theta}{\lambda(u)} \cdot \pi_{\ell}(u,t) = \frac{\theta}{\lambda(u)} \cdot \sum_{v \in N_{out}(u)} \frac{(1-\alpha)\pi_{\ell-1}(v,t)}{d_{out}(u)}
	% \end{align}

	% Therefore, 
	% \begin{align}
	% \Var \left[ \epi_{\ell}(u,t) \right] \leq \frac{\theta}{\lambda(u)} \cdot \sum_{v \in N_{out}(u)} \frac{\alpha \theta}{\lambda(u)}=\frac{\alpha \theta^2}{\left( \sqrt{d_{out}(u)} \right)^2} \cdot d_{out}(u) \leq \alpha \theta^2
	% \end{align}

	% Furthermore, $\Var \left[ \epi_{0}(x,t) \right]=0 \leq
        % \frac{\theta}{\lambda(u)} \cdot \pi_{0}(x,t) $ holds for
        % $\forall x \in V$, which finishes this proof.

The following lemma analyzes the expected query cost of 
the RBS algorithm.
\begin{lemma}
	\label{lem:vbs2_cost}
	Let $C_{total}$  denote the total cost during the whole push process from level $0$ to level $(L-1)$, 
	the expected time cost of algorithm~\ref{alg:vbs2} can be expressed as that
	\begin{align*}
	\E \left[C_{total}\right] \leq \frac{1}{\alpha \theta} \sum_{u \in V} \lambda(u) \cdot \pi (u,t).
	\end{align*}
\end{lemma}
Note that $\lambda(u)$ is an adjustable sampling function that balances the
variance and time cost. In our case, it is a 
function of node $u$. Hence, we do not extract $\lambda(u)$ from the last summation symbol.
% \begin{proposition}
% \label{pro:lambda1}
% 	For relative error, we set $\lambda(u)=1$ and $\theta = \delta$ in
%         Algorithm~\ref{alg:vbs2}. 
% 	For each $s \in V$ and $\ell<L$, the variance of estimator
%         $\epi_{\ell}(s,t)$ is bounded by $\delta \pi_{\ell}(s,t)$ with
%         time cost $\tilde{O}\left(\frac{n \pi(t)}{\alpha \delta} \right)$.	
% 	If the target node $t$ is chosen uniformly at random from $V$, the time cost can be rewritten as $\frac{1}{\alpha \theta}$. 
% \end{proposition}
With the help of
Lemma~\ref{lem:vbs2_unbiasedness},~\ref{lem:vbs2_variance} ,
and~\ref{lem:vbs2_cost}, we are  able to prove
Theorem~\ref{thm:relative} and~\ref{thm:additive}. For the sake of
readability, we defer all proofs to the appendix.

% \begin{proposition}
% \label{pro:lambda2}
% 	As for algorithm~\ref{alg:vbs2}, set $\lambda(u)=\sqrt{d_{out}(u)}$ corresponding the push operation through edge $(u,v)$. 
% 	For each $s \in V$ and $\ell<L$, the variance of estimator $\epi_{\ell}(s,t)$ is less than $\alpha \e^2$ with time cost $\frac{1}{\alpha \theta} \sum_{u \in V} \sqrt{d_{out}(u)}\cdot \pi(u,t)$.	
% 	If the target node $t$ is chosen uniformly at random from $V$, the time cost can be rewritten as $\frac{\sqrt{d}}{\alpha \theta}$, where $d$ represents nodes' averaged degree in graph $G$. 
% \end{proposition}

%%% Local Variables:
%%% mode: latex
%%% TeX-master: "paper"
%%% End:

%\vspace{-1mm}
\section{applications} \label{sec:applications}
In this section, we discuss how the RBS algorithm improves the three
concrete applications mentioned in Section~\ref{sec:intro}: heavy hitters PPR
query, single-source SimRank computation, and approximate PPR matrix
computation.

\header{\bf Heavy hitters PPR computation.}
Following the definition in~\cite{wang2018heavyhitters}, we define the
$c$-approximate heavy hitter as follows.

%\vspace{-1mm}
\begin{definition}[$c$-approximate heavy hitter]
  Given a real value \(0<\phi<1,\) a constant real value \(0<c<1,\) two nodes \(s, t\) in \(V,\) we
  say that \(s\) is:
  \begin{itemize}
\item  a \(c\) -absolute \(\phi\) -heavy hitter of \(t\) if \(\pi(s,
  t)>(1+c) \phi \cdot n \pi(t)\);
\item a \(c\) -permissible \(\phi\)-heavy hitter of \(t\) if \((1-c)
  \phi \cdot n \pi(t) \leq \pi(s, t) \leq\)
\((1+c) \phi \cdot n \pi(t)\);
\item not a \(c\) -approximate \(\phi\) -heavy hitter of \(t,\)
  otherwise.
  \end{itemize}
\end{definition}
%\vspace{-1mm}

Given a target node $t$, a heavy hitter algorithm is required to return all \(c\) -absolute
\(\phi\) -heavy hitters and to exclude all nodes that are not a \(c\)
-approximate \(\phi\) -heavy hitter of
\(t\). Wang et al. utilizes the traditional Backward Search to derive the $c$-approximate heavy hitter\cite{wang2018heavyhitters}. 
The time complexity is  $O \left( \sum_{u \in V}
  \frac{d_{out}(u) \cdot \pi(u,t)}{\phi n \pi(t)} \right)$ in the worst case.
%For a random target node $t$, the time complexity can be simplified to $O(\frac{\bar{d}}{\phi})$.

On the other hand, by running the RBS algorithm with relative error threshold $\delta = c\phi n \pi(t)$, we can return the $c$-approximate heavy hitter for node $t$ with high  probability. 
By Theorem~\ref{thm:relative}, the running time of RBS algorithm is bounded
by $\tilde{O}\left( {n \pi(t) \over \delta}\right) = \tilde{O}\left(
  {1 \over \phi}\right)$. Note that this complexity is optimal up to
log factors, as there may be  $O\left(
  {1 \over \phi}\right)$ heavy hitters for node $t$.  Therefore, RBS
achieves optimal {\em worst-case} query complexity for the $c$-approximate heavy hitter problem.

\header{\bf Single-source SimRank computation.} 
% SimRank is an important metric to evaluate the similarities between nodes in a graph. 
% The main idea of SimRank is that two nodes are similar if they are connected with similar nodes.
% The most similar node is the node itself.
% The definition formula given in~\cite{JW02} is recursive and cannot be applied to compute the SimRank on large graphs.
% Hence, most of the work tries to rewrite the computation formula to
% avoid the recursive nature.
Recall that if we revert the direction of every edge in the graph, the SimRank similarity $s(u,v)$ of node $u$ and $v$ equals to the probability that  two $\alpha$-discounted random walks from $u$ and $v$ visit at the same node $w$ with the same steps.
PRSim~\cite{wei2019prsim} and SLING~\cite{TX16}, two state-of-the-art
SimRank algorithms, formulate the SimRank in terms of $\ell$-hop PPR:
% between node $u$ and $v$ with decay factor $c$ based on the below formula:
%\vspace{-5mm}
\begin{align}
%\vspace{-3mm}
s(u,v)=\frac{1}{(1-\sqrt{c})^2}\sum_{\ell=0}^{\infty}\sum_{w \in V} \pi_\ell(u,w)\pi_\ell(v,w)\eta(w), 
%\vspace{-5mm}
\end{align}
where $\pi_\ell(u,w)$ is the $\ell$-hop PPR with decay factor $\alpha= 1-\sqrt{c}$, and $\eta(w)$ is a value called the last meeting
probability. SLING~\cite{TX16} proposes to use Backward Search to precompute an 
approximation of $\pi_\ell(v,w)$ with additive error $\e$ for each $w,
v \in V, \ell =0,\ldots, \infty$, while  PRSim~\cite{wei2019prsim}
only precomputes an approximation of $\pi_\ell(v,w)$ for node $w$ with
large PageRanks. 
Recall that the running time of the Backward Search
algorithm on a random node $t$ is $O\left( {\bar{d} \over \e}
\right)$. 
It follows that  the total  precomputation cost for SLING or
PRSim is bounded by $O\left( {m \over \e} \right)$. 
However, according to Theorem~\ref{thm:additive},
if we replace the Backward Search algorithm with the new RBS algorithm with
additive error, the running time for a random target node $t$ is
improved to  $\tilde{O}\left( {\sqrt{\bar{d}} \over \e} \right)$. 
And thus the total 
precomputation time is improved to $\tilde{O}\left( {n \sqrt{\bar{d}} \over \e} \right)$.

%Because PRSim focuses on single-source SimRank, $\pi_\ell(u,% w)$ can be derived by applying single-source PPR computation method with given source node $u$.
% With the given source node $u$, PRSim samples abundant random walks and records the stopping node $w$ simultaneously.
% For each target node $w$, %single-target PPR $\pi(v,w)$ is required. 
% PRSim use Backward Search to calculate $\pi(v,w)$ with large PageRank of $w$ in the preprocessing stage.
% Hence, the time complexity is $O(\frac{d}{\e}) $ for each target node $w$. 
% However, if we substitute the Backward Search part with RBS, 
% the time complexity can be reduced to $O(\frac{1}{\e})$ per target node with $\s$ additive error. 
% Because the large PageRank nodes usually have large degrees in the meantime, 
% the preprocessing time can be reduced greatly on some large graphs like twitter. 

\header{\bf Approximate PPR matrix.}
As mentioned in Section~\ref{sec:intro}, computing the approximate PPR
matrix is the computational bottleneck for various graph embedding and
graph neural network algorithms, such as HOPE~\cite{ou2016asymmetric},
STRAP~\cite{wei2019strap}, PPRGo~\cite{Xu2019PPRGo} and
GDC~\cite{klicpera2019GDC}. 
However, computing the PPR matrix is costly; 
Applying the Power Method to $n$ nodes takes $\tilde{O}(mn)$ time, which is infeasible on large graphs. PPRGo~\cite{Xu2019PPRGo}
proposes to apply Forward Search to each source node $s\in V$ to
construct the approximate PPR matrix; 
While STRAP~\cite{wei2019strap}
employs Backward Search to each target node $t \in V$ to compute
$\pi(s,t), s\in V$ and then put $\pi(s,t)$ into an inverted list
indexed by $s$. 
For the former approach, recall that the Forward Search only guarantees an
additive error of $\e d_{out}(t)$ for the estimator of $\pi(s,t)$~\cite{AndersenCL06},
which is undesirable for nodes with high degrees. On the other hand,
the latter approach incurs a running time of $O\left( {m \over
    \e}\right)$ for computing an approximate PPR matrix with additive
error $\e$. By replacing the Backward Search algorithm with the new
RBS algorithm with additive error, we can improve the complexity to
$\tilde{O}\left( {n \sqrt{\bar{d}} \over \e} \right)$, which is
sub-linear to the number of edges $m$. % The new complexity
\section{Experiments} \label{sec:exp}
\begin{table}[t]
	%\vspace{-2mm}
	\centering
	\tblcapup
	\caption{Data Sets.}
	\vspace{-2mm}
	\tblcapdown
%	\begin{small}
		\begin{tabular}{|l|l|r|r|} %p{1.3in}|}
			\hline
			{\bf Data Set} & {\bf Type} & {\bf $\boldsymbol{n}$} & {\bf $\boldsymbol{m}$}	 \\ \hline
			ca-GrQc (GQ) & undirected & 5,242 & 28,968\\
			AS-2000(AS) & undirected & 6,474 & 25,144\\
			%CA-HepTh(HT) & undirected & 9,877 & 51,946\\
			%Wikivote (WV) & directed & 7,115 & 103,689\\
			%CA-HepPh (HP) & undirected & 12008 & 236978\\
			% Wiki-Vote(WV)	& 	directed &	7,155	&	103,689 \\
			% {HepTh(HT)}	    & 	undirected &	9,877	&	25,998		\\
			% {AS-Caida(AC)}	    &	directed &	26,475	&	106,762		\\
			% {HepPh(HP)}	        &	directed &	34,546	&	421,578 \\
			% Cnr-2000 (CN) & directed & 325,557 &	3,216,152 \\
			% {Web-Google(WG)} & directed & 875,713 &	5,105,039 \\)
			%As-Skitter (AS) & undirected & 1,696,415	& 11,095,298 \\
			%      {\color{red} In-2004(IN) } & directed & 1,382,908	& 16,917,053
			% \\
			DBLP-Author (DB) & undirected & 5,425,963 & 17,298,032 \\
			% {Com-LiveJournal(CL)} & undirected & 3,997,962	& 34,681,189 \\
		%	LiveJournal (LJ) &	directed & 4,847,571 & 68,475,391\\
			IndoChina (IC)	& 	directed &	7,414,768  &	191,606,827		\\
			Orkut-Links (OL) & undirected & 3,072,441 & 234,369,798 \\
			
			%DBpediaLink (DL) & directed & 18,268,992 & 172,183,984 \\
			%WikiLink(WL) & directed & 12,150,976 & 378,142,420 \\
			%Twitter(TW) & directed & 41,652,230 & 1,468,365,182 \\
			% { Web-Base}   & { directed} & {
			% 118,142,155} & { 1,019,903,190} \\
			%Web-Base (WB) & directed & 118,142,155& 1,019,903,190 \\
			It-2004 (IT)	&	directed & 41,290,682 & 1,135,718,909\\
			
			Twitter (TW) & directed & 41,652,230& 1,468,364,884 \\
			%SK-2005 (SK) & directed & 50,636,154	& 1,949,412,601 \\
			%Friendster (FD)  & directed & 68,349,466 & 2,586,147,869 \\
			%UK-Union (UK) & directed & 133,633,040 & 5,507,679,822 \\
			\hline
		\end{tabular}
%	\end{small}
	\label{tbl:datasets}
	%\tbldown
	\vspace{-3mm}
\end{table}

This section experimentally evaluates the performance of RBS against
state-of-the-art methods.
%The datasets we used are shown in Table~\ref{tbl:datasets}.
Section~\ref{subsec:stQuery} presents the empirical study for single-target PPR queries.
Section~\ref{subsec:threeapplications} applies RBS to three concrete applications to show its effectiveness.
The information of the datasets we used is listed in
table~\ref{tbl:datasets}. All datasets are obtained from \cite{snapnets,LWA}.
All experiments are conducted on a machine with an Intel(R) Xeon(R) E7-4809 @2.10GHz CPU and 196GB memory.

\begin{figure*}[!t]
%	\begin{small}
		\centering
		\vspace{-1mm}
		%    \begin{footnotesize}
		\begin{tabular}{ccccc}
			%\multicolumn{4}{c}{\hspace{-4mm} \includegraphics[height=5mm]{./Figs/legend_large.eps}} \vspace{-1mm} \\
			\hspace{-4mm} \includegraphics[height=27mm]{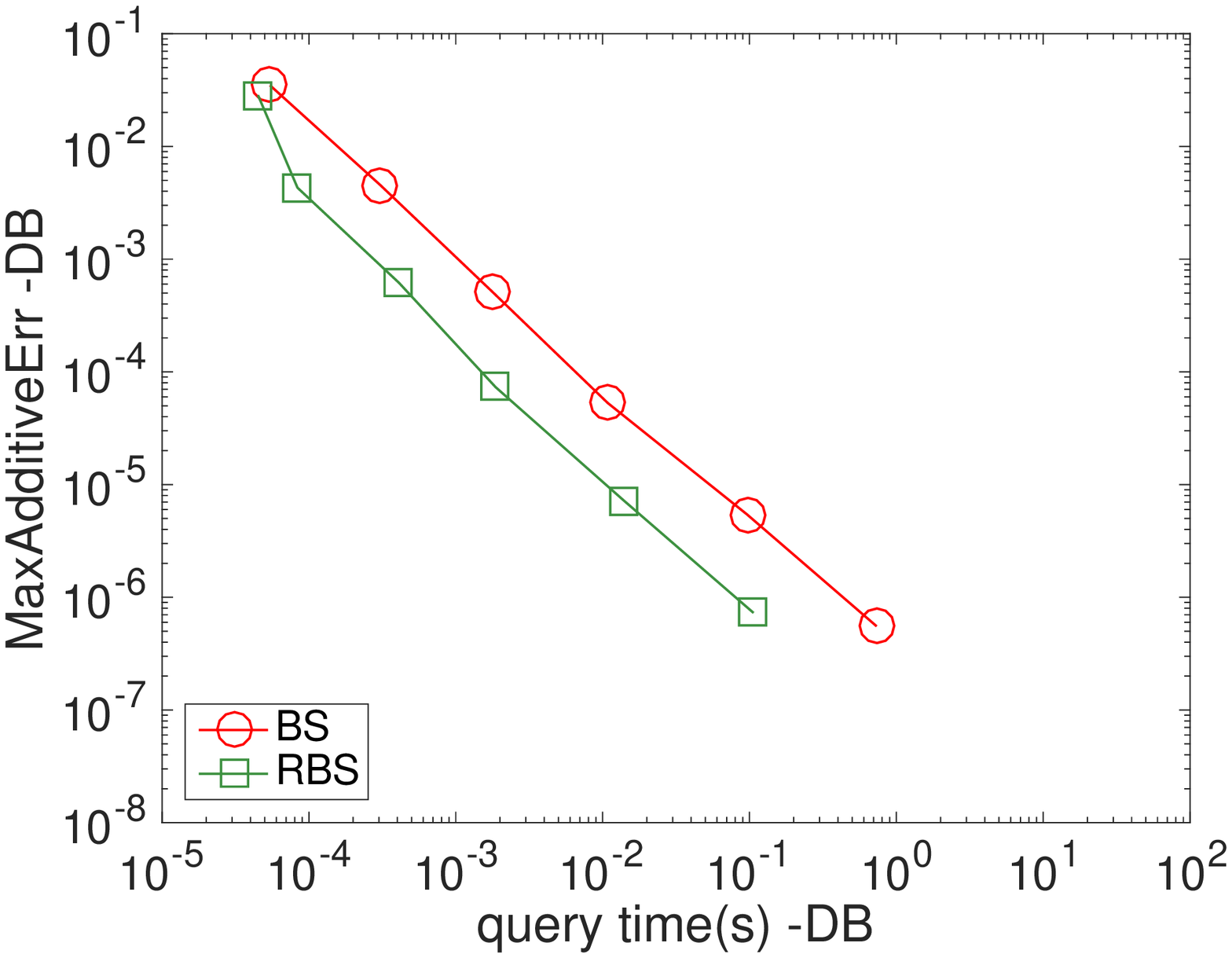} &
			%\hspace{-3mm} \includegraphics[height=25mm]{./Figs/maxerr-query-LJ.eps} &
			\hspace{-3mm} \includegraphics[height=27mm]{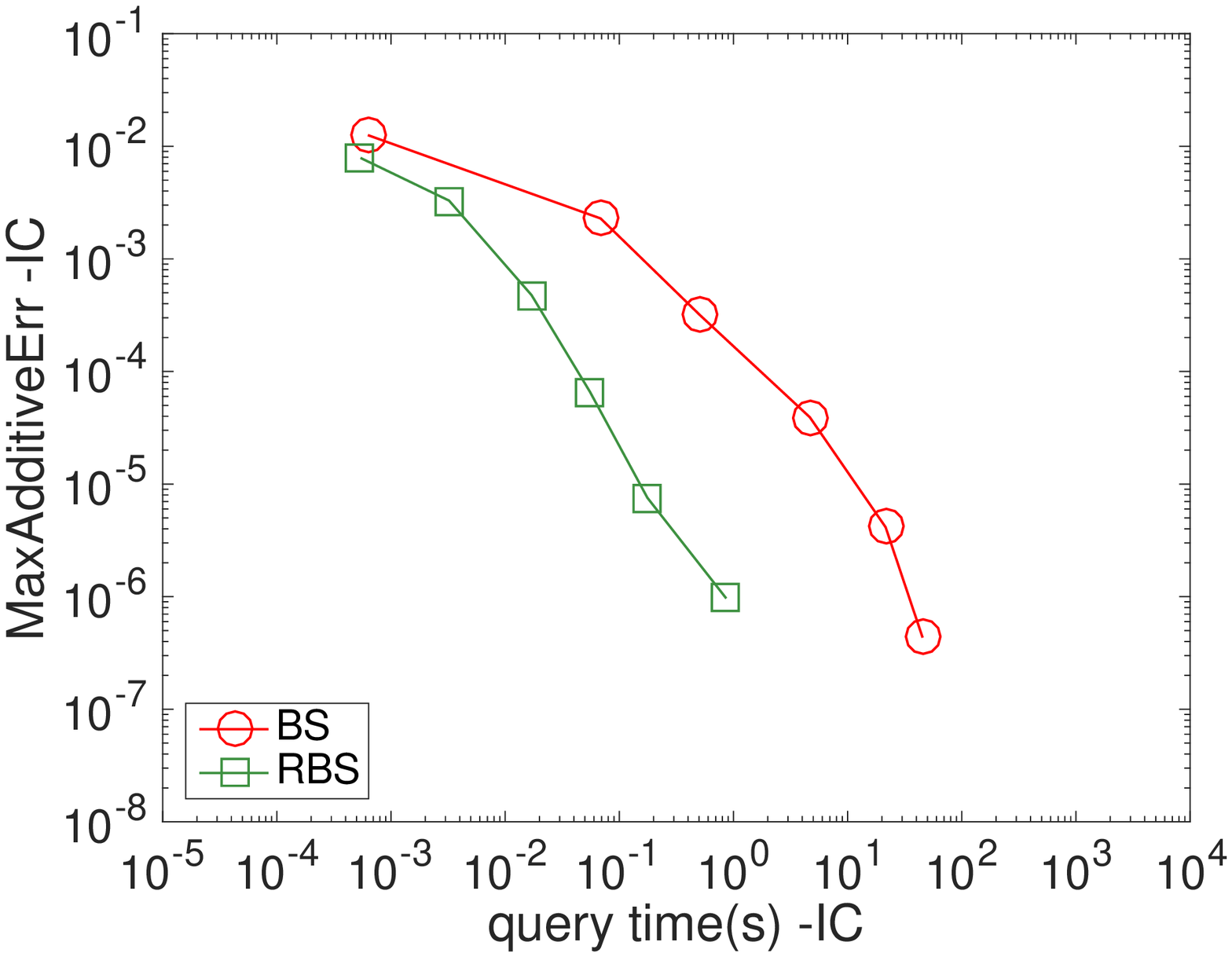} &
			\hspace{-3mm} \includegraphics[height=27mm]{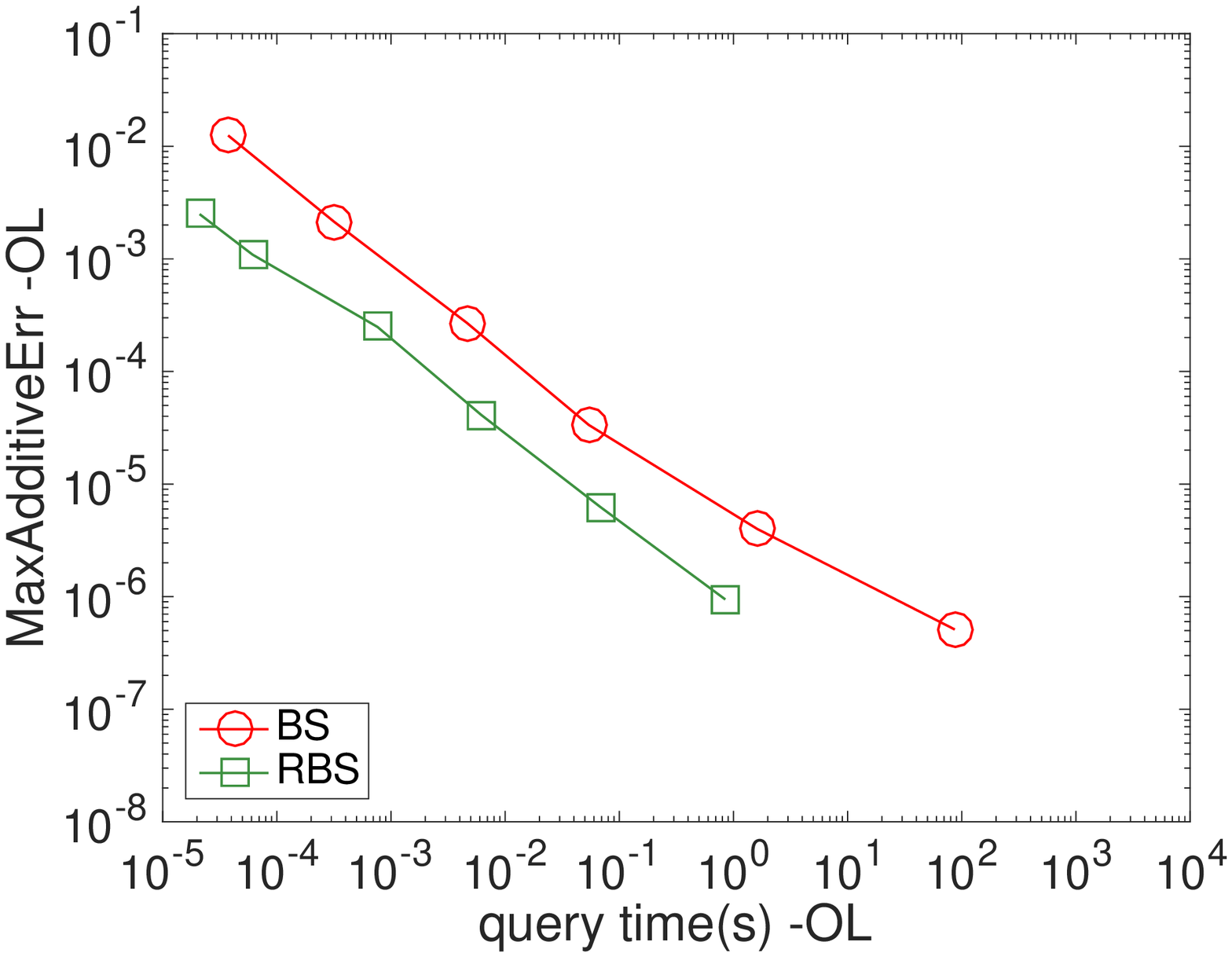} &
			\hspace{-3mm} \includegraphics[height=27mm]{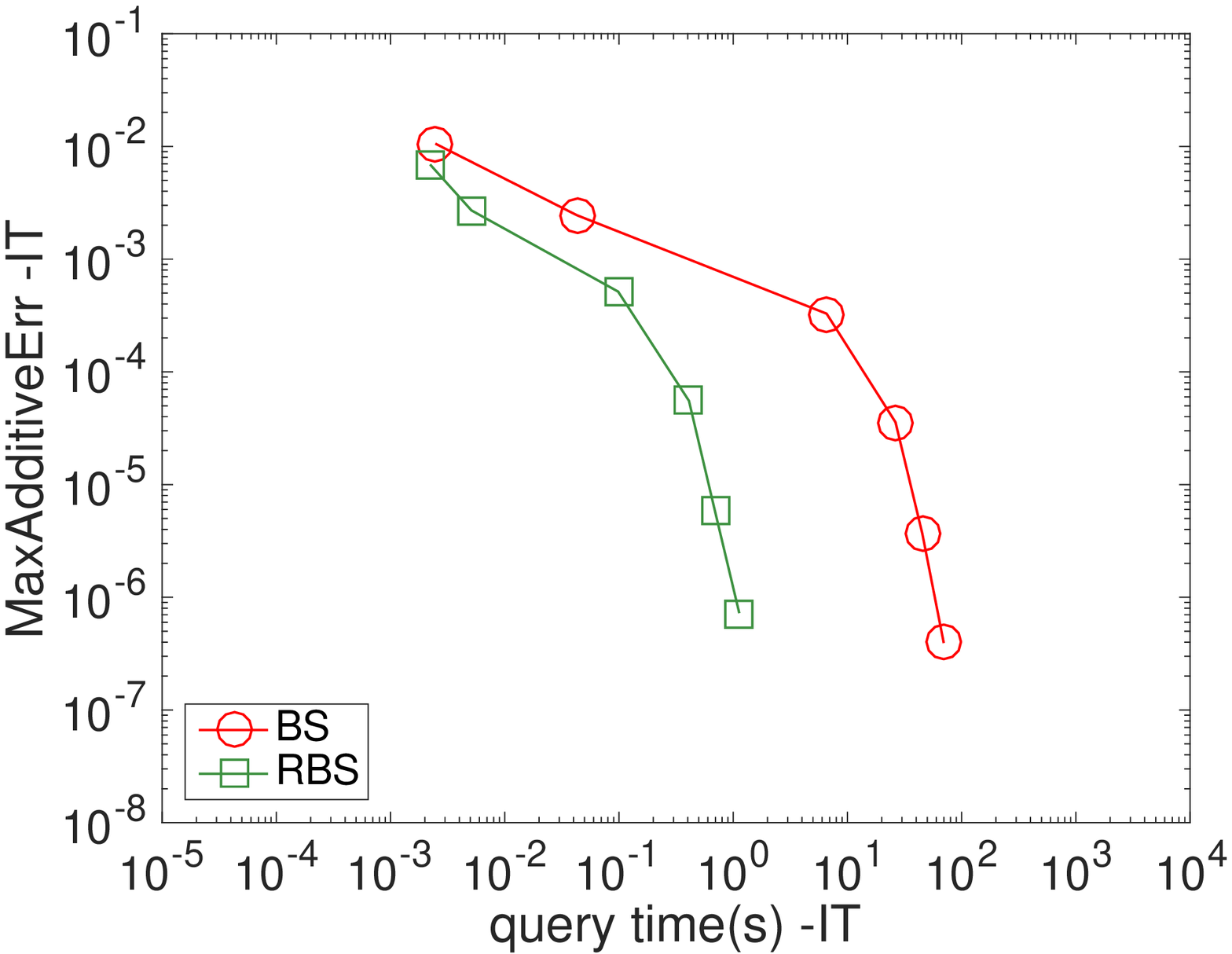} &
			\hspace{-3mm} \includegraphics[height=27mm]{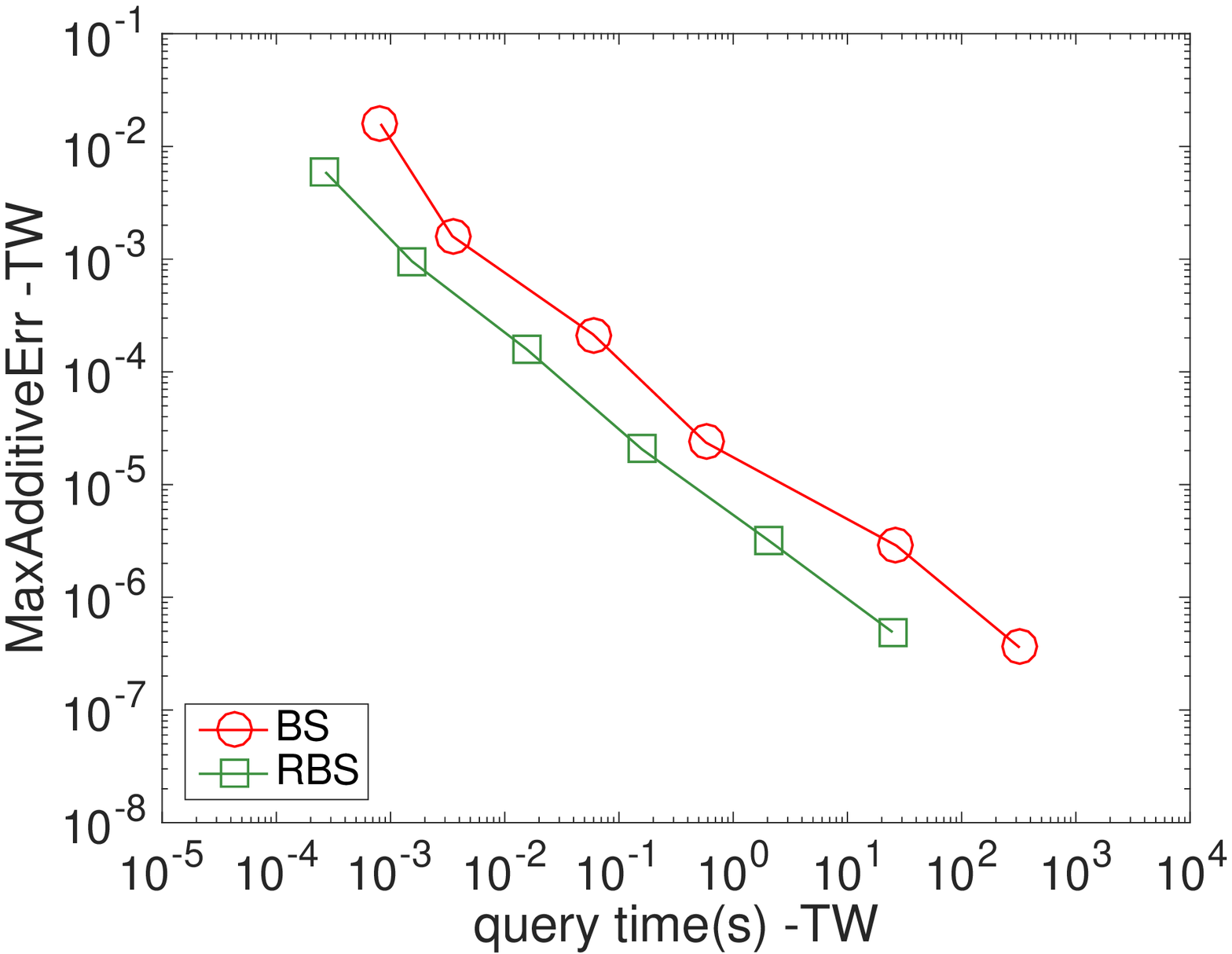}
		\end{tabular}
		\vspace{-3mm}
		\caption{Tradeoffs between  {\em MaxAdditiveErr} and query time.}
		\label{fig:Absmaxerr-query}
		%\vspace{-1mm}
%	\end{small}
\end{figure*}

% \begin{figure*}[!t]
% %	\begin{small}
% 		\centering
% 		\vspace{0mm}
% 		%    \begin{footnotesize}
% 		\begin{tabular}{ccccc}
% 			%\multicolumn{4}{c}{\hspace{-4mm} \includegraphics[height=5mm]{./Figs/legend_large.eps}} \vspace{-1mm} \\
% 			\hspace{-4mm} \includegraphics[height=27mm]{./Figs/Relaerr-query-DB.eps} &
% 			%\hspace{-3mm} \includegraphics[height=25mm]{./Figs/maxerr-query-LJ.eps} &
% 			\hspace{-3mm} \includegraphics[height=27mm]{./Figs/Relaerr-query-IC.eps} &
% 			\hspace{-3mm} \includegraphics[height=27mm]{./Figs/Relaerr-query-OL.eps} &
% 			\hspace{-3mm} \includegraphics[height=27mm]{./Figs/Relaerr-query-IT.eps} &
% 			\hspace{-3mm} \includegraphics[height=27mm]{./Figs/Relaerr-query-TW.eps}
% 		\end{tabular}
% 		\vspace{0mm}
% 		\caption{Tradeoffs between {\em MaxRelativeErr} and query time.}
% 		\label{fig:Relamaxerr-query}
% 		%\vspace{-1mm}
% %	\end{small}
% \end{figure*}

\begin{figure*}[!t]
%	\begin{small}
		\centering
		\vspace{-1mm}
		%    \begin{footnotesize}
		\begin{tabular}{cccccc}
			%\multicolumn{4}{c}{\hspace{-4mm} \includegraphics[height=5mm]{./Figs/legend_large.eps}} \vspace{-1mm} \\
			\hspace{-6mm} \includegraphics[height=27mm]{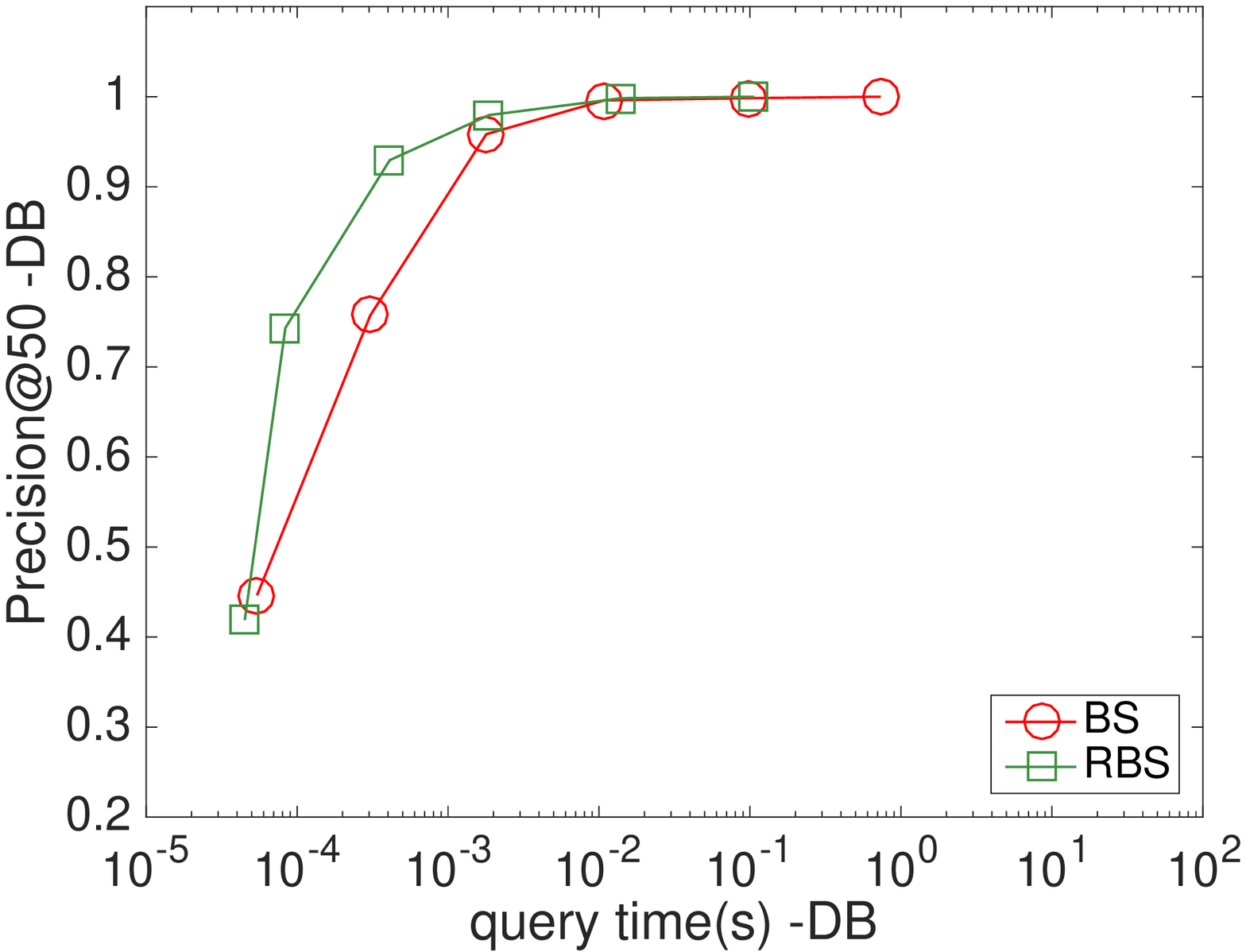} &
			%\hspace{-3mm} \includegraphics[height=24mm]{./Figs/precision-query-LJ.eps} &
			\hspace{-3mm} \includegraphics[height=27mm]{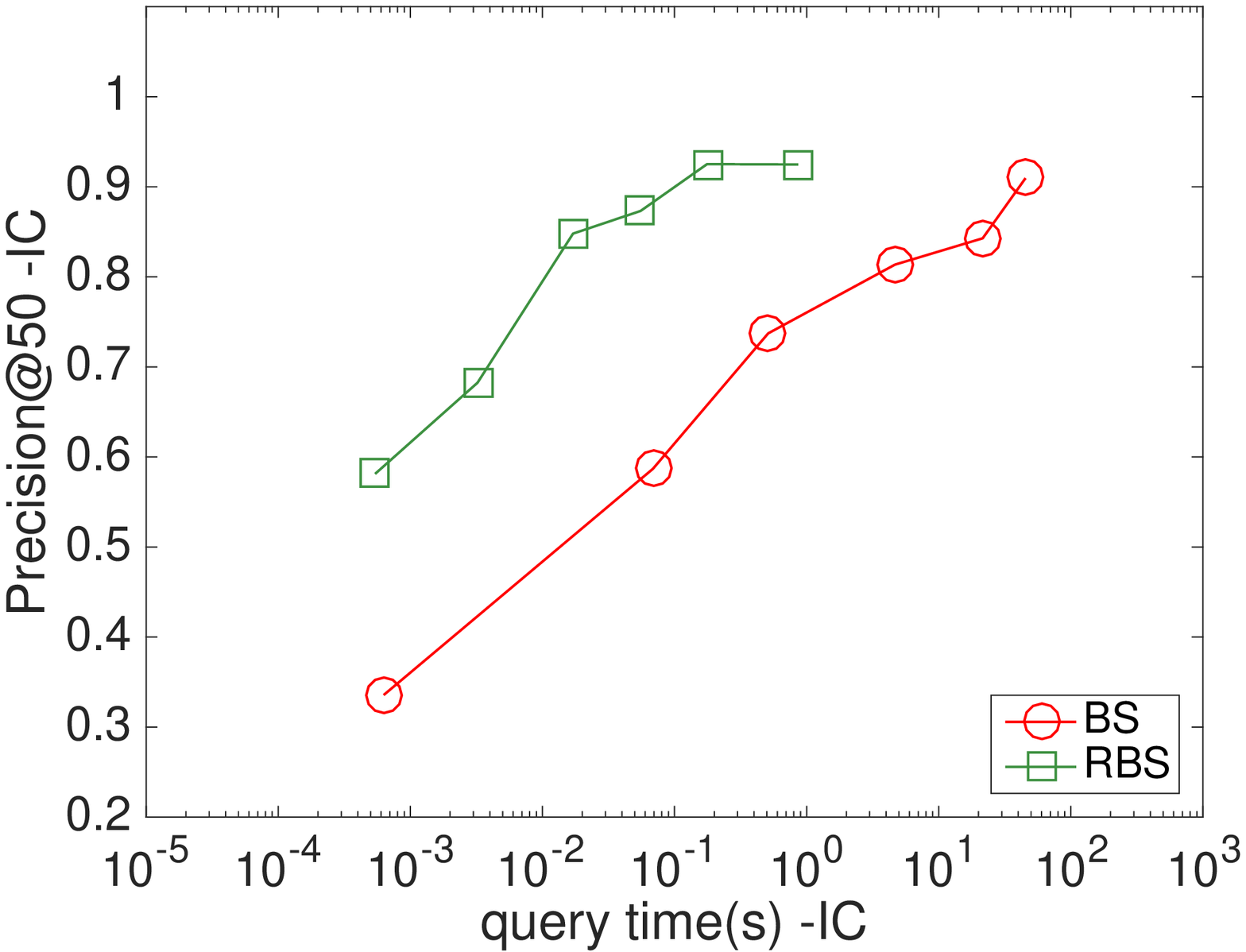}&
			\hspace{-3mm} \includegraphics[height=27mm]{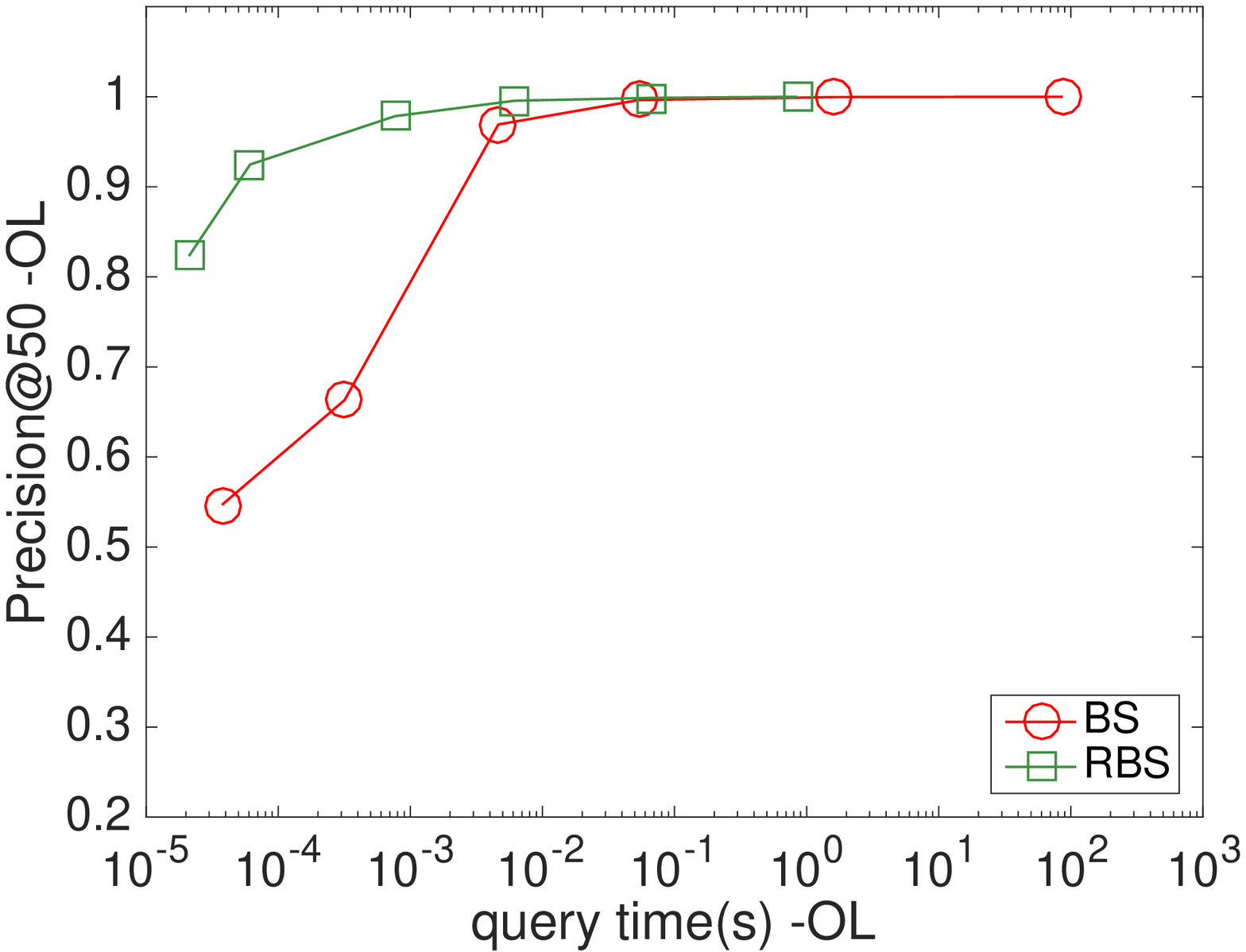}&
			\hspace{-3mm} \includegraphics[height=27mm]{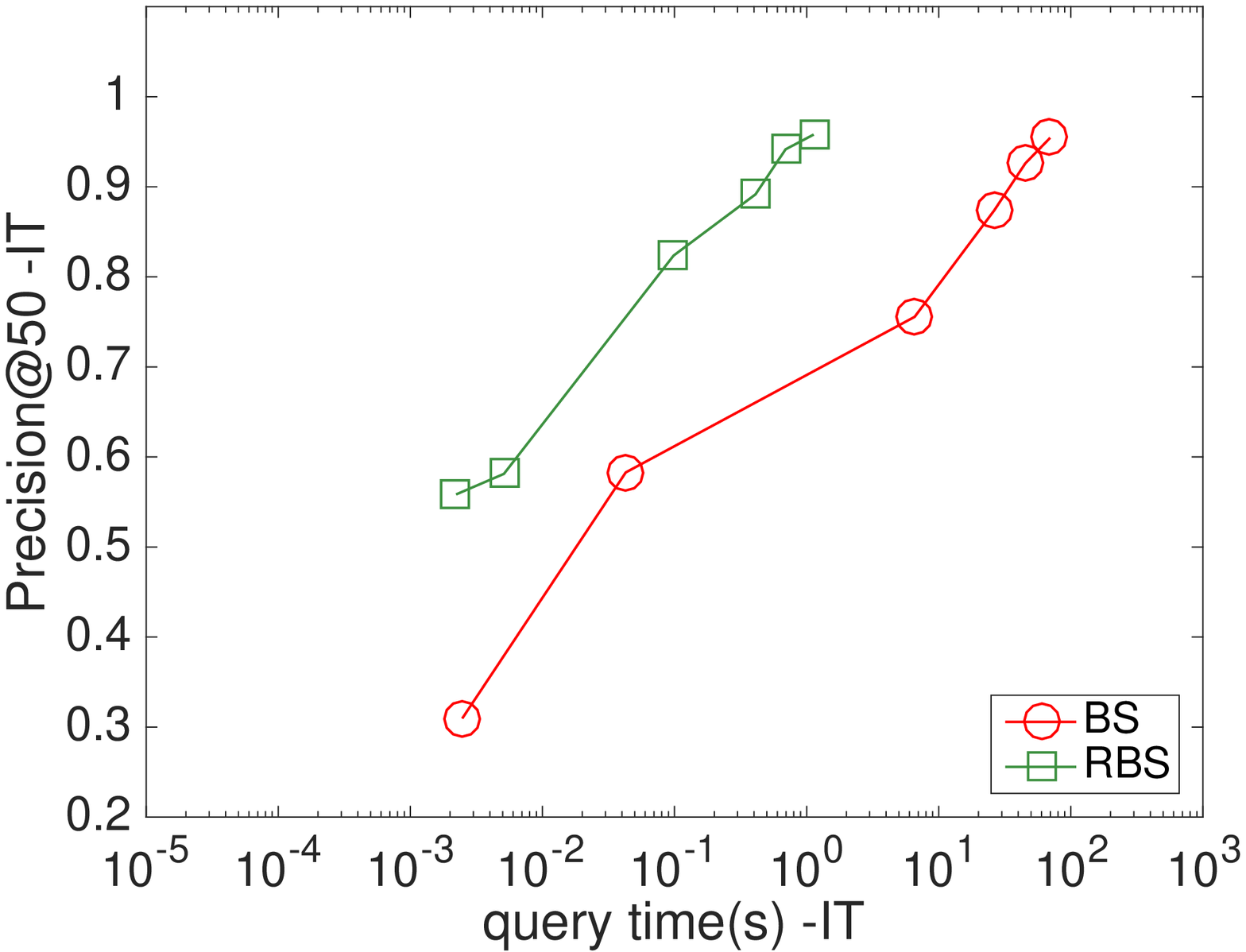} &
			\hspace{-3mm} \includegraphics[height=27mm]{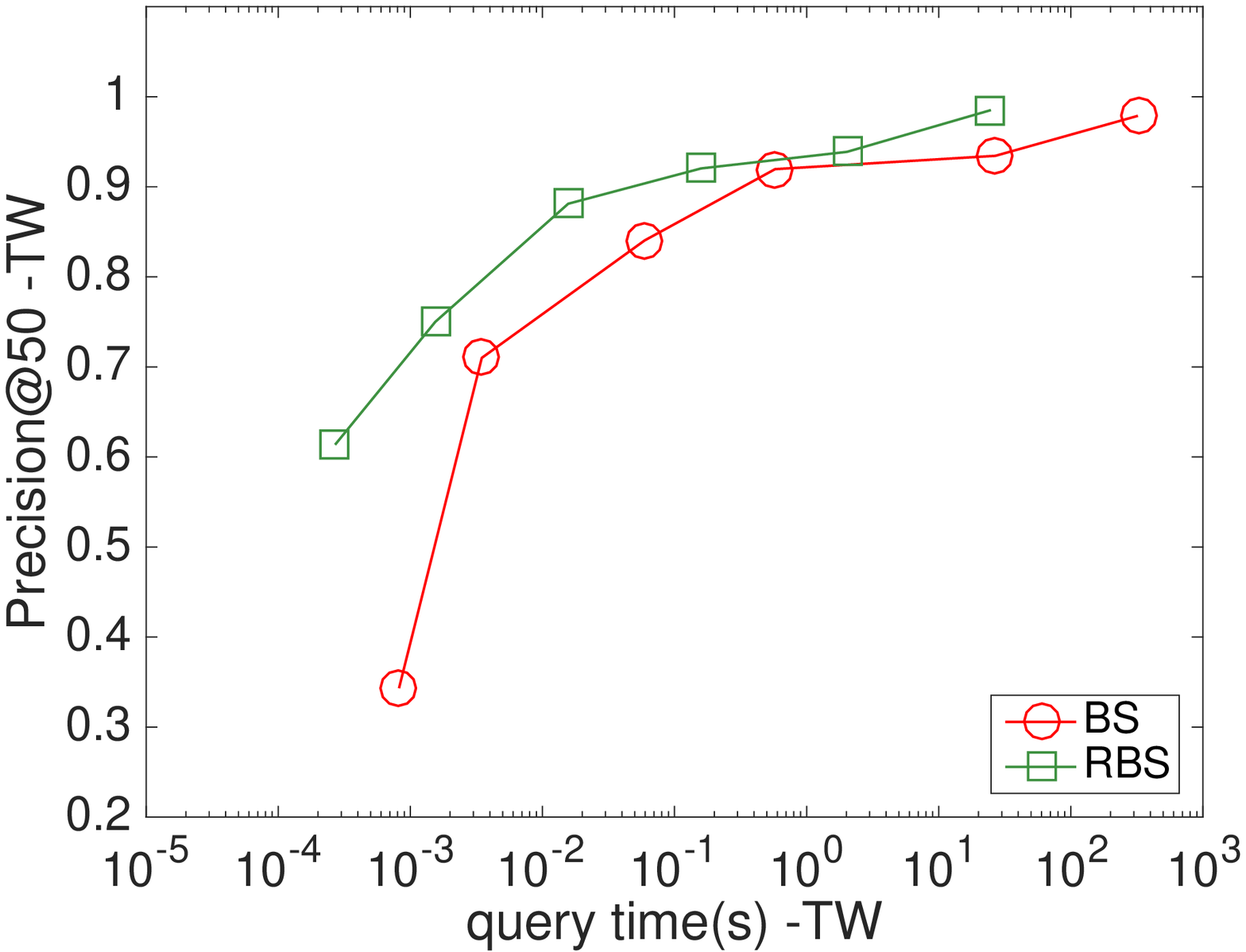}
		\end{tabular}
		\vspace{-3mm}
		\caption{ Tradeoffs between {\em Precision@50} and query time.}
		\label{fig:precision-query}
		%\vspace{-1mm}
%	\end{small}
\end{figure*}

\vspace{-1mm}
\subsection{Single-Target Query} \label{subsec:stQuery}
%\header{\bf Comparison with Backward
%Search~\cite{lofgren2013personalized}.}

\header{\bf Metrics and experimental setup.}
%Specifically, node $u \in V$ is selected as query node with probability $\frac{d_{in}(u)+d_{out}(u)}{2m}$.
For a given query node, we apply the Power
Method~\cite{page1999pagerank} with $L=\lceil\log_{1-\alpha}(10^{-7})\rceil$
iterations to obtain the ground truths of the single-target queries based
on the following formula: $\vec{\pi_t}=(1-\alpha)\vec{\pi_t}\cdot \mathbf{P}^\top+\alpha \cdot \vec{e_t}$. 
%\begin{align}
%\label{eqn:power_method_st_2}
%\vec{\pi_t}=(1-\alpha)\vec{\pi_t}\cdot \mathbf{P}^\top+\alpha \cdot \vec{e_t}. 
%\end{align}
To evaluate the additive error, we consider {\em MaxAdditiveErr}, which is defined that: %as follows:
%$MaxAdditiveErr=\max_{v_i \in V}\left|\pi(v_i,t)-\epi(v_i,t)\right|$, 
\begin{equation}
\begin{aligned}
\vspace{-4mm}
MaxAdditiveErr &=\max_{v_i \in V}\left|\pi(v_i,t)-\epi(v_i,t)\right|
%MaxRelativeErr &=\max_{v_i \in V, \pi(v_i,t) >\delta}\frac{\left|\pi(v_i,t)-\epi(v_i,t)\right|}{\pi(v_i,t)},
\end{aligned}
\vspace{-1mm}
\end{equation}
where $\epi(v_i,t)$ is the estimator of PPR value $\pi(v_i,t)$.
For relative error, there lacks a practical metric that evaluates the
relative error threshold $\delta$. Hence, we consider the  {\em Precision@k}, which evaluates the
quality of the single-target top-$k$ queries. More precisely,
let  $V_k$ and $\hat{V}_k$ denote the set containing the nodes with
single-target top-$k$ queries returned by the  ground truth and the
approximation  methods,  respectively.
{\em Precision@k}  is defined as the percentage of nodes in
$\hat{V}_k$ that coincides with the actual top-$k$ results $V_k$. In
our experiment, we set $k=50$.
On each dataset, we sample $100$ target query nodes according to their
degrees and report the averages of the {\em MaxAdditiveErr} and {\em Precision@k} for each evaluated methods.

% then the specific calculating formulas are shown below.
% %We use the below formula to calculate the three metrics.

% Here $t$ represents one of query node and we set $k=50$.
% Take average value between each query node to return the final scores.
% Besides, we use {\em Precision@k} to check the precision of top-k nodes returned by approximation algorithms.

\header{\bf Results. }
We evaluate  the performance of RBS against Backward
Search~\cite{lofgren2013personalized} for the single-target PPR query.
For RBS, we set
$\lambda(u)=\sqrt{d_{out}(u)}$ and $\theta = \e$ for additive error, 
$\lambda(u)=1$ and $\theta = \delta$
for relative error.  For
Backward Search (BS), we set $\e = \delta$ for relative error.
We vary the additive error parameter $\e$ and relative threshold
$\delta$ from $0.1$ to $10^{-6}$ in
both experiments. Following previous work~\cite{WangTXYL16}, we set the decay factor
$\alpha$ to be $0.2$.
% In the algorithms of Backward Search, it requires to pick the node with largest residue for each iteration.
% We use priority queue to cope with this function.

%The detailed explanation for Power Method is given in section~\ref{sec:pre}.
%According to the setting of iteration numbers, we

% In order to evaluate the results' additive error and relative error respectively.

% We set $\lambda(u)=\sqrt{d_{out}(u)}$ in RBS and

Figure~\ref{fig:Absmaxerr-query} shows the tradeoffs between the {\em
  MaxAdditiveErr} and the query time for the additive error
experiments. Figure~\ref{fig:precision-query} presents the tradeoffs
between {\em Precision@k} and the query time for the relative error experiments.
In general, we  observe that under the same approximation quality, RBS outperforms BS by
orders of magnitude in terms of query time. We also observe that RBS
offers a more significant advantage over BS when we need high-quality
estimators. In particular, to obtain an additive error of $10^{-6}$ on
IT, we observe a 100x query time speedup for RBS.
From
Figure~\ref{fig:precision-query}, we also observe that the precision
of RBS with relative error approaches $1$ more rapidly, which concurs with  our
theoretical analysis.

% also observe
% from
% The results suggest that
% the precision of RBS approaches $1$

% The closer to $1$ means more accurate to find the top-$50$ nodes.
% %For convenience to observation, we omits the precision results below $0.6$.
% The experiments shows RBS can becomes $1$ faster than BS, which means RBS can find the exact top-$k$ nodes in less time.

\vspace{-2mm}
\subsection{Applications} \label{subsec:threeapplications}
We now evaluate RBS in three concrete applications:
heavy hitters query, single-source SimRank computation and approximate
PPR matrix approximation.

\header{\bf Heavy hitters PPR query.}
Recall that in section~\ref{sec:applications},
the heavy hitter of target node $t$ is defined as node $s$ with
$\pi(s,t)> \phi \cdot n \pi(t)$, where $\phi$ is a parameter.
Following~\cite{wang2018heavyhitters}, we fix $\phi=10^{-5}$ in our
experiments. Since the number of true heavy hitters is oblivious to
the algorithms, we use the {\em F1 score} instead of precision to
evaluate the approximate algorithms, where the {\em F1 score} is
defined as $F1=\frac{2\cdot precision \cdot recall}{precision+recall}$.
We set
$\lambda(u)=1$ and $\theta = \delta$ for RBS, and $\e = \delta$ for
BS, and set $ \delta$  from  $0.1$ to $10^{-6}$ to illustrate how the {\em
  F1 score} varies with the query time.% for RBS and BS. 

%we can pick out the heavy hitters of target node $t$ that $\pi(s,t)> \phi \cdot n \pi(t)$.
%Set $\phi=10^{-5}$ according to the conclusion given
%in~\cite{wang2018heavyhitters}.

% Based on RBS and BS's single-target PPR results and ground truths,
% we can return the node set of heavy hitters respectively.
% In order to compare the accuracy of the two methods,
% we choose {\em F1score} as the metric to do the evaluation.
% %And use PageRank's definition formula to derive $\pi(t)$ for each query node.
% %The we can obtain the node set of approximating heavy hitters of each methods separately.
% %Here we still use the Power Method results as ground truth, the same as section~\ref{subsec:stQuery}.
% Denote $H(t)$ and $\hat{H}(t)$ as the heavy-hitter sets returned by ground truths or approximation methods for target node $t$.
% %Then we can calculate {\em F1score} to assess the accuracy of every methods.
% {\em F1score} should be computed as below.
% %The definition of {\em F1score} is shown below.

% \begin{equation}
% \begin{aligned}
% F1score=\frac{2\cdot precision \cdot recall}{precision+recall}
% \end{aligned}
% \end{equation}
% where $precision=\frac{\left| H(t) \cap \hat{H}(t) \right|}{\left|\hat{H}(t) \right|}$ and $recall=\frac{\left| H(t) \cap \hat{H}(t) \right|}{\left|H(t) \right|}$.
% Notation $\left| \cdot \right|$ represents the number of elements in the corresponding set.
%the  {\em F1 score} and query time for RBS and BS. 
In general, RBS
returns a higher  {\em F1 score}  than BS does, given the same amount
of query time. In particular, to achieve an {\em F1 score} of 1 on the
IC dataset, RBS requires a query time that is 80x less than BS
does. The results suggest that by replacing BS with RBS, we can
improve the performance of heavy hitters PPR queries.

% We can observe that RBS can return higher {\em F1 score} than Backward
% Search during the same query time.
% Higher {\em F1score} reveals the higher accuracy for finding results.
% Hence, the import of RBS in heavy hitters problems do enhance the searching efficiency.

\begin{figure*}[!t]
%	\begin{small}
		\centering
		\vspace{-1mm}
		%    \begin{footnotesize}
		\begin{tabular}{ccccc}
			%\multicolumn{4}{c}{\hspace{-4mm} \includegraphics[height=5mm]{./Figs/legend_large.eps}} \vspace{-1mm} \\
			\hspace{-6mm} \includegraphics[height=27mm]{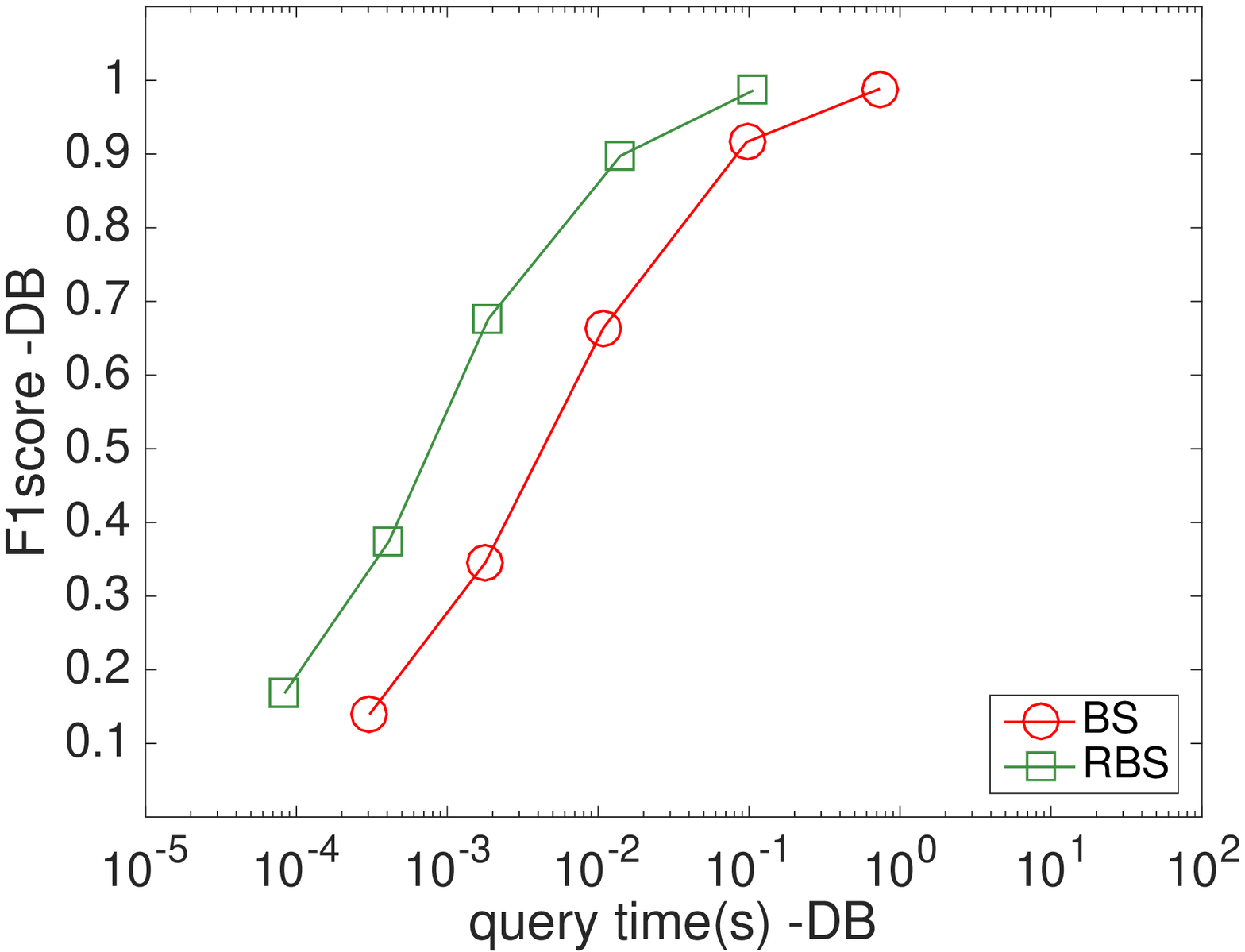} &
			\hspace{-3mm} \includegraphics[height=27mm]{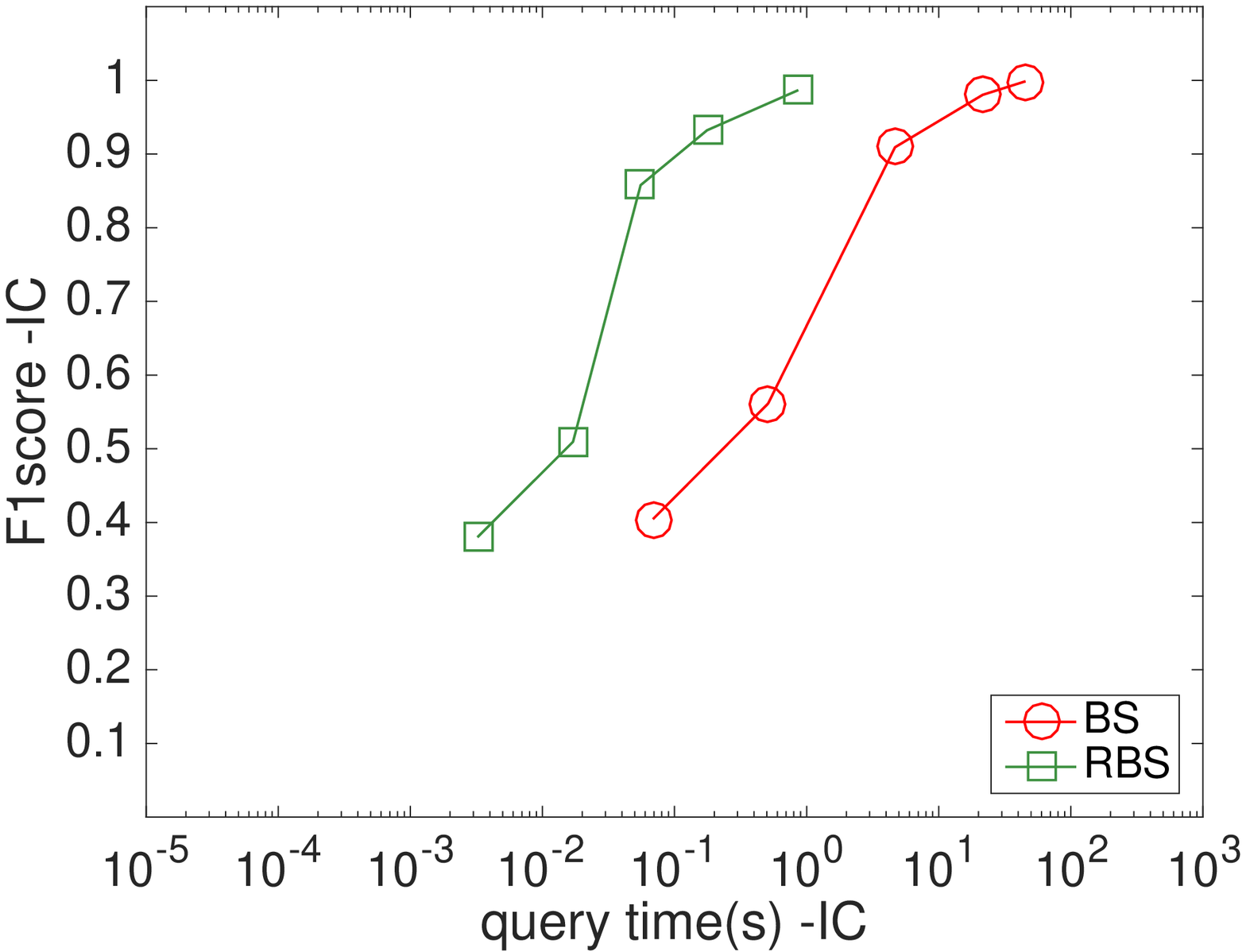}&
			\hspace{-3mm} \includegraphics[height=27mm]{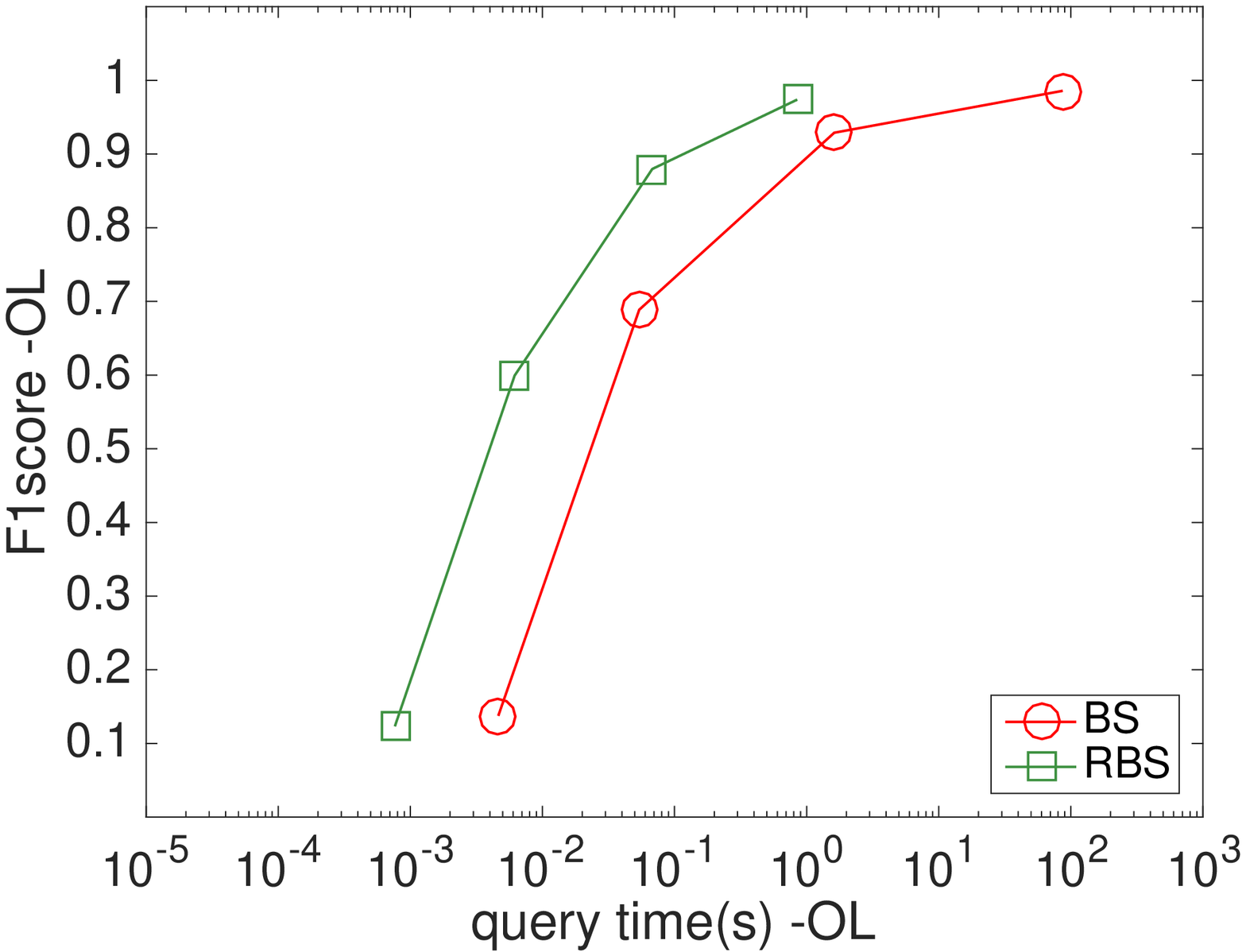}&
			\hspace{-3mm} \includegraphics[height=27mm]{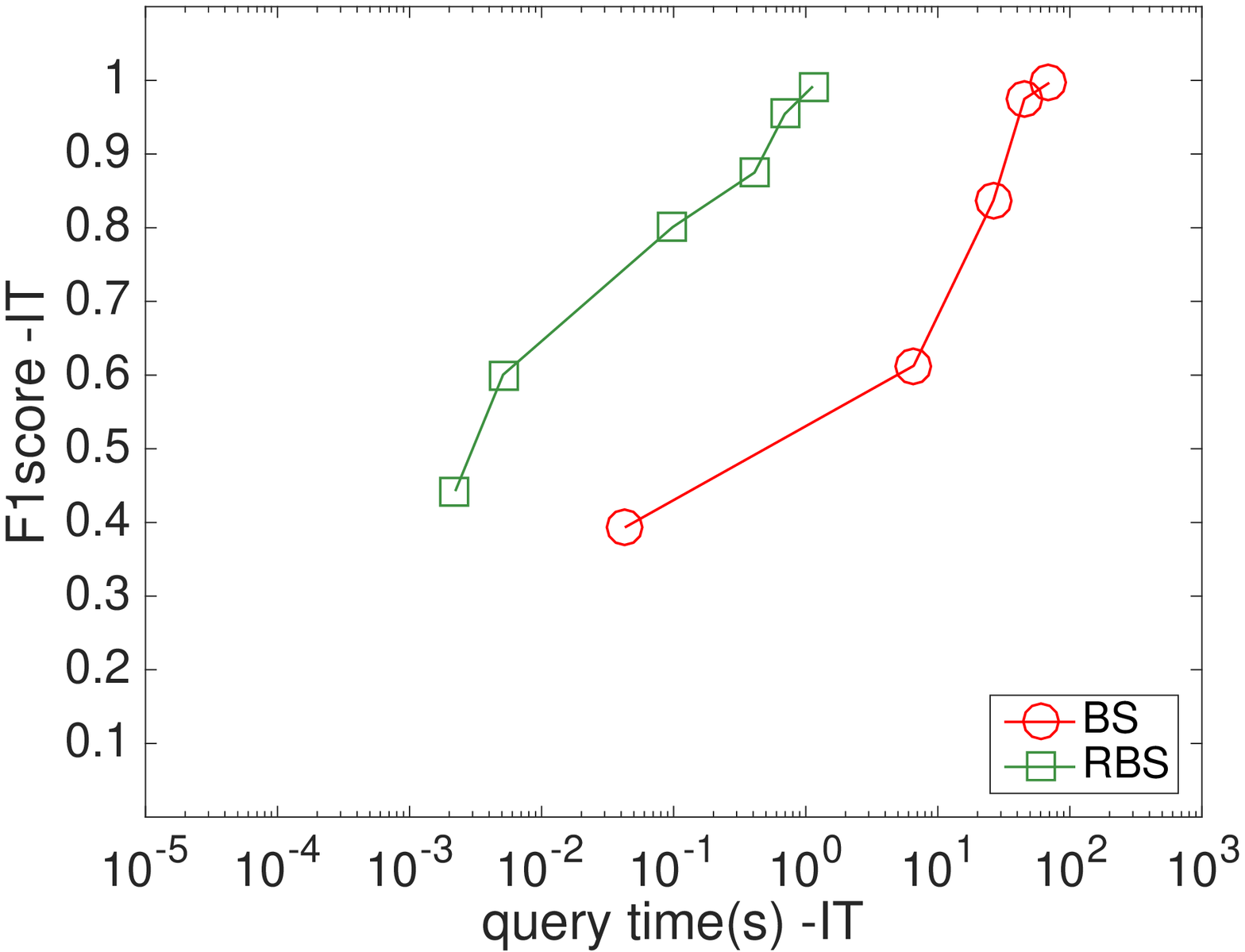}&
			\hspace{-3mm} \includegraphics[height=27mm]{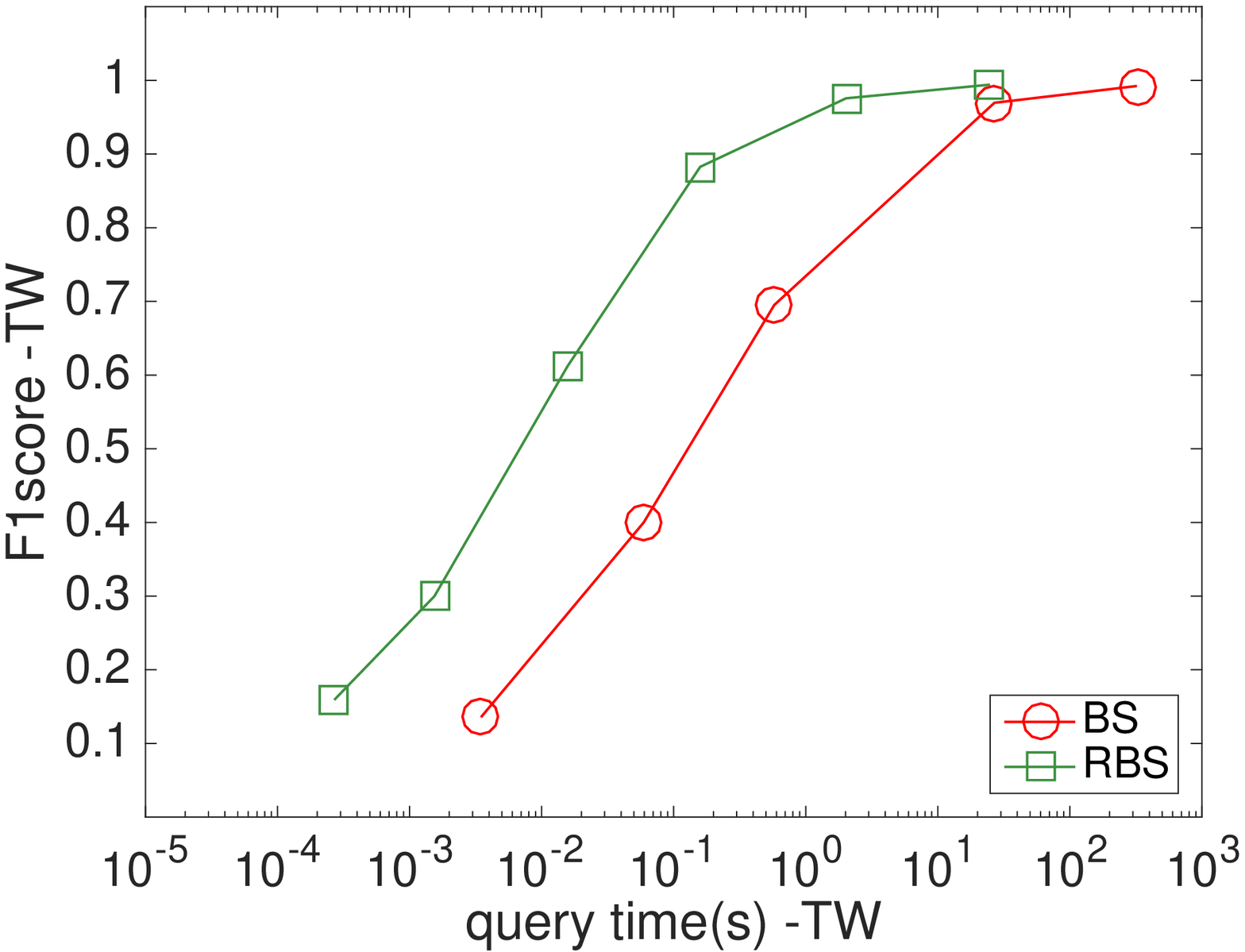}
		\end{tabular}
		\vspace{-3mm}
		\caption{ Heavy hitters: Tradeoffs between {\em F1 score} and query time.}
		\label{fig:f1score-query}
		\vspace{-1 mm}
%	\end{small}
\end{figure*}

\header{\bf Single-source SimRank computation.}
% \header{\bf Improvement in PRSim~\cite{wei2019prsim}.}
As claimed in section~\ref{sec:applications}, SimRank can be expressed in terms of PPR values as below:
%$s(u,v)=\frac{1}{(1-\sqrt{c})^2}\sum_{\ell=0}^{\infty}\sum_{w \in V} \pi_\ell(u,w)\pi_\ell(v,w)\eta(w).$
\vspace{-2mm}
\begin{align}
\vspace{-3mm}
s(u,v)=\frac{1}{(1-\sqrt{c})^2}\sum_{\ell=0}^{\infty}\sum_{w \in V} \pi_\ell(u,w)\pi_\ell(v,w)\eta(w).
\vspace{-6mm}
\end{align}
The state-of-the-art single-source SimRank methods, SLING~\cite{TX16} and
PRSim~\cite{wei2019prsim}, pre-compute the $\ell$-hop PPR values $\pi_\ell(v,w)$ with the
Backward Search algorithm
and store them in index.
%The computing optimizing for $\pi_\ell(u,w)$ can
%Because PRSim only compute $\pi_\ell(u,w)$ with large
We take PRSim %~$\footnote{ https://github.com/wzskytop/PRSim-Code}$ 
as example to show the benefit from replacing BS with RBS.

\begin{figure}[!t]
%	\begin{small}
		\centering
		%\vspace{-2mm}
		%    \begin{footnotesize}
		\begin{tabular}{cccc}
			%\multicolumn{4}{c}{\hspace{-4mm} \includegraphics[height=5mm]{./Figs/legend_large.eps}} \vspace{-1mm} \\
			%\hspace{-3mm} \includegraphics[height=30mm]{./Figs/PRSim-maxerr-query-DB.eps} &
			%\hspace{-3mm} \includegraphics[height=30mm]{./Figs/PRSim-maxerr-query-LJ.eps} &
			\hspace{-3mm} \includegraphics[height=33mm]{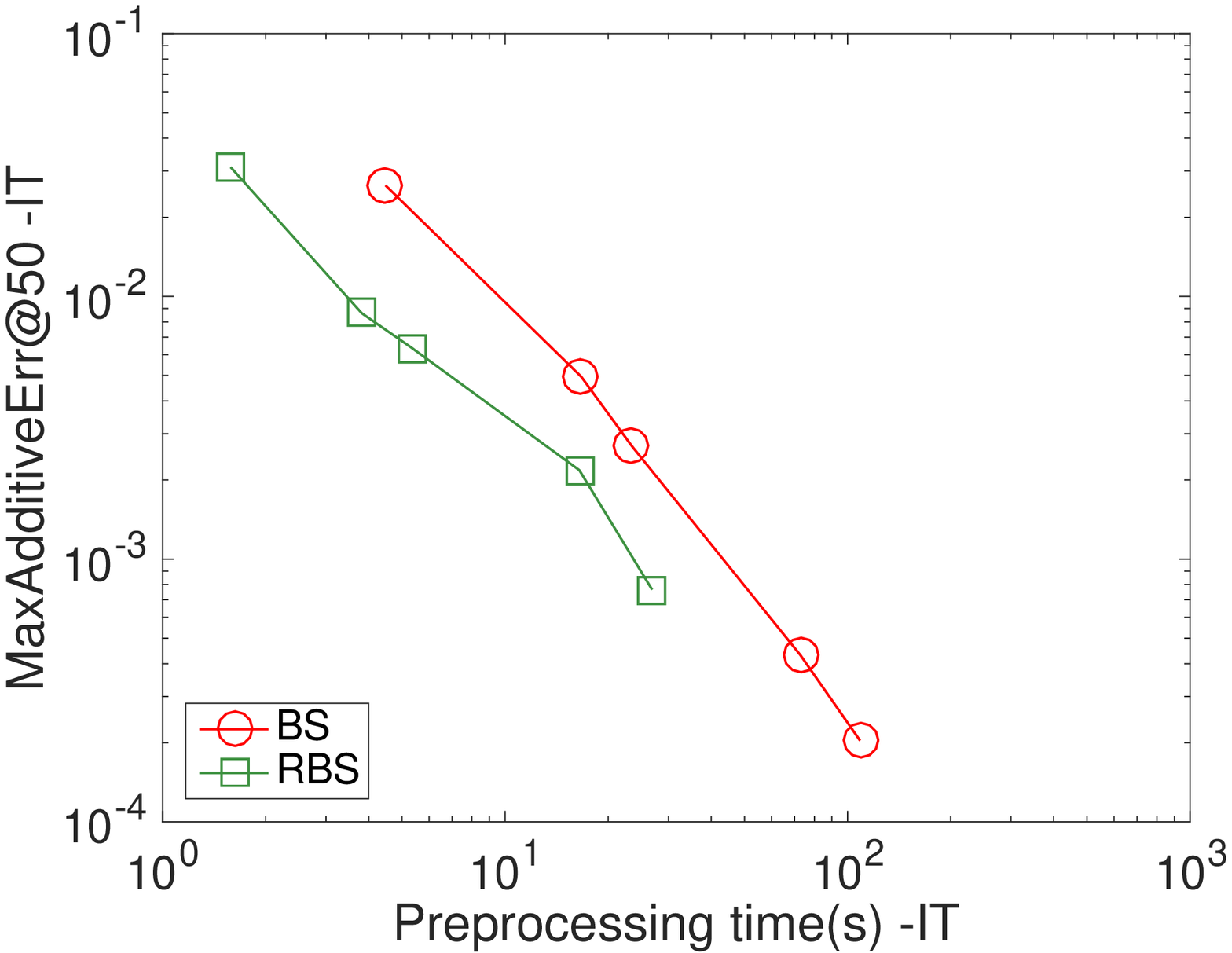} &
			\hspace{-3mm} \includegraphics[height=33mm]{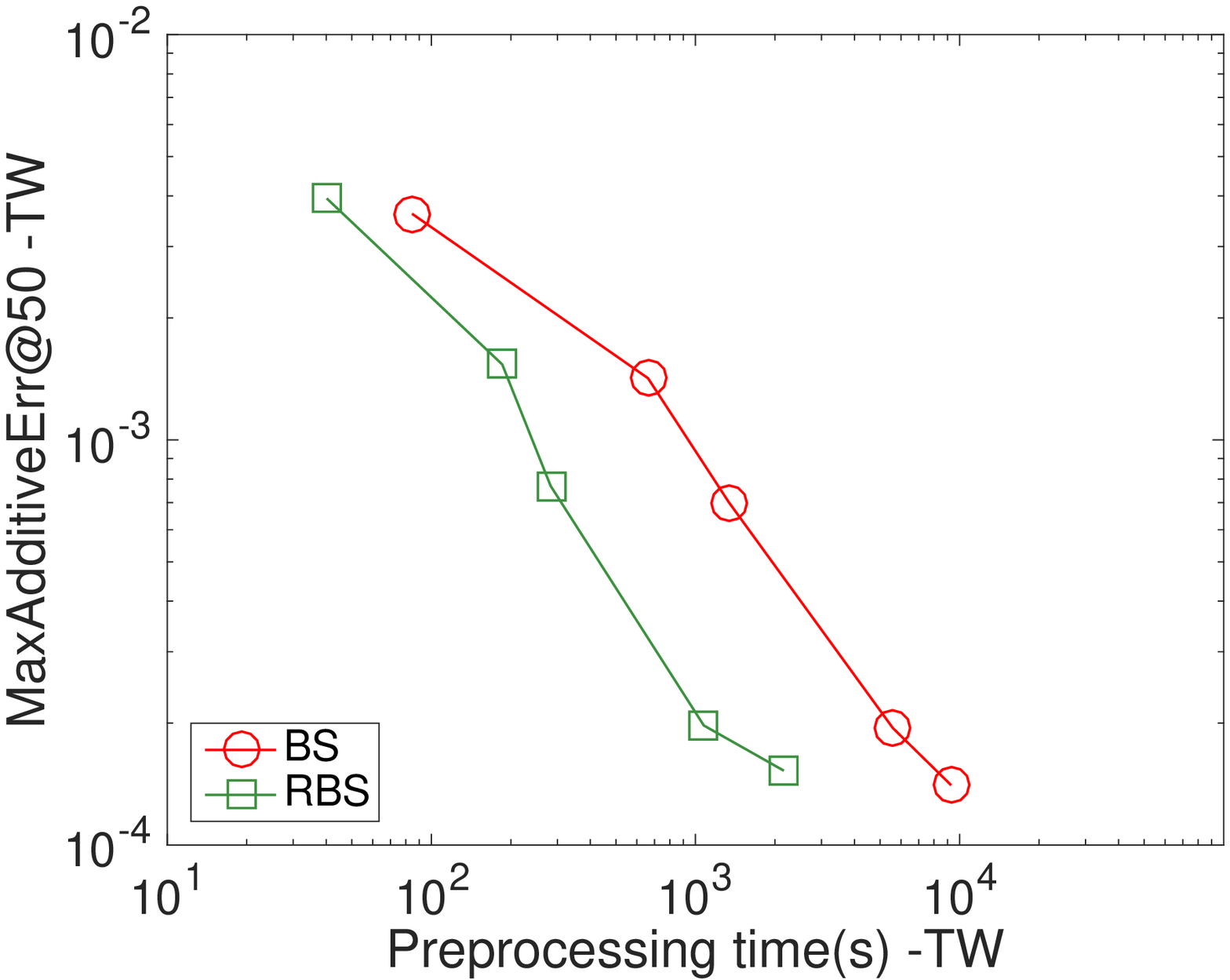}
		\end{tabular}
		\vspace{-3mm}
		\caption{ Single-source SimRank: Tradeoffs between
                  {\em MaxAdditiveErr@50} and preprocessing time.}
		\label{fig:PRSim-maxerr-query}
		\vspace{-3mm}
%	\end{small}
\end{figure}

Following~\cite{wei2019prsim}, we evaluate the tradeoffs between the preprocessing time
with {\em MaxAdditiveErr@50}, the maximum additive error of
single-source top-$50$ SimRank values for a given query node. PRSim has one error parameter $\e$. 
We vary it in $\{0.5,0.1,0.05,0.01,0.005\}$ to plot the tradeoffs. We sample $100$ query nodes
uniformly and use the {\em pooling} method  in~\cite{wei2019prsim}
to derive the actual top-$50$ SimRank values for each query node, and
return the average of the {\em MaxAdditiveErr@50} for each approximate
method. Figure~\ref{fig:PRSim-maxerr-query} illustrates the tradeoffs between the preprocessing time
with {\em MaxAdditiveErr@50}. We observe that by replacing BS with
RBS, we can achieve a significantly lower preprocessing time without
increasing the approximation quality of the single-source SimRank
results. This suggests RBS also outperforms BS  for computing
$\ell$-hop PPR to a target node.

% Originally, it use Backward Search to derive $\pi_\ell(v,w)$ for large PageRank node $w$.
% Replace BS with RBS and observe the change of preprocessing time along with results error.
% The trade-off plots are shown in figure~\ref{fig:PRSim-maxerr-query}.
% PRSim uses the same metric {\em MaxAdditiveErr@50} explained in section~\ref{subsec:stQuery}.

\header{\bf Approximate PPR matrix.}
An approximate PPR matrix consists of PPR estimators for all pairs of
nodes, and is widely used in graph learning.
Recall that
PPRGo~\cite{Xu2019PPRGo}
proposes to apply {\em Forward Search (FS)} to each source node $s\in V$ to
construct the approximate PPR matrix, while STRAP~\cite{wei2019strap}
employs Backward Search (BS) to each target node $t \in V$ to compute
$\pi(s,t), s\in V$ and then put $\pi(s,t)$ into an inverted list
indexed by $s$. We evaluate RBS against FS and BS in terms of additive
error and running time for computing the approximate PPR matrix.

\begin{figure}[!t]
%	\begin{small}
		\centering
		%\vspace{-1mm}
		%    \begin{footnotesize}
		\begin{tabular}{cc}
			%\multicolumn{4}{c}{\hspace{-4mm} \includegraphics[height=5mm]{./Figs/legend_large.eps}} \vspace{-1mm} \\
			\hspace{-3mm} \includegraphics[height=33mm]{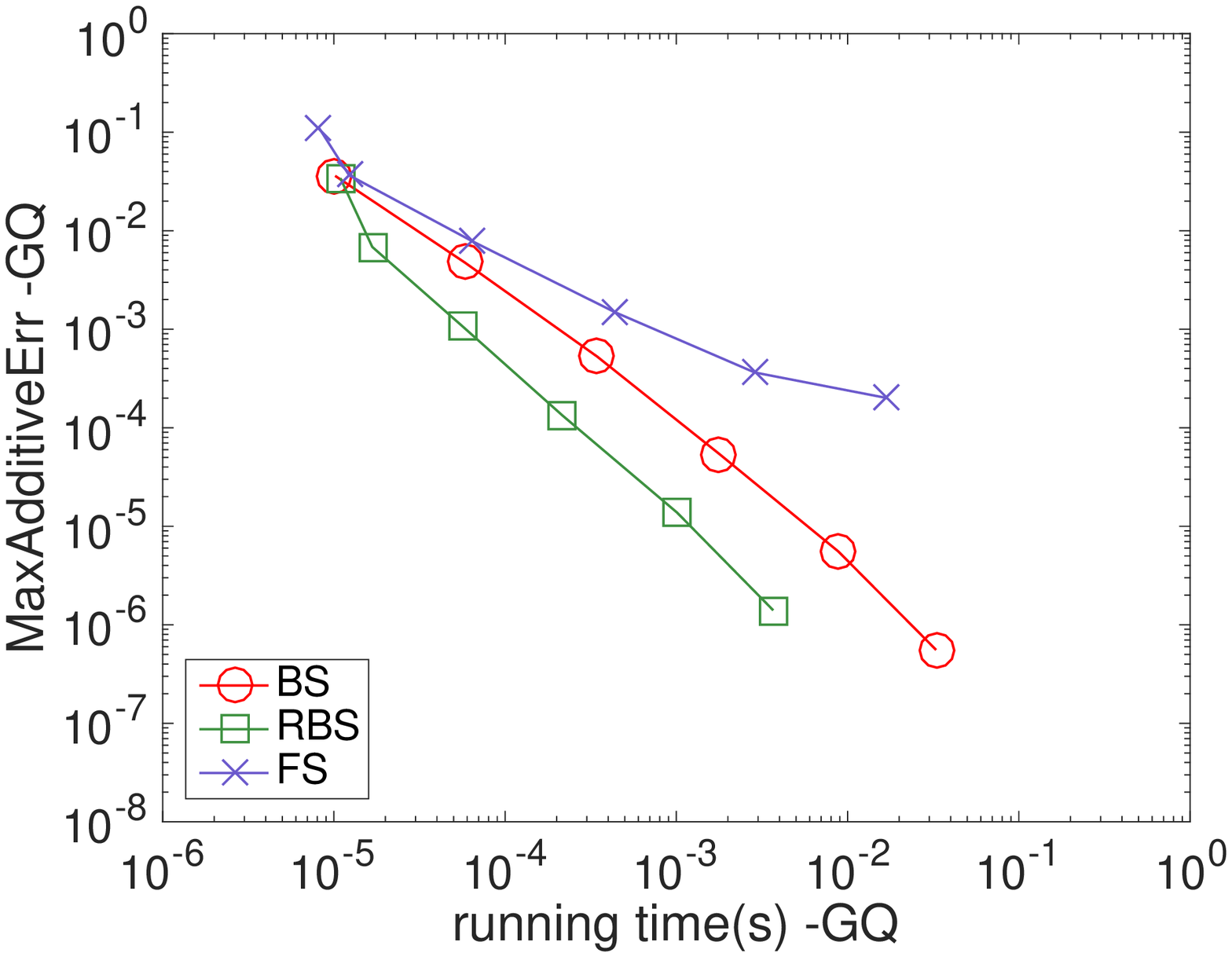} &
			\hspace{-3mm} \includegraphics[height=33mm]{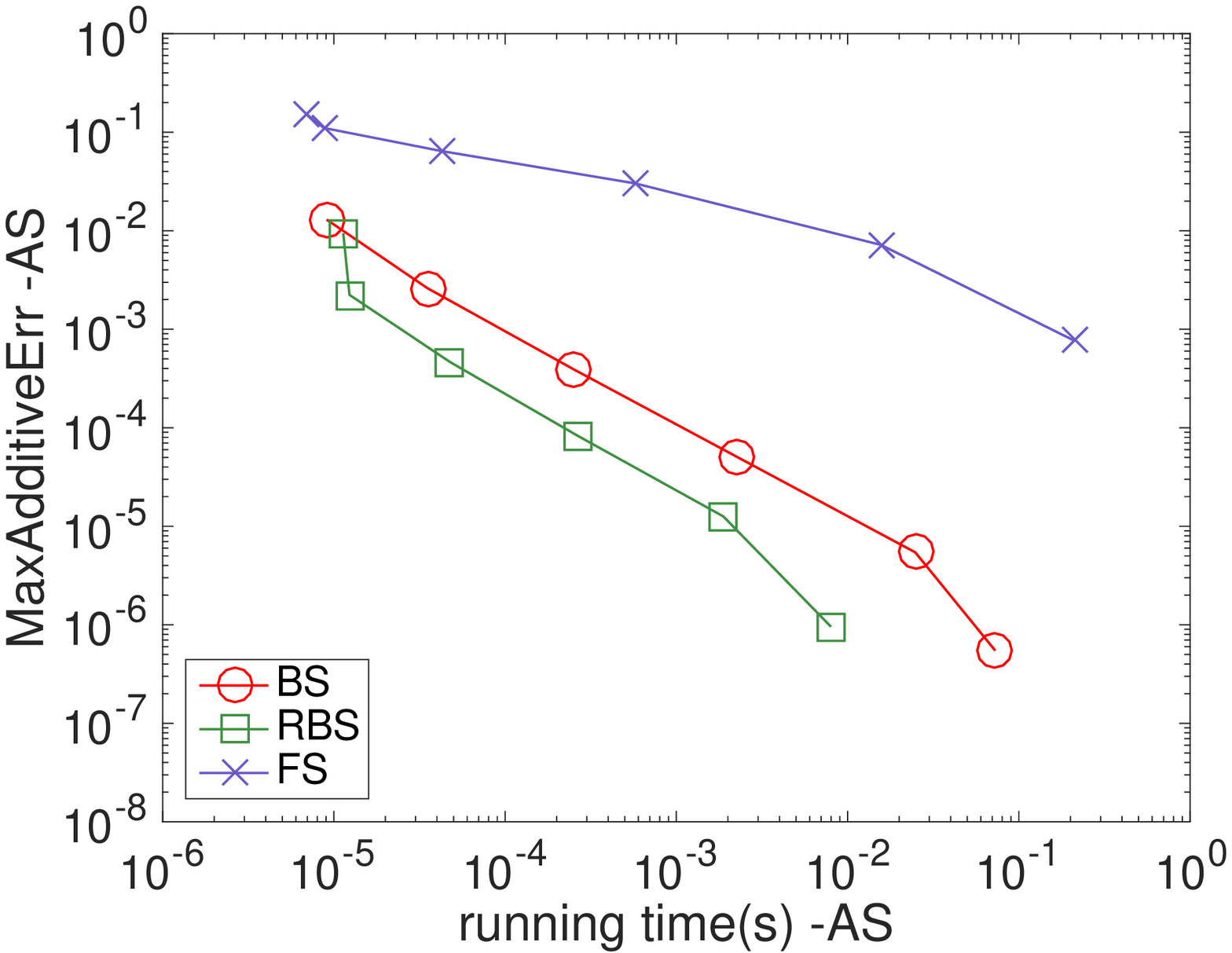}
			%\hspace{-0mm} \includegraphics[height=30mm]{./Figs/Allpairs-maxerr-query-GQ.jpg} &
			%\hspace{-3mm} \includegraphics[height=30mm]{./Figs/Allpairs-maxerr-query-AS.jpg}
		\end{tabular}
		\vspace{-3mm}
		\caption{ Approximate PPR matrix: tradeoffs between
                  {\em MaxAdditiveErr} and running time.}
		\label{fig:Allpairs-maxerr-query}
		\vspace{-3mm}
%	\end{small}
\end{figure}

Due to the scalability limitation of the Power Method, we conduct this
experiment on two small datasets: GQ and AS.
% Because of the infeasibility to apply Power Method for all-pairs PPR's ground truths on large graphs,
% we only conduct PPR matrix experiments on two small graph datasets (GQ and AS).
We set $\lambda(u)=\sqrt{d_{out}(u)}$ and $\theta = \e$ for RBS, and
vary the additive error parameter $\e$ in RBS from $0.1$ to
$10^{-6}$. Similarly,  we vary the parameter $\e$ of FS and BS from $0.1$ to $10^{-6}$.
%Compare {\em MaxAdditiveErr@50} along with query time.
Figure~\ref{fig:Allpairs-maxerr-query} shows the tradeoffs between
{\em MaxAdditiveErr} of PPR values of all node pairs and the running time for the three methods. We first
observe that given the same error budget, BS outperforms FS in terms
of running time. This result concurs with our theoretical analysis that FS only guarantees an
additive error of $\e d_{out}(t)$ while BS guarantees an additive
error of $\e$. Therefore, it may be worthy of taking the extra step to
convert the single-target PPR results into inverted lists indexed by
the source nodes. On the other hand, by replacing BS with RBS, we can
further improve the tradeoffs between the running time  and the
approximation quality, which demonstrates the superiority of ours.

% The calculation equation of {\em MaxAdditiveErr@50} is the same as before. %section~\ref{subsec:stQuery}.
% From figure~\ref{fig:Allpairs-maxerr-query}, we can observe that RBS performs best in the three methods and the gap between Backward Search and Forward Search exists indeed, which highlights the importance of Backward Search.

% % \header{\bf Improvement in PRSim~\cite{wei2019prsim}.}
% A PPR matrix contains the PPR for any pair of nodes and is widely used in graph embedding, GNN and so on.
% According to the analysis in section~\ref{sec:applications},
% replacing Backward Search with RBS can speed up PPR matrix' approximation.
% So we conduct corresponding experiments to verify this conclusion.
% Besides, even though we can also use Forward Search to obtain PPR matrix, the performance of the two methods can be different due to their diverse error bounds.
% Our experiments want to check this conclusion in the mean time.

%Hence, the improvement aiming at Backward Search is more sensible and

% \subsection{Improvement in Heavy hitters~\cite{wang2018heavyhitters}}
% %\header{\bf Improvement in Heavy hitters~\cite{wang2018heavyhitters}.}

%%% Local Variables:
%%% mode: latex
%%% TeX-master: "paper"
%%% End:

\vspace{-1mm}
\section{Conclusion} \label{sec:conclusion}
In this paper, we study  the {\em single-target PPR query}, which
measures the importance of a given target node $t$ to every node $s$
in the graph. We present an algorithm RBS to compute approximate
single-target PPR query with optimal computational complexity. We
show that RBS improves three
concrete applications in graph mining:  heavy hitters PPR
query, single-source SimRank computation, and scalable graph neural
networks. The experiments suggest that RBS 
outperforms the state-of-the-art algorithms in terms of both
efficiency and precision on real-world benchmark datasets. For future
work, we note that a few works combine the Backward Search algorithm with the
Monte-Carlo algorithm to obtain near-optimal query cost for
single-pair queries~\cite{lofgren2014fast,lofgren2015personalized}. An
interesting open problem is whether we can replace the Backward Search
algorithm with RBS to further improve the complexity of these algorithms.

% is whether we can combine RBS with 
% the Monte-Carlo algorithm to improve the complexity of single-pair PPR
% query~\cite{lofgren2014fast,lofgren2015personalized}. 

%%% Local Variables:
%%% mode: latex
%%% TeX-master: "paper"
%%% End:

%\vspace{-2mm}
\section{ACKNOWLEDGEMENTS}
This research is supported by National Natural Science Foundation of China (No. 61832017, No. 61972401, No. 61932001, No.U1936205), by Beijing Outstanding Young Scientist Program NO. BJJWZYJH012019100020098, and by the Fundamental Research Funds for the Central Universities and the Research Funds of Renmin University of China under Grant 18XNLG21. 
Junhao Gan is supported by Australian Research Council (ARC) DECRA DE190101118. 
Sibo Wang is also supported by Hong Kong RGC ECS Grant No. 24203419. 
Zengfeng Huang is supported by Shanghai Science and Technology Commission Grant No. 17JC1420200, and by Shanghai Sailing Program Grant No. 18YF1401200.
%\vspace{-2mm}

%%% Local Variables:
%%% mode: latex
%%% TeX-master: "paper"
%%% End:

%\bibliographystyle{ACM-Reference-Format}
%\bibliography{paper}

\begin{small}
	\bibliographystyle{plain}
	\bibliography{paper}
\end{small}

\appendix
%\clearpage
%\vspace{-2 mm}
\section{appendix} \label{sec:appendix}

% \begin{comment}
% \subsection{Chebyshev's Inequality} \label{sec:chebyshev}
% \vspace{-1mm}\begin{lemma}[Chebyshev's inequality] \label{lmm:chebysev}
% 	Let $X$ be a random variable, then $\Pr\left[\left| X -E[X]\right| \geq \e\right] \le {\Var[X] \over \e^2 }. $
% \end{lemma}
% \vspace{-2mm}
% \subsection{Median Trick} \label{sec:median-of-mean}
% \vspace{-1mm}\begin{lemma}[\cite{charikar2002finding}]\label{lmm:median}
% 	Let $X_1, \ldots, X_k$ be $k \ge 3\log {1\over \delta} $ i.i.d. random variables, such
% 	that $\Pr\left[\left| X_i -E[X_i]\right| \geq \e\right] \le {1\over 3}$.
% 	Let $X =\textrm{Median}_{1 \le i \le k}X_i$, then
% 	$\Pr\left[\left| X -E[X]\right| \geq \e\right] \le \delta$.
% \end{lemma}
% \end{comment}

%\vspace{-1 mm}

\subsection{Proof of Lemma~\ref{lem:BS_bound}}
\begin{proof}
For sake of completeness, we first show that the time complexity of Algorithm~\ref{alg:bp} is bounded by $O \left( \sum_{v \in V} \frac{d_{in}(v)\cdot \pi(v,t)}{\e} \right)$. Recall that in Algorithm~\ref{alg:bp}, 
we pick the node $v$ with the largest residue $r^b(v,t)$, transfer a probability mass of $\alpha r^b(v,t)$ to its reserve $\pi^b(v,t)$, and push the remaining probability mass of $(1-\alpha)r^b(v,t)$ to its in-neighbors. For an in-neighbor $u \in N_{in}(v)$ with out-degree $d_{out}(u)$, the residue $r^b(u,t)$ is increased by $\frac{(1-\alpha)\cdot r^b(v,t)}{d_{out}(u)}$.
Therefore, each push on $v$ adds at least $\alpha r^b(v,t) \ge \alpha \e$ to its reserve  $\pi^b(v,t)$. By equation~\eqref{eqn:BS_error}, we have
\begin{equation*}
	\begin{aligned}
	\pi(v,t)
	=\pi^b(v,t) + \sum_{x \in V}r^b(x,t)\cdot \pi(v,x)\ge \pi^b(v,t). 
	\end{aligned}
\end{equation*}
It follows that the number of pushes on $v$ can be bounded by $\frac{\pi^b(v,t)}{\alpha \e} \le  \frac{\pi(v,t)}{\alpha \e} $. We also observe that each push on $v$ touches all its in-neighbors and thus costs $O(d_{in} (v))$. Consequently, the cost for pushes on $v$ is bounded by $O\left(\frac{d_{in}(v)\cdot \pi(v,t)}{\alpha \e}\right)$, and the total cost of pushes is bounded by $O \left( \sum_{v \in V} \frac{d_{in}(v)\cdot \pi(v,t)}{\e} \right)$. Here we ignore the constant $\alpha$ in the Big-Oh.

Next, we prove the running time of Algorithm~\ref{alg:bp} is also bounded by $O\left(\sum_{u \in V}\frac{d_{out}(u)\cdot \pi(u,t)}{\alpha \e}\right)$. In particular, we use $RP(u)$ to denote the number of times that $u$ receives a push from one of its out-neighbor. Recall that the cost of a push on node $v$ is $O(d_{in} (v))$, so we can charge a cost of $O(1)$ to each of its in-neighbors. Therefore, the running time of Algorithm~\ref{alg:bp} can also be bounded by $O\left(\sum_{u\in V}RP(u)\right)$, the total number of pushes received by any nodes. To derive a upper bound on $RP(u)$, let $r^b(u,t)$ and $\pi^b(u,t)$ denote the residue and reserve of node $u$ when Algorithm~\ref{alg:bp} terminates. By equation~\eqref{eqn:BS_error}, we have
\begin{equation}
\label{eqn:invariant2}
	\begin{aligned}
	\pi(u,t)
	&=\pi^b(u,t)+r^b(u,t)\cdot \pi(u,u)+\sum_{x\neq u}r^b(x,t)\cdot \pi(u,x)\\
	&\ge \pi^b(u,t)+\alpha \cdot r^b(u,t). 
	\end{aligned}
\end{equation}
The last inequality uses the fact that $\pi(u,u)$, the PPR value of a node $u$ to itself, is at least $\alpha$.

Now consider one extra push operation on $u$, even if the residue $r^b(u,t) < \e$. After this push operation, the residue of $u$ becomes $0$, and the reserve of $u$ becomes $\pi^b(u,t)+\alpha \cdot r^b(u,t)$. We observe that for each push that $u$ receives from its out-neighbor $v$, the residue of $u$ is increased by at least $\frac{(1-\alpha)\cdot r^b(v,t)}{d_{out}(u)} \ge \frac{(1-\alpha)\e}{d_{out}(u)} $, and thus the reserve is increased by at least  $\frac{\alpha(1-\alpha)\e}{d_{out}(u)} $. Consequently, the the number of pushes that $u$ receives from any out-neighbors can be bounded as 
\begin{equation*}
RP(u) \le \left.\left(\pi^b(u,t)+\alpha \cdot r^b(u,t)\right) \middle/\frac{\alpha(1-\alpha)\e}{d_{out}(u)} \right.\le \frac{d_{out}(u)\pi(u,t)}{\alpha(1-\alpha)\e}.
\end{equation*}
We use equation~\eqref{eqn:invariant2} in the last inequality.
Consequently, the running time of Algorithm~\ref{alg:bp} is bounded by $\sum_{u \in V} RP(u) = O \left(\sum_{u \in V} \frac{d_{out}(u)\pi(u,t)}{\e}\right)$. Here we ignore the constant $\alpha(1-\alpha)$ in the Big-Oh.
\end{proof}

\vspace{-1 mm}
\subsection{Proof of Lemma~\ref{lem:vbs2_unbiasedness}}
\begin{proof}
	During a push operation through edge $(u,v)$ from level $\ell$ to level $\left(\ell+1\right)$, we denote $X_{\ell+1}(u,v)$ as $\epi_{\ell+1}(u,t)$'s increments caused by this push.
	%When the push process comes to level $\ell$ and all of level $i$ ($i < \ell$) have been finished
	According to Algorithm~\ref{alg:vbs2}, 
	$X_{\ell+1}(u,v)$ is assigned as $\frac{(1-\alpha)\epi_{\ell}(v,t)}{d_{out}(u)}$ deterministically if $\frac{(1-\alpha)\epi_{\ell}(v,t)}{d_{out}(u)} \ge \frac{\alpha \theta}{\lambda(u)}$. 
	Otherwise, $X_{\ell+1}(u,v)$ will be $\frac{ \alpha \theta }{\lambda(u)}$ with probability $\frac{\lambda(u) \cdot (1-\alpha)\epi_{\ell}(v,t)}{\alpha \theta \cdot d_{out}(u)}$, or $0$ with the left probability.
	If we use $\{ \epi_{\ell} \}$ to denote the set of $\epi_{\ell}(v, t)$ for all $v\in V$, 
	the expectation of $X_{\ell+1}(u,v)$ conditioned on all estimators $\{ \epi_{\ell} \}$ can be derived that
	\vspace{-1mm}\begin{equation}%\nonumber
	\label{eqn:incre2_expectation}
	\begin{aligned}
	\E \left[ X_{\ell+1}(u,v) \mid \{\epi_{\ell}\} \right]
	&=\left\{
	\begin{array}{ll}
	\frac{(1-\alpha)\epi_{\ell}(v,t)}{d_{out}(u)}, \quad if \frac{(1-\alpha)\epi_{\ell}(v,t)}{d_{out}(u)} \ge \frac{ \alpha \theta }{\lambda(u)}\\
	\frac{ \alpha \theta }{\lambda(u)} \cdot \frac{\lambda(u) (1-\alpha)\epi_{\ell}(v,t)}{\alpha \theta \cdot d_{out}(u)}, \quad otherwise
	\end{array} 
	\right.\\
	&= \frac{(1-\alpha)\epi_{\ell}(v,t)}{d_{out}(u)}.
	\end{aligned}
	\end{equation}
	% \begin{comment}
	% 	So the increments from level $\ell$ to level $(\ell+1)$ through $u$ to $v$ is determined by $\epi_\ell(u,t)$'s specific value.
	% Conditioned on $\epi_\ell(u,t)$, the expectation of $X_{l}(u,v)$  $\E\left[ X_{l}(u,v) \mid \{\epi_\ell\} \right]$ is settled as $\frac{(1-\alpha)\epi_\ell(u,t)}{d_{out}(v)}$.
	% But if $\frac{(1-\alpha)\epi_\ell(u,t)}{\alpha d_{out}(v)} < \theta$, $X_{l}(u,v)$'s value will be influenced by the random number $r$. 
	% When $r \leq \frac{(1-\alpha)\epi_\ell(u,t)}{\alpha \theta d_{out}(v)} $, namely with $\frac{(1-\alpha)\epi_\ell(u,t)}{\alpha \theta d_{out}(v)} $ probability, we set $X_{l}(u,v)$ as $\alpha \e$, or $0$ with left probability. Hence, in this scenario, 
	% \begin{align}
	% \label{eqn:expectation_x2}
	% \E\left[ X_{l}(u,v) \mid \{\epi_\ell\} \right] = \alpha \theta \cdot \frac{(1-\alpha)\epi_\ell(u,t)}{\alpha \theta d_{out}(v)} = \frac{(1-\alpha)\epi_\ell(u,t)}{d_{out}(v)}
	% \end{align}
	% In general, $\E\left[ X_{l}(u,v) \mid \{\epi_\ell\} \right] = \frac{(1-\alpha)\epi_\ell(u,t)}{d_{out}(v)}$ both com from the two situations.
	% \end{comment}
	%Based on this, we can derive $\epi_{\ell}(u,t)$'s conditional expectation utilizing the fact that $\epi_{\ell}(u,t)=\sum_{v \in N_{out}(u)} X_{\ell}(u,v)$.
	%In the following, we use $\{ \epi_{\ell-1} \}$ to denote the set of $\epi_{\ell-1}(v, t)$ for all $v\in V$, 
	%and $\E\left[\epi_{\ell}(u,t) \mid \{ \epi_{\ell-1} \} \right]$ denote the conditional expectation of the estimator
	Because $\epi_{\ell+1}(u,t)=\sum_{v \in N_{out}(u)} X_{\ell+1}(u,v)$, 
	the conditional expectation of $\epi_{\ell+1}(u,t)$ conditioned on the value of all estimators $\{ \epi_{\ell} \}$ at $\ell$-th level can be derived that 
	\vspace{-1mm}\begin{equation}\nonumber
	\begin{aligned}
		\E \left[\epi_{\ell+1}(u,t) \mid \{ \epi_{\ell} \} \right]
	&= \E \left[ \sum_{v \in N_{out}(u)} X_{\ell+1}(u,v) \mid \{ \epi_{\ell} \}\right]\\
	&= \sum_{v \in N_{out}(u)} \E \left[ X_{\ell+1}(u,v) \mid \{ \epi_{\ell} \}\right].
	\end{aligned}
	\end{equation}
	%Applying the conclusion in equation~\eqref{eqn:incre2_expectation}, we have % $	\E
	Based on equation~\eqref{eqn:incre2_expectation}, we have % $	\E
        % \left[\epi_{\ell+1}(u,t) \mid \{ \epi_{\ell}\} \right]=\sum_{v
        %   \in N_{out}(u)}
        % \left(\frac{(1-\alpha)\epi_{\ell}(v,t)}{d_{out}(u)}\right)$. 
	\begin{align}
	\label{eqn:vbs2_conditional_exp}
	\E \left[\epi_{\ell+1}(u,t) \mid \{ \epi_{\ell}\} \right]=\sum_{v \in N_{out}(u)} \left(\frac{(1-\alpha)\epi_{\ell}(v,t)}{d_{out}(u)}\right).
	\end{align}
	\begin{comment}
	Moreover, with the help of%Applying the fact that
	\begin{align}
	\E \left[ \E \left[\epi_{\ell+1}(v,t) \mid\{ \epi_\ell\}\right] \right]= \E \left[ \epi_{\ell+1}(v,t)  \right]
	\end{align}
	We can change the conditional expectation to $\E \left[\epi_{\ell+1}(v,t) \right]$.
	\end{comment}
	Because $\E \left[ \epi_{\ell+1}(u,t)  \right] =\E \left[ \E \left[\epi_{\ell+1}(u,t) \mid \{ \epi_{\ell} \} \right] \right]$, 
	it follows that
	\vspace{-1mm}\begin{equation}\nonumber
	\begin{aligned}
	\E \left[ \epi_{\ell+1}(u,t)  \right] 
	= \hspace{-2mm}\sum_{v \in N_{out}(u)} \left(\frac{(1-\alpha) \E \left[\epi_{\ell}(v,t)\right]}{d_{out}(u)} \right).
	\end{aligned}
	\end{equation}
	$\E \left[\epi_{i}(x,t)\right]=\pi_{i}(x,t)$ holds for $\forall x \in V$ and $i = 0$ in the initial state, 
	because $\E \left[ \epi_{0}(t,t) \right] = \pi_{0}(t,t)= \alpha$ and $\E \left[ \epi_{0}(u,t) \right]= \pi_{0}(u,t)= 0 $ ($u \neq t$). 
 	%Next, we perform mathematical induction on $i$ and 
 	Assume $\E \left[\epi_{i}(x,t)\right]=\pi_{i}(x,t)$ holds for $\forall x \in V$ and $i \leq \ell$. 
%	Because $\E \left[ \epi_{0}(t,t) \right] = \pi_{0}(t,t)= \alpha$ and $\E \left[ \epi_{0}(u,t) \right]= \pi_{0}(u,t)= 0 $ ($u \neq t$), 
%	this assumption is satiable at least in the initial state.
%	Using the mathematical induction we can get
	We can derive that 
	\vspace{-1mm}\begin{equation}\nonumber
	\begin{aligned}
		\E \left[ \epi_{\ell+1}(u,t)  \right] &= \sum_{v \in N_{out}(u)} \left(\frac{(1-\alpha) \E \left[\epi_{\ell}(v,t)\right]}{d_{out}(u)} \right) \\
	&=\sum_{v \in N_{out}(u)} \left(\frac{(1-\alpha) \pi_{\ell}(v,t)}{d_{out}(u)} \right) = \pi_{\ell+1}(u,t), 
	\end{aligned}
	\end{equation}
	which testifies the unbiasedness.
\end{proof}

\vspace{-1 mm}
\subsection{Proof of Lemma~\ref{lem:vbs2_variance}}
\begin{proof}
During each push operation, the randomness comes from the second scenario that $\frac{(1-\alpha)\epi_{\ell}(v,t)}{d_{out}(u)} < \frac{ \alpha \theta }{\lambda(u)}$.
	Focus on this situation, 
	\vspace{-1mm}\begin{equation}\nonumber
	\begin{aligned}
	&\Var\left[ X_{\ell+1}(u,v) \mid \{ \epi_{\ell} \} \right] 
	\leq \E \left[ X_{\ell+1}^2(u,v) \mid \{ \epi_{\ell} \}\right]\\
	&=\left( \frac{\alpha \theta}{\lambda(u)} \right)^2 \cdot \frac{\lambda(u) \cdot (1-\alpha)\epi_{\ell}(v,t)}{\alpha \theta \cdot d_{out}(u)} 
	= \frac{\alpha \theta}{\lambda(u)} \cdot \frac{(1-\alpha)\epi_{\ell}(v,t)}{d_{out}(u)}.
	\end{aligned}
	\end{equation}
	Note that for each $v \in N_{out}(u) $, $X_{\ell+1}(u,v)$ is independent with each other because of the independent generation for the random number $r$ in Algorithm~\ref{alg:vbs2}. 
	Applying $\epi_{\ell+1}(u,t)=\sum_{v \in N_{out}(u)} X_{\ell+1}(u,v)$, the conditional variance of $\epi_{\ell+1}(u,t)$ is followed that
	\vspace{-1mm}\begin{equation}
	\label{eqn:vbs2_conditional_var}
	\begin{aligned}
		\Var \left[\epi_{\ell+1}(u,t) \mid \{ \epi_{\ell}\} \right]
	%= \Var \left[\sum_{v \in N_{out}(u)} X_{\ell}(u,v) \mid  \epi_{\ell-1}(v,t) \right]\\
	&= \sum_{v \in N_{out}(u)} \Var[X_{\ell+1}(u,v) \mid \{ \epi_{\ell} \}] \\
	&\leq \frac{\alpha \theta}{\lambda(u)} \cdot \sum_{v \in
          N_{out}(u)} \frac{(1-\alpha) \epi_{\ell}(v,t)}{d_{out}(u)}.  
	\end{aligned}
	\end{equation}
	% \begin{comment}
	% \vspace{-1mm}\begin{equation}\nonumber
	% \begin{aligned}
	% &\Var \left[\epi_{\ell+1}(v,t) \mid \epi_\ell(u,t)\right]\\ 
	% &= \Var \left[\sum_{u \in N_{out}(v)} X_{\ell}(u,v) \mid  \epi_\ell(u,t) \right]\\
	% &= \sum_{u \in N_{out}(v)} \Var[X_{\ell}(u,v) \mid  \epi_\ell(u,t)]\\
	% &= \sum_{u \in N_{out}(v)} \left( \E \left[ X_{\ell}^2(u,v) \mid \epi_\ell(u,t)\right]-\E^2\left[X_{\ell}(u,v) \mid \epi_\ell(u,t)\right] \right)\\
	% &\leq \sum_{u \in N_{out}(v)} \E \left[X_{\ell}^2(u,v) \mid \epi_\ell(u,t)\right]\\
	% %&=\left\{
	% %\begin{array}{ll}
	% %\sum_{u \in N_{out}(v)} \frac{(1-\alpha)^2\epi_\ell^2(u,t)}{d_{out}^2(v)}-\frac{(1-\alpha)^2\epi_\ell^2(u,t)}{d_{out}^2(v)} &, if \quad  \frac{(1-\alpha)\epi_\ell(u,t)}{\alpha \sqrt{d_{out}(v)}} %\ge \theta\\
	% %\sum_{u \in N_{out}(v)} \frac{\alpha^2 \theta^2}{d_{out}(v)}\cdot \frac{(1-\alpha)\epi_\ell(u,t)}{\alpha \theta \sqrt{d_{out}(v)}}- \frac{(1-\alpha)^2\epi_\ell^2(u,t)}{d_{out}^2(v)} %&, otherwise
	% %\end{array} 
	% %\right.\\
	% & = \sum_{u \in N_{out}(v)} \alpha^2 \theta^2 \cdot \frac{(1-\alpha)\epi_\ell(u,t)}{\alpha \theta d_{out}(v)}\\
	% & = \alpha \theta \cdot \sum_{u \in N_{out}(v)} \frac{(1-\alpha) \epi_\ell(u,t)}{d_{out}(v)} 
	% \end{aligned}
	% \end{equation}
	% \end{comment}
	By the total variance law, $\Var \left[\epi_{\ell+1}(u,t)\right] = \E \left[ \Var \left[\epi_{\ell+1}(u,t) \mid \{ \epi_{\ell} \} \right] \right] + \Var\left[\E\left[\epi_{\ell+1}(u,t) \mid \{ \epi_{\ell} \}  \right]\right]$.
	% \vspace{-1mm}\begin{equation}
	% \begin{aligned}
	% &\Var \left[\epi_{\ell}(u,t)\right] \\
	% &= \E \left[ \Var \left[\epi_{\ell}(u,t) \mid \{ \epi_{\ell-1} \} \right] \right] + \Var\left[\E\left[\epi_{\ell}(u,t) \mid \{ \epi_{\ell-1} \}  \right]\right] \\
	% %&\leq \alpha \theta \pi_{\ell+1}(v,t) +\frac{(1-\alpha)^2}{d_{out}^2(v)} \cdot \Var \left[\sum_{u \in N_{out}(v)} \epi_\ell(u,t)\right]
	% \end{aligned}
	% \end{equation}
	Based on equation~\eqref{eqn:vbs2_conditional_var} and the unbiasedness of $\epi_{\ell}(v,t)$ proven in Lemma~\ref{lem:vbs2_unbiasedness}, 
	we have
	\vspace{-1mm}\begin{equation}
	\label{eqn:vbs2_eps_totalvar}
	\begin{aligned}
	&\E \left[ \Var \left[\epi_{\ell+1}(u,t) \mid \{ \epi_{\ell} \} \right] \right]
	\leq \frac{\alpha \theta}{\lambda(u)} \cdot \sum_{v \in N_{out}(u)} \frac{(1-\alpha) \E\left[ \epi_{\ell}(v,t) \right]}{d_{out}(u)} \\
	&=\frac{\alpha \theta}{\lambda(u)} \cdot \sum_{v \in N_{out}(u)} \frac{(1-\alpha) \pi_{\ell}(v,t) }{d_{out}(u)} 
	= \frac{\alpha \theta}{\lambda(u)} \cdot \pi_{\ell+1}(u,t).
	\end{aligned}
	\end{equation}
	Meanwhile, applying equation~\eqref{eqn:vbs2_conditional_exp}, we can derive
	\vspace{-1mm}\begin{equation}\nonumber
	\begin{aligned}
	\Var\left[\E\left[\epi_{\ell+1}(u,t) \mid \{ \epi_{\ell}\} \right]\right] 
	&= \Var \left[ \sum_{v \in N_{out}(u)} \left(\frac{(1-\alpha)\epi_{\ell}(v,t)}{d_{out}(u)}\right)  \right]\\
	&=\frac{(1-\alpha)^2}{d^2_{out}(u)} \cdot \Var\left[ \sum_{v \in N_{out}(u)} \epi_{\ell}(v,t)\right].
	%&\leq \frac{(1-\alpha)^2}{d_{out}(u)} \cdot \sum_{v \in N_{out}(u)} \Var\left[ \epi_{\ell-1}(v,t)\right]
	\end{aligned}
	\end{equation}
	The convexity of variance implies that: %$	\Var \left[\sum_{v \in N_{out}(u)} \epi_{\ell}(v,t)\right] \leq d_{out}(u) \cdot \sum_{v \in N_{out}(u)} \Var\left[\epi_{\ell}(v,t)\right]$.
	\begin{align}\nonumber
	\Var \left[\sum_{v \in N_{out}(u)} \epi_{\ell}(v,t)\right] \leq d_{out}(u) \cdot \sum_{v \in N_{out}(u)} \Var\left[\epi_{\ell}(v,t)\right]. 
	\end{align}
	Therefore, we can rewrite $\Var\left[\E\left[\epi_{\ell+1}(u,t) \mid \{ \epi_{\ell}\} \right]\right] $ as below:
	\vspace{-1mm}\begin{equation}
	\label{eqn:vbs2_var_totalvar}
	\begin{aligned}
	\Var\left[\E\left[\epi_{\ell+1}(u,t) \mid \{ \epi_{\ell} \} \right]\right] 
	\leq \frac{(1-\alpha)^2}{d_{out}(u)} \cdot \sum_{v \in N_{out}(u)} \Var\left[ \epi_{\ell}(v,t)\right].
	\end{aligned}
	\end{equation}
	Applying equation~\eqref{eqn:vbs2_eps_totalvar} and equation~\eqref{eqn:vbs2_var_totalvar}, we can derive that
	\begin{align}\nonumber
	\Var \left[\epi_{\ell+1}(u,t)\right]\leq \frac{\alpha \theta}{\lambda(u)} \cdot \pi_{\ell+1}(u,t) + \frac{(1-\alpha)^2}{d_{out}(u)} \cdot \hspace{-2mm} \sum_{v \in N_{out}(u)}\hspace{-2mm} \Var\left[\epi_{\ell}(v,t)\right].
	\end{align}
	\begin{comment}
	\vspace{-1mm}\begin{equation}\nonumber
	\begin{aligned}
	&\Var \left[\epi_{\ell+1}(v,t)\right]\\
	& = \E \left[ \Var \left[\epi_{\ell+1}(v,t) \mid \epi_\ell\right] \right] + \Var\left[\E\left[\epi_{\ell+1}(v,t) \mid \epi_\ell\right]\right] \\
	& \leq \E \left[\alpha \theta \cdot \sum_{u \in N_{out}(v)} \frac{(1-\alpha) \epi_\ell(u,t)}{d_{out}(v)} \right] + \Var \left[\sum_{u \in N_{out}(v)} \left(\frac{(1-\alpha)\epi_\ell(u,t)}{d_{out}(v)} \right)\right] \\
	& = \alpha \theta \epi_{\ell+1}(v,t) +\frac{(1-\alpha)^2}{d_{out}^2(v)} \cdot \Var \left[\sum_{u \in N_{out}(v)} \epi_\ell(u,t)\right] \\
	& \leq  \alpha \theta \epi_{\ell+1}(v,t) + \frac{(1-\alpha)^2}{d_{out}^2(v)} \cdot d_{out}(v) \cdot \sum_{u \in N_{out}(v)} \Var\left[\epi_\ell(u,t)\right]\\
	& = \alpha \theta \epi_{\ell+1}(v,t) + \frac{(1-\alpha)^2}{d_{out}(v)} \cdot \sum_{u \in N_{out}(v)} \Var\left[\epi_\ell(u,t)\right]
	\end{aligned}
	\end{equation}
	\end{comment}
	$\Var[\epi_{i}(x,t)] \leq \frac{\theta}{\lambda(u)} \cdot \pi_{i}(x,t)$ holds for $\forall x \in V$ when $i=0$, 
	because $\Var[\epi_{0}(x,t)]=0$. 
	Assume $\Var[\epi_{i}(x,t)] \leq \frac{\theta}{\lambda(u)} \cdot \pi_{i}(x,t)$ holds for $\forall x \in V$ and $i < \ell$.  
	Using mathematical induction, we can derive that
	\vspace{-1mm}\begin{equation}%\nonumber
	\label{eqn:variance2}
	\begin{aligned}\nonumber
	&\Var\left[\epi_{\ell+1}(u,t)\right]
	%\leq \alpha \theta \epi_{\ell+1}(v,t) + \frac{(1-\alpha)^2}{d_{out}(v)} \cdot \sum_{u \in N_{out}(v)} \theta \epi_{\ell}(u,t)\\
	\leq \frac{\alpha \theta}{\lambda(u)} \cdot \pi_{\ell+1}(u,t) + \frac{(1-\alpha)^2 \theta}{\lambda(u)\cdot d_{out}(u)} \cdot \hspace{-2mm} \sum_{v \in N_{out}(u)} \hspace{-2mm}\pi_{\ell}(v,t)\\
	&= \frac{\alpha \theta}{\lambda(u)}  \cdot \pi_{\ell+1}(u,t) + \frac{(1-\alpha) \theta}{\lambda(u)}  \cdot \pi_{\ell+1}(u,t) =\frac{ \theta }{\lambda(u)} \cdot \pi_{\ell+1}(u,t).
	\end{aligned}
      \end{equation}

For relative error, we set $\lambda(u)=1$ that
$\Var \left[ \epi_{\ell+1}(u,t) \right] \leq  \theta \cdot \pi_{\ell+1}(u,t).$

For additive error, we set $\lambda(u) = \sqrt{d_{out}(u)}$, and it follows that
\vspace{-1mm}\begin{equation}\nonumber
\label{eqn:additive_conditional_var}
\begin{aligned}
&\Var\left[\epi_{\ell+1}(u,t)\right]
\leq \frac{ \theta }{\lambda(u)} \cdot \pi_{\ell+1}(u,t)
=\frac{ \theta }{\lambda(u)} \cdot \sum_{v \in N_{out}(u)} \frac{(1-\alpha) \pi_\ell(v,t)}{d_{out}(u)}.
%= \Var \left[\sum_{v \in N_{out}(u)} X_{\ell}(u,v) \mid  \epi_{\ell-1}(v,t) \right]\\
%= \sum_{v \in N_{out}(u)} \Var[X_{\ell+1}(u,v) \mid \{ \epi_{\ell} \}] \\
%&\leq  \frac{\alpha\theta}{\lambda(u)} \cdot \sum_{v \in N_{out}(u)} \frac{\alpha \theta}{\lambda(u)}=\frac{\alpha^2 \theta^2}{\left( \sqrt{d_{out}(u)} \right)^2} \cdot d_{out}(u) \leq \alpha^2 \theta^2.
\end{aligned}
\end{equation}
Note that $\frac{(1-\alpha) \pi_\ell(v,t)}{d_{out}(u)}<\frac{\alpha \theta}{\lambda(u)}$ in the randomized scenario. 
So,  
\begin{equation}\nonumber
	\begin{aligned}
		\Var\left[\epi_{\ell+1}(u,t)\right]
		\leq \frac{\theta}{\lambda(u)} \cdot \sum_{v \in N_{out}(u)}\frac{\alpha \theta}{\lambda(u)}
		=\frac{\alpha \theta^2}{\lambda^2(u)}\cdot d_{out}(u)
		= \alpha \theta^2. 
	\end{aligned}
\end{equation}

\begin{comment}
	The first inequality is due to the fact that we only introduce randomness if
$\frac{(1-\alpha)\epi_{\ell-1}(v,t)}{d_{out}(u)} \leq
        \frac{\alpha \theta}{\lambda(u)}$. Similar to the proof of the
        relative error, we inductively assume $\Var
        \left[\epi_{\ell-1}(u,t)\right] \le \sum_{i=0}^{\ell-2}
        (1-\alpha)^{2i}\alpha^2 \theta^2$. Combining
        Equation~\eqref{eqn:additive_conditional_var}
        and~\eqref{eqn:vbs2_var_totalvar}, we have 
$$	\Var\left[\epi_{\ell}(u,t)\right] \le  \alpha^2 \theta^2 + (1-\alpha)^2\sum_{i=0}^{\ell-2}
        (1-\alpha)^{2i}\alpha^2 \theta^2 \le \sum_{i=0}^{\ell-1}
        (1-\alpha)^{2i}\alpha^2 \theta^2.$$
The Lemma follows by observing         $\sum_{i=0}^{\ell-1}
        (1-\alpha)^{2i} \le 1/\alpha$.
\end{comment}
\vspace{-5 mm}
\end{proof}

%\vspace{-2 mm}
\subsection{Proof of Lemma~\ref{lem:vbs2_cost}}
\begin{proof} 
	Let $C_{\ell+1}(u,v)$ denote the cost of one push operation through edge $(u,v)$ from level $\ell$ to $\ell+1$.
	%So $C_\ell(u,v)$ equals the frequency of the backward push from node u to v. 
	If $\frac{(1-\alpha)\epi_{\ell}(v,t)}{ d_{out}(u)} \ge \frac{\alpha \theta}{\lambda(u)}$, the push operation will be guaranteed once. 
	Otherwise, the push happens with probability $\frac{\lambda(u) \cdot (1-\alpha)\epi_{\ell}(v,t)}{\alpha \theta \cdot d_{out}(u)}$.
	Hence, we can derive the expectation of $C_{\ell+1}(u,v)$ conditioned on the value of all estimators $\{ \epi_{\ell} \}$ at $\ell$-th level$\{\epi_{\ell}\}$ that
	\vspace{-1mm}\begin{equation}%\nonumber
	\label{eqn:incre2_conditional_cost}
	\begin{aligned}\nonumber
	\E \left[ C_{\ell+1}(u,v) \mid \{\epi_{\ell}\} \right]
	&=\left\{
	\begin{array}{ll}
	1, \quad if \quad \frac{(1-\alpha)\epi_{\ell}(v,t)}{d_{out}(u)} \ge \frac{\alpha \theta}{\lambda(u)},\\
	1 \cdot \frac{\lambda(u) \cdot (1-\alpha)\epi_{\ell}(v,t)}{\alpha \theta \cdot d_{out}(u)}, \quad otherwise.
	\end{array} 
	\right.\\
	%&\leq \frac{\lambda(u) \cdot (1-\alpha)\epi_{\ell}(v,t)}{\alpha \theta \cdot d_{out}(u)}.
	\end{aligned}
	\end{equation}
	Note that if $\frac{(1-\alpha)\epi_{\ell}(v,t)}{d_{out}(u)} \ge \frac{\alpha \theta}{\lambda(u)}$,  
	the conditional expectation $\E \left[ C_{\ell+1}(u,v) \mid \{\epi_{\ell}\} \right]$ satisfies that $\E \left[ C_{\ell+1}(u,v) \mid \{\epi_{\ell}\} \right]=1 \leq \frac{\lambda(u) \cdot (1-\alpha)\epi_{\ell}(v,t)}{\alpha \theta \cdot d_{out}(u)}$ .
	Thus, we can derive that $\E \left[ C_{\ell+1}(u,v) \mid \{\epi_{\ell}\} \right] \leq \frac{\lambda(u) \cdot (1-\alpha)\epi_{\ell}(v,t)}{\alpha \theta \cdot d_{out}(u)}$ always holds.
	\begin{comment}
	\vspace{-1mm}\begin{equation}\nonumber
	%\label{eqn:incre_vbs1}
	\begin{aligned}
	C_\ell(u,v)
	&=\left\{
	\begin{array}{ll}
	1 &,if \quad  \frac{(1-\alpha)\epi_\ell(u,t)}{\alpha d_{out}(v)} \ge \theta\\
	\left\{
	\begin{array}{ll}
	1 & w.p \quad \frac{(1-\alpha)\epi_\ell(u,t)}{\alpha \theta d_{out}(v)}  \\
	0 &w.p \quad 1-\frac{(1-\alpha)\epi_\ell(u,t)}{\alpha \theta d_{out}(v)} 
	\end{array} 
	\right.
	&,otherwise
	\end{array} 
	\right.
	\end{aligned}
	\end{equation}
	\end{comment}
	Applying the unbiasedness of $\epi_{\ell}(v,t)$ according to Lemma~\ref{lem:vbs2_unbiasedness}, we have
	\begin{align}\nonumber
	\E \left[ C_{\ell+1}(u,v) \right]=\E \left[ \E \left[C_{\ell+1}(u,v) \mid \{\epi_{\ell}\}\right] \right] \leq \frac{\lambda(u) \cdot (1-\alpha)\pi _{\ell}(v,t)}{\alpha \theta \cdot d_{out}(u)}.
	\end{align}
	Recall that $C_{total}$ denotes the total cost in the whole process and $C_{total}=\sum_{i=1}^{L} \sum_{u \in V} \sum_{v \in N_{out}(u)} C_i(u,v)$. 
	The expectation of $C_{total}$ can be derived that 
	%Utilizing the fact that $C_{total}=\sum_{\ell=1}^{L} \sum_{u \in V} \sum_{v \in N_{out}(u)} C_\ell(u,v)$,
	%we can derive the expectation of $C_{total}$ as below.
	\vspace{-1mm}\begin{equation}\nonumber
	%\label{eqn:incre_vbs1}
	\begin{aligned}
	&\E \left[C_{total}\right]
	%=\E \left[\sum_{\ell=0}^{\infty} \sum_{v \in V} \sum_{u \in N_{out}(v)} C_\ell(u,v)\right]\\
	=\sum_{i=1}^{L} \sum_{u \in V} \sum_{v \in N_{out}(u)} \E\left[C_{i}(u,v)\right]\\
	&\hspace{-1mm}\leq \sum_{i=1}^{\infty} \sum_{u \in V} \hspace{-1mm}\sum_{v \in N_{out}(u)} \hspace{-2mm} \frac{\lambda(u) \cdot (1-\alpha)\pi _{i-1}(v,t)}{\alpha \theta \cdot d_{out}(u)}
	%= \sum_{\ell=0}^{\infty} \sum_{v \in V} \frac{1}{\alpha \theta } \cdot \pi _{\ell}(v,t)\\
\hspace{-1mm}	= \hspace{-1mm} \frac{1}{\alpha \theta} \hspace{-0.7mm}\cdot \hspace{-0.7mm}\sum_{u \in V} \lambda(u) \hspace{-0.7mm} \cdot \hspace{-0.7mm}\sum_{i=1}^{\infty} \pi _{i}(u,t).
	\end{aligned}
	\end{equation}
	According to the property of $\ell$-hop PPR that $\sum_{i=0}^{\infty} \pi _{i}(u,t)=\pi (u,t)$,
	%the above equation can be simplified as
	\begin{align}\nonumber
	\E \left[C_{total}\right] \leq \frac{1}{\alpha \theta} \sum_{u \in V} \lambda(u) \cdot \pi (u,t),
	\end{align}
        which proves the lemma.
\end{proof}

%\vspace{-2 mm}
\subsection{Proof of Theorem~\ref{thm:relative}}
\begin{proof}
We first show that by truncating at the $L = \log_{1-\alpha} \theta$
hop, we only introduce an additive error of $\theta$. More precisely,
note that $\sum_{i=L+1}^{\infty}\alpha(1-\alpha)^i \leq
(1-\alpha)^{(L+1)} \leq \theta$. By setting a $\theta$ that is
significantly smaller than the relative error threshold $\delta$ or
the additive error bound $\e$, we can accomodate the $\theta$ additive
error without increasing the asymptotic query time.

	According to Lemma~\ref{lem:vbs2_variance}, we have $\Var \left[ \epi_{\ell}(s,t) \right] \leq \theta \pi_{\ell}(s,t)$.
        By Chebyshev inequality, we have
        $$\Pr\left[ \left| \epi_{\ell}(s,t)  - \pi_\ell(s,t )\right|
          \ge \sqrt{3\theta \pi_{\ell}(s,t)} \right] \le 1/3. $$
        We claim that this variance implies an $\e_r$-relative error
        for all $ \pi_{\ell}(s,t)  \ge 3\theta /\e_r^2$. For a proof,
        note that $ \theta  \le \e_r^2
        \pi_{\ell}(s,t)  /3$ and consequently $ \sqrt{3\theta
          \pi_{\ell}(s,t)}  \le  \sqrt{ \e_r^2
          \pi_{\ell}(s,t)^2} =  \e_r  \pi_{\ell}(s,t)$. It follows
        that $\Pr\left[ \left| \epi_{\ell}(s,t)  - \pi_\ell(s,t )\right|
          \ge \e_r  \pi_{\ell}(s,t) \right] \le 1/3$ for all $
        \pi_{\ell}(s,t)  \ge 3\theta /\e_r^2$.  By setting $\theta =
       { \e_r^2\delta \over 3L}$, we obtain a constant relative error
        guarantee for all $\pi_{\ell}(s,t) \ge \delta /L$, and
        consequently a constant relative error for  $\pi(s,t) \ge \delta $.  To obtain a high
        probability result, we can apply the Median-of-Mean
        trick~\cite{charikar2002finding}, which takes the median of $O(\log n)$ independent
        copies of $\epi_{\ell}(s,t) $ as the final estimator to
        $\pi_{\ell}(s,t)$. This trick brought the failure probability
        from $1/3$ to $1/n^2$ by increasing the running time by a
        factor of $O(\log n)$. Applying the union bound to $n$ source
        nodes $s\in V$ and $\ell = 0,\ldots, L$, the failure
        probability becomes $1/n$. Finally, by setting $\lambda(u) =1$
        in 
        Lemma~\ref{lem:vbs2_cost}, we can rewrite the time cost as
        below. 
	\begin{align}\nonumber
	\E \left[C_{total}\right] \leq \frac{1}{\alpha \theta} \sum_{u \in V} \lambda(u) \cdot \pi (u,t) = \frac{1}{\alpha \theta} \sum_{u \in V}\pi (u,t) = \frac{n \pi(t)}{\alpha \theta},
	\end{align}
	where $\pi(t)$ represents $t$'s PageRank and $n\pi (t)=\sum_{u
          \in V} \pi _{\ell}(u,t)$ according to PPR's definition. By
        setting $\theta =
       { \e_r^2\delta \over 3L}$ and running $O(\log n)$ independent
       copies of Algorithm~\ref{alg:vbs2}, the time complexity
       can be bounded by $O\left(  \frac{n \pi(t) L \log n}{\alpha
           \e_r^2 \delta}\right) = \tilde{O} \left( \frac{n \pi(t)}{\delta}\right)$.        
	If we choose the target node $t$ uniformly at random from set
        $V$, then $\E \left[\pi(t)\right]=\frac{1}{n}$, and the
        running time becomes $\tilde{O} \left(\frac{1}{ \delta} \right)$.

\end{proof}

%\vspace{-3 mm}
\subsection{Proof of Theorem~\ref{thm:additive}}
\begin{proof}
  Applying  Lemma~\ref{lem:vbs2_variance}, we have $\Var \left[
    \epi_{\ell}(s,t) \right] \leq \alpha \theta^2$. Consequently, we
  have 
$\Var \left[
    \epi(s,t) \right] = \Var \left[
    \sum_{\ell=0}^{L} \epi_{\ell}(s,t) \right]  \le \alpha L
  \theta^2. $ 
By Chebyshev's
  inequality, we have
  $\Pr\left[ \left| \epi(s,t)  - \pi(s,t )\right|
          \ge \sqrt{3L\alpha}\theta \right] \le 1/3. $
        By setting $\theta = \e/\sqrt{3L\alpha}$, it follows that
        $\epi(s,t)$ is an $\e$ additive error
        for all $ \pi(s,t)$. Similar to the proof of
        Theorem~\ref{thm:relative}, we can use the median of $O(\log
        n)$ independent copies of $\epi(s,t)$ as the estimator to
        reduce the failure probability from $1/3$ to $1/n$ for all
        source nodes $s\in V$. 

	For the time cost, Lemma~\ref{lem:vbs2_cost} implies that 
	\begin{align}\nonumber
	\E \left[C_{total}\right] \leq \frac{1}{\alpha \theta} \sum_{u \in V} \lambda(u) \cdot \pi (u,t)= \frac{1}{\alpha \theta} \sum_{u \in V} \sqrt{d_{out}(u)} \cdot \pi (u,t).
	\end{align}
Recall that we set $\theta = \e/\sqrt{3L\alpha}$ and run $O(\log n)$
independent copies of Algorithm~\ref{alg:vbs2}, it follows that the running time can
be bounded by $\tilde{O} \left(\frac{1}{\e} \sum_{u \in V}
  \sqrt{d_{out}(u)} \cdot \pi (u,t) \right)$. 
	If $t$ is chosen uniformly at random, we have  $\sum_{t \in V}
        \pi (u,t)=1$. Ignoring the $\tilde{O}$ notation, we have
	\vspace{-1mm}\begin{equation}
	\begin{aligned}\nonumber
	& \E \left[C_{total}\right] \leq  \frac{1}{ \e} \cdot \frac{1}{n} \cdot \sum_{t \in V} \sum_{u \in V} \sqrt{d_{out}(u)} \cdot \pi (u,t) \\
	& = \frac{1}{\e} \cdot \frac{1}{n} \cdot \sum_{u \in V} \sqrt{d_{out}(u)} \sum_{t \in V} \pi (u,t) = \frac{1}{\e} \cdot \frac{1}{n} \cdot \sum_{u \in V} \sqrt{d_{out}(u)} .
	\end{aligned}
	\end{equation}
	By the AM-GM inequality, we have $	\frac{1}{n} \cdot \sum_{u \in V} \sqrt{d_{out}(u)} \leq \sqrt{\frac{\sum_{u \in V} d_{out}(u)}{n}}=\sqrt{\bar{d}}$,
	% \begin{align}
	% \frac{1}{n} \cdot \sum_{u \in V} \sqrt{d_{out}(u)} \leq \sqrt{\frac{\sum_{u \in V} d_{out}(u)}{n}}=\sqrt{d}
	% \end{align}
	Hence, $\E \left[C_{total}\right] \leq \frac{1}{\e} \cdot \frac{1}{n} \cdot \sum_{u \in V} \sqrt{d_{out}(u)} \leq \frac{\sqrt{\bar{d}}}{\e},$
	% \vspace{-1mm}\begin{equation}
	% \begin{aligned}
	% & \E \left[C_{total}\right] \leq \frac{1}{\e} \cdot \frac{1}{n} \cdot \sum_{u \in V} \sqrt{d_{out}(u)} \leq \frac{\sqrt{d}}{\e},
	% \end{aligned}
	% \end{equation}
	and the theorem follows.
\end{proof}

%%% Local Variables:
%%% mode: latex
%%% TeX-master: "paper"
%%% End:

\end{document}